\RequirePackage{fix-cm}
\documentclass[smallextended]{svjour3}       

\smartqed 
\usepackage{listings}
\usepackage{verbatim}
\usepackage[skip=1pt,labelfont=bf]{caption}
\usepackage{xcolor} 
\usepackage{multirow}
\usepackage{rotating}
\usepackage{algorithm}

\usepackage{amsmath,amsfonts}

\usepackage{graphicx}
\usepackage{cite}

\usepackage{amsmath, amsfonts}

\usepackage{amsmath,amsfonts}
\usepackage{amssymb}
\usepackage{array}
\usepackage{balance}

\usepackage{calligra}
\usepackage{color}
\usepackage{colortbl}
\usepackage{courier}
\usepackage{csvsimple}
\usepackage{enumitem}
\usepackage{fancybox}
\usepackage{fontenc}
\usepackage{longtable}
\usepackage{lscape}
\usepackage{moreverb}
\usepackage{multicol}
\usepackage{multirow}
\usepackage{pifont}
\usepackage{rotating}
\usepackage{setspace}
\usepackage{subfigure}

\usepackage{booktabs}
\usepackage{makecell}
\usepackage[most]{tcolorbox}
\usepackage{threeparttable}
\usepackage{tikz}
\usepackage[normalem]{ulem}
\usepackage{url}
\usepackage{soul}
\usepackage{xspace}
\usepackage{lineno}
\usepackage{pifont}

\usepackage{natbib}

\usepackage{etoolbox}
\usepackage{bbding}
\usepackage{fontawesome5} 
\usepackage[misc]{ifsym}

\usepackage{cite}
\usepackage{amsmath,amssymb,amsfonts}
\usepackage{graphicx}
\usepackage{textcomp}

\usepackage{algpseudocode}

\def\BibTeX{{\rm B\kern-.05em{\sc i\kern-.025em b}\kern-.08em
    T\kern-.1667em\lower.7ex\hbox{E}\kern-.125emX}}
\usepackage{graphicx}
\usepackage{tabularx}
\usepackage{enumitem}
\usepackage{siunitx} 
\usepackage{array} %
\newcolumntype{P}[1]{>{\centering\arraybackslash}p{#1}} %
\usepackage{comment}
\usepackage{xspace}
\usepackage{tcolorbox}
\usepackage{url}

\usepackage[hyperfigures,breaklinks,colorlinks,linkcolor=blue,citecolor=blue,urlcolor=blue]{hyperref}

\makeatletter

\patchcmd{\NAT@citex}
  {\@citea\NAT@hyper@{%
     \NAT@nmfmt{\NAT@nm}%
     \hyper@natlinkbreak{\NAT@aysep\NAT@spacechar}{\@citeb\@extra@b@citeb}%
     \NAT@date}}
  {\@citea\NAT@nmfmt{\NAT@nm}%
   \NAT@aysep\NAT@spacechar\NAT@hyper@{\NAT@date}}{}{}

\patchcmd{\NAT@citex}
  {\@citea\NAT@hyper@{%
     \NAT@nmfmt{\NAT@nm}%
     \hyper@natlinkbreak{\NAT@spacechar\NAT@@open\if*#1*\else#1\NAT@spacechar\fi}%
       {\@citeb\@extra@b@citeb}%
     \NAT@date}}
  {\@citea\NAT@nmfmt{\NAT@nm}%
   \NAT@spacechar\NAT@@open\if*#1*\else#1\NAT@spacechar\fi\NAT@hyper@{\NAT@date}}
  {}{}

\makeatother

\definecolor{darkpastelred}{rgb}{0.76, 0.23, 0.13}
\definecolor{ao(english)}{rgb}{0.0, 0.5, 0.0}
\definecolor{darkpastelred}{rgb}{0.76, 0.23, 0.13}
\definecolor{ao(english)}{rgb}{0.0, 0.5, 0.0}
\definecolor{yellow}{RGB}{255,255,153}
\definecolor{grey}{RGB}{224,224,224}

\newcommand{\highlight}[1]{\begin{tcolorbox}[leftrule=0mm,rightrule=0mm,toprule=0mm,bottomrule=0mm,left=2pt,right=2pt,top=2pt,bottom=2pt]
  #1
  \end{tcolorbox}
}

\newboolean{showcomments}
\setboolean{showcomments}{true}
\ifthenelse{\boolean{showcomments}}
 { \newcommand{\mynote}[2]{
      \fbox{\bfseries\sffamily\scriptsize#1}
        {\small$\blacktriangleright$\textsf{\emph{#2}}$\blacktriangleleft$}}}
        { \newcommand{\mynote}[2]{}}
        
\definecolor{DarkOrange}{rgb}{0.8,0.3,0.0}
\definecolor{DarkCyan}{rgb}{0.0, 0.55, 0.55}
\definecolor{DarkCyel}{rgb}{1.0, 0.49, 0.0}
\definecolor{yellow-green}{rgb}{0.6, 0.8, 0.2}

\newcolumntype{?}{!{\vrule width 1pt}}

\newcommand{\name}{\textsc{BRMiner}\xspace}

\begin{document}

\title{Enriching Automatic Test Case Generation by Extracting Relevant Test Inputs from Bug Reports}

\author{{Wendkûuni C. Ouédraogo} \textsuperscript{1}        \and \\
        {Laura Plein} \textsuperscript{1}                   \and
        {Kader Kaboré} \textsuperscript{1}                  \and
        {Andrew Habib} \textsuperscript{1}                  \and
        {Jacques Klein} \textsuperscript{1}                 \and \\
        {David Lo} \textsuperscript{2}                      \and 
        {Tegawend\'e F. Bissyand\'e} \textsuperscript{1}
}

\institute{
    \begin{itemize}
        \item[\Letter] {Wendkûuni C. Ouédraogo} \\
                \email{wendkuuni.ouedraogo@uni.lu} \\
        \item[] {Laura Plein} \\
                \email{laura.plein@uni.lu} \\
        \item[] {Kader Kaboré} \\
                \email{abdoulkader.kabore@uni.lu} \\
        \item[] {Andrew Habib} \\
                \email{andrew.a.habib@gmail.com} \\
        \item[] {Jacques Klein} \\
                \email{jacques.klein@uni.lu} \\
        \item[] {David Lo} \\
                \email{davidlo@smu.edu.sg} \\
        \item[] {Tegawend\'e F. Bissyand\'e} \\
                \email{tegawende.bissyande@uni.lu} \\ 
        \at
        \item[\textsuperscript{1}] SnT Centre, University of Luxembourg
        \item[\textsuperscript{2}] Singapore Management University
    \end{itemize}
}

\date{Received: date / Accepted: date}

\maketitle
\begin{abstract} 
The quality of software is closely tied to the effectiveness of the tests it undergoes. Manual test writing, though crucial for bug detection, is time-consuming, which has driven significant research into automated test case generation. However, current methods often struggle to generate relevant inputs, limiting the effectiveness of the tests produced. To address this, we introduce \name, a novel approach that leverages Large Language Models (LLMs) in combination with traditional techniques to extract relevant inputs from bug reports, thereby enhancing automated test generation tools. In this study, we evaluate \name using the Defects4J benchmark and test generation tools such as EvoSuite and Randoop. Our results demonstrate that \name achieves a Relevant Input Rate (RIR) of 60.03\% and a Relevant Input Extraction Accuracy Rate (RIEAR) of 31.71\%, significantly outperforming methods that rely on LLMs alone. The integration of BRMiner's input enhances EvoSuite ability to generate more effective test, leading to increased code coverage, with gains observed in branch, instruction, method, and line coverage across multiple projects. Furthermore, \name facilitated the detection of 58 unique bugs, including those that were missed by traditional baseline approaches. Overall, \name's combination of LLM filtering with traditional input extraction techniques significantly improves the relevance and effectiveness of automated test generation, advancing the detection of bugs and enhancing code coverage, thereby contributing to higher-quality software development. 
\end{abstract}


\section{Introduction}
\label{sec:intro}
Software testing is as old as the software industry. It requires test suites, which are often incomplete since developer-provided test cases are typically non-exhaustive as they are not systematically produced in most projects for the whole code~\citep{kochhar2015understanding,kochhar2013empirical,kochhar2013adoption}. Thus, automated software test generation has early attracted significant attention in the research and practice communities due to its numerous potential benefits, including the reduction of manual effort and the increase of test coverage~\citep{ASE-2017-ToffolaSP}. To that end, a large body of literature has proposed several techniques for the automatic generation of test inputs, including Search-Based Software Testing (SBST)~\citep{5342440}, feedback-directed random test generation~\citep{pacheco2007feedback,artzi2011framework,pradel2012fully}, Dynamic Symbolic Execution (DSE)~\citep{king1976symbolic,godefroid2005dart,xie2005symstra,majumdar2007directed,cadar2013symbolic}, and concolic execution~\citep{sen2005cute,cadar2008exe,sen2006cute}. While the tools based on those techniques have shown effectiveness, they often face limitations that hinder their adoption in real-world projects.

For example, while prior research has underscored the efficacy of Search-Based Software Testing (SBST) techniques in attaining notable code coverage~\citep{panichella2015reformulating, panichella2017automated}, \citet{perera2020defect} argue that high code coverage alone does not guarantee the optimal detection of bugs. In line with this, \citet{shamshiri2015automatically} and \citet{almasi2017industrial} assessed the performance of EvoSuite~\citep{fraser2011evosuite}, a leading SBST tool, against alternative test generation methods and tools like Randoop~\citep{pacheco2007feedback}, focusing on open-source and proprietary software projects, respectively. Their findings suggest a superior performance by EvoSuite over its counterparts, but also demonstrate that it falls short in effectively detecting genuine bugs. Notably, EvoSuite could identify only about 23\% of bugs from the Defects4J dataset~\citep{just2014defects4j}, even when the testing criterion was set at 100\% branch coverage.

One of the main reasons behind this limitation is that these generators primarily rely on either randomly-generated inputs or inputs from a fixed pool, which may not match the project specifications~\citep{ASE-2017-ToffolaSP}. Additionally, test generators tend to focus on the source code and its dynamic execution, generating less natural and readable values~\citep{shelke2014generation}.

To address these challenges, researchers have proposed various approaches: Some have leveraged test corpora~\citep{ASE-2017-ToffolaSP}, while others have explored web search queries~\citep{shelke2014generation,mcminn2012search,shahbaz2012automated}, user session data~\citep{elbaum2003improving}, web services~\citep{bozkurt2011automatically}, or the web of data~\citep{mariani2014link} to generate realistic test inputs. Regarding mobile application testing, \citet{liu2017automatic} leveraged the Google news corpus. In~\citep{milani2014leveraging}, researchers leveraged existing tests to automate test generation for web applications. Nevertheless, despite the existence of these approaches, the potential of bug reports as a valuable source of relevant inputs, including strings for test data generation, remains largely untapped.

Researchers have previously explored extracting test cases from bug reports, such as analyzing multicore dumps to facilitate concurrency bug reproduction~\citep{weeratunge2010analyzing}, reproducing system-level concurrency failures~\citep{yu2017descry}, translating bug reports into test cases for mobile apps~\citep{fazzini2018automatically}, and enriching compiler testing with real programs from bug reports~\citep{zhong2022enriching}. Our work builds on these foundations by introducing a new angle: we propose \name, a technique that mines literal inputs from bug reports to improve the relevance and effectiveness of automated test case generation.

Unlike previous studies, which often focus on generating test cases as a whole, our approach emphasizes the extraction of specific input values from bug reports, which are crucial for enhancing test coverage and bug detection capabilities. Furthermore, to address the complexity and ambiguity often present in bug reports, \name incorporates a filtering mechanism using a Large Language Model (LLM), specifically GPT-3.5-turbo\footnote{https://platform.openai.com/docs/models/gpt-3-5-turbo/[accessed 2024-05-21]}. This LLM component is capable of understanding the semantic nuances of bug reports, allowing it to refine the extracted literals and improve their relevance for triggering bugs during test execution~\citep{fan2023large, wang2024software}.

Additionally, \name integrates seamlessly with the EvoSuite test case generation tool, allowing for the direct incorporation of these LLM-refined literals into automatically generated test cases. This integration not only enhances the traditional test generation process by ensuring more contextually relevant inputs but also addresses challenges in test oracle generation by increasing the likelihood that the generated test cases will expose latent defects~\citep{liu2024testing}.

\textbf{This paper.} Bug reports offer a diverse range of example inputs which remain largely untapped in the extraction of inputs for automatic test case generation. Leveraging bug reports for this purpose offers two significant advantages. Firstly, they provide a wealth of valid inputs for various string types, enhancing test coverage and bug detection. Secondly, bug reports typically contain human-readable inputs, which prove invaluable when manual confirmation is necessary due to the absence of an automated oracle.

In this study, we propose a simple, yet effective approach for automatically extracting real test inputs from bug reports, aimed at addressing the problem of automatic test case generation containing relevant test inputs. Our approach involves exploring bug reports to identify and extract relevant inputs found in bug reports. We developed a tool called \name that enables us to automatically extract these relevant inputs, both qualitatively and quantitatively. To further enhance the effectiveness of the extracted inputs, we integrate a filtering step using GPT-3.5-turbo, an LLM capable of understanding the semantic context of bug reports and refining the inputs accordingly~\citep{tsigkanos2023variable, yang2023white}.

To generate test cases incorporating these inputs, we modified the source code of EvoSuite~\citep{fraser2011evosuite}, a widely known search-based test suite generator tool used for automatic test case generation. Additionally, we extended our evaluation by incorporating Randoop~\citep{pacheco2007feedback}, a feedback-directed random testing tool, to compare the effectiveness of test cases generated with \name's extracted inputs. To evaluate our generated test cases, we use the Defects4J benchmark~\citep{just2014defects4j}.

The main contributions of this paper are as follows:
\begin{enumerate}
\item We present \name, an automatic test input extraction approach. This approach offers several advantages, such as the ability to efficiently extract realistic test inputs for testing projects with semantic information, and the improvement of the efficiency of automatic test case generation for inputs that cannot be effectively generated randomly.
\item We showcase findings from practical case studies conducted on Defects4J projects. These were designed to assess the efficacy of our methodology in generating test cases based on bug reports, in contrast to the baseline approach of EvoSuite and Randoop which operate without external inputs.
\item We emphasize the significance of integrating external inputs into EvoSuite and Randoop, particularly inputs extracted from bug reports. For EvoSuite, we modified its source code to enable the use of external inputs, as it does not support this feature by default. In contrast, Randoop inherently supports external input files, so we directly fed it with the extracted inputs. These inputs offer crucial context, bolstering the performance of both tools, particularly when used in conjunction with Dynamic Symbolic Execution (DSE) for EvoSuite.
\item We demonstrate the effectiveness of our approach by showcasing the finding of valid, well-formed inputs from bug reports, resulting in improved bug detection and enhanced test coverage for a program's source code. To foster further research in this area, we have publicly shared the implementation of our approach and all relevant experimental data in an anonymous repository. The repository is accessible via the following link:
\begin{center}
    \url{https://anonymous.4open.science/r/BRMiner-C853/}
\end{center}
\end{enumerate}

The rest of this paper is organized as follows. Section~\ref{sec:brminer} presents our approach for relevant input extraction from bug reports. Section~\ref{sec:setup} provides an overview of our experimental setup, including implementation details, evaluation of technical challenges, and presentation of results (Section~\ref{sec:results}). Section~\ref{sec:discussion} discusses existing research on test input extraction and generation in the field of automatic test generation, along with the limitations of our approach and future research directions. Finally, Section~\ref{sec:conclusion} summarizes the paper, draws conclusions, and mentions future work.

\section{Background}
\label{sec:background}
This section presents the foundational concepts and methodologies that support our approach, including Dynamic Symbolic Execution (DSE), EvoSuite, LLM-based test input generation, and prompt engineering techniques.

\subsection{DSE and EvoSuite}
\label{sec:dse-evosuite}
Dynamic Symbolic Execution (DSE) is among the widely adopted techniques for automated test case generation~\citep{valle2022mutation}. This method simultaneously explores multiple paths of the source code using symbolic values in place of concrete ones. Employing a symbolic execution engine enables efficient control over the paths traversed within the source code~\citep{baldoni2018survey}. Its applications include the automated generation of unit tests and the identification of actual defects or memory overflows~\citep{cadar2013symbolic}.

While Search-based testing excels at formulating unit test suites for object-oriented code, it occasionally falters in producing specific values for intricate code parts. In contrast, Dynamic Symbolic Execution (DSE) shines in creating these precise values but finds complexity when faced with data types necessitating a chain of calls. \citet{galeotti2013improving} paved the way for a combined technique, leveraging the advantages of both methods. They refined the Genetic Algorithm (GA) within EvoSuite to smoothly incorporate DSE and, guided by search feedback, determined the optimal moments to employ DSE. This modification led to a jump in average branch coverage. Furthermore, when compared to standalone DSE tools, EvoSuite combined with DSE proved superior in achieving coverage. Empirical analyses underscored the practical advantages of integrating SBST and DSE, with negligible downsides. 

In subsequent research, \citet{galeotti2014extending} highlighted that this fusion not only augments coverage and effectiveness but also upholds SBST's flexibility in tailoring test suites for various metrics and non-functional factors. Crucially, this approach ensures that all test cases correspond to legitimate object conditions. As a testament to its capabilities, EvosuiteDSE made a notable debut at the ninth unit testing competition during SBST 2021, registering a commendable score of 47.14.

\subsection{LLM-based Test Input Generation}
\label{sec:llmbased-test-input}
LLMs have recently gained traction in the field of software testing, particularly in generating or refining test inputs that are both contextually relevant and semantically aligned with the software under test~\citep{fan2023large,wang2024software}. These models, built primarily on the Transformer architecture~\citep{vaswani2017attention}, are trained on vast corpora of text and code using techniques such as Masked Language Modeling and Causal Language Modeling~\citep{shanahan2024talking,naveed2023comprehensive}. The resulting models, including BERT~\citep{devlin2018bert}, GPT~\citep{brown2020language,yenduri2024gpt}, and T5~\citep{raffel2020exploring}, have been enhanced with methods like Reinforcement Learning from Human Feedback (RLHF)~\citep{ziegler2019fine,bai2022training}, which improves their alignment with human intent.

In the context of test input generation, LLMs have demonstrated significant potential in generating diverse and realistic inputs~\citep{liu2024testing,yang2023white}. These models can analyze natural language descriptions, such as bug reports or user stories, to extract meaningful inputs that can be used to trigger specific software behaviors~\citep{tsigkanos2023variable}. By leveraging the extensive knowledge embedded in LLMs, these inputs are not only syntactically correct but also semantically relevant, thereby improving the effectiveness of test cases in uncovering bugs.

Recent advancements in the field have seen LLMs being integrated with traditional testing tools to enhance the coverage and precision of generated test cases. For instance, in the \name approach, an LLM like GPT-3.5-turbo is leveraged to refine extracted inputs from bug reports, ensuring that they are contextually appropriate and likely to trigger relevant software faults. This refinement process is crucial for improving the overall quality of the test cases and increasing the likelihood of exposing latent defects that might be missed by conventional test generation methods.

\subsection{Prompt Engineering}
\label{sec:prompt-engineering}
Prompt engineering involves creating prompts that enable LLMs to effectively address specific tasks~\citep{amatriain2024prompt,siddiq2024quality,vogelsang2024using,naveed2023comprehensive}. This technique is particularly valuable in complex problem-solving scenarios where a single line of reasoning might not be sufficient~\citep{amatriain2024prompt,kojima2022large,wei2022chain,reynolds2021prompt}. By guiding LLMs to consider multiple possibilities before reaching a conclusion, prompt engineering allows these models to adopt a more human-like approach to reasoning.

Prompt engineering is a rapidly evolving field that focuses on crafting prompts to guide LLMs in generating desired outputs, ranging from broad descriptions to detailed specifications. The importance of prompt engineering is twofold: firstly, highly specific prompts yield precise responses, while more general prompts may produce less relevant outputs~\citep{si2022prompting}. Secondly, by carefully adjusting the instructions within prompts, it is possible to control the output format of LLMs, thereby minimizing the need for post-processing and improving the performance evaluation~\citep{macedo2024exploring}.

Various studies~\citep{fan2023large,wang2024software} have leveraged LLMs and prompt engineering for automated software engineering tasks, traditionally employing approaches such as Zero-shot~\citep{kojima2022large} and Few-shot learning~\citep{brown2020language}. However, recent advancements have broadened the scope of prompt engineering to include more sophisticated techniques such as Chain-of-Thought~\citep{wei2022chain} and Tree-of-Thoughts~\citep{yao2024tree, long2023large}.

\textbf{Zero-shot learning} enables models to generalize to new tasks without specific examples but may falter with complex tasks~\citep{sahoo2024systematic,kojima2022large}. \textbf{Few-shot learning}, or \textbf{in-context learning}, involves presenting the model with a few examples to improve its performance on novel tasks~\citep{brown2020language,reynolds2021prompt}. \textbf{Chain-of-Thought} (CoT), introduced by Wei et al.~\citep{wei2022chain}, structures the model's reasoning into intermediate steps, which can be enhanced by combining CoT with few-shot prompts for complex reasoning tasks. 

\textbf{Tree-of-Thoughts} (ToT), proposed by Yao et al.~\cite{yao2024tree} and ~\cite{long2023large}, builds on Chain-of-Thought by employing a systematic exploration of a "tree" of reasoning steps. This method integrates search algorithms, allowing the model to explore different branches of reasoning more strategically, which is particularly useful for solving complex problems.

In the context of \name, we used the Tree-of-Thoughts (ToT) technique to enhance the refinement of inputs extracted from bug reports. By guiding GPT-3.5-turbo through a structured exploration of possible inputs and their relevance to the reported bug, ToT helps in generating more contextually accurate and useful test inputs. This advanced prompt engineering method enables \name to better align LLM behavior with the specific demands of test input generation, leading to more effective and targeted test cases.
\section{Bug reports and relevant input}
\label{sec:bugreportsandrelevantinputs}
This section explores the significance of bug reports in software testing and defines the concept of relevant inputs within the context of these reports, highlighting their critical role in enhancing automated test case generation.

\subsection{Bug reports}
\label{subsec:valueofbugreports}
Bug reports are integral to the software development lifecycle, serving as essential artifacts that document software issues, their context, and potential resolutions. These reports often provide a wealth of detailed information, including the observed behavior, the specific conditions under which the issue arose, and, crucially, the inputs that triggered the reported bugs. Typically, a bug report includes a thorough description of the problem, outlining the steps required to reproduce it, the environment in which the issue occurred, and sometimes the expected versus actual results. Moreover, bug reports may include supplementary materials such as code snippets, stack traces, and user logs, offering developers a comprehensive perspective on the problem and facilitating a deeper understanding necessary for effective debugging and resolution.

\subsection{Relevant inputs} 
\label{subsec:rel-inputs}
Relevant inputs are defined by several key characteristics, as highlighted in various studies~\citep{ASE-2017-ToffolaSP,shahbaz2012automated,mcminn2012search,mariani2014link,bozkurt2011automatically,liu2017automatic}. These inputs bypass domain-specific sanity checks, contribute to branch coverage, and are both syntactically and semantically valid. They are realistic, context-appropriate, and crucially, they facilitate effective bug detection.

\begin{figure*}[!ht]
    \centering
    \flushleft
    \subfigure[Bug report]{
    \includegraphics[scale=0.65]{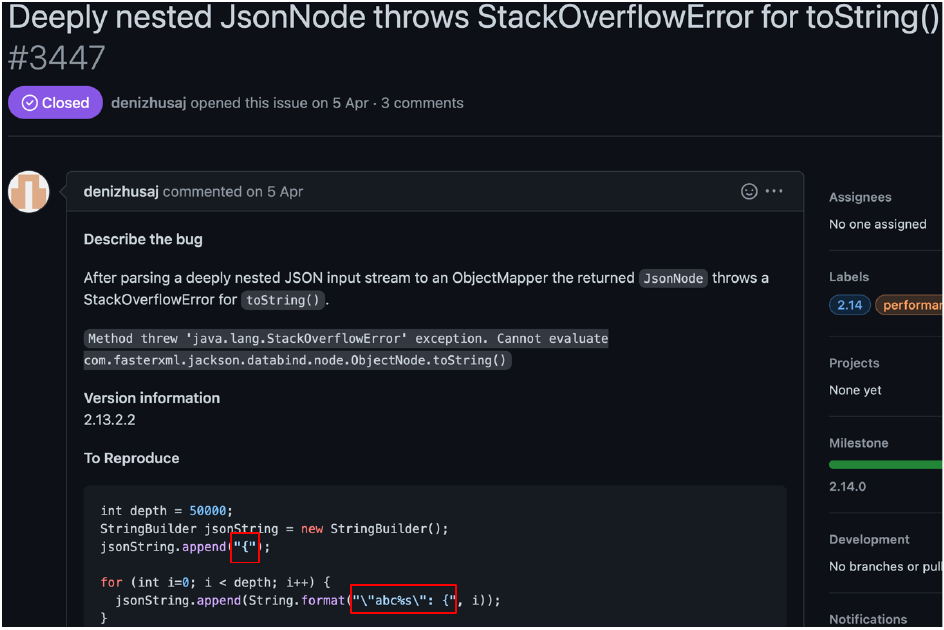}%
    }
    \subfigure[Developer-written test case]{
    \includegraphics[scale=0.65]{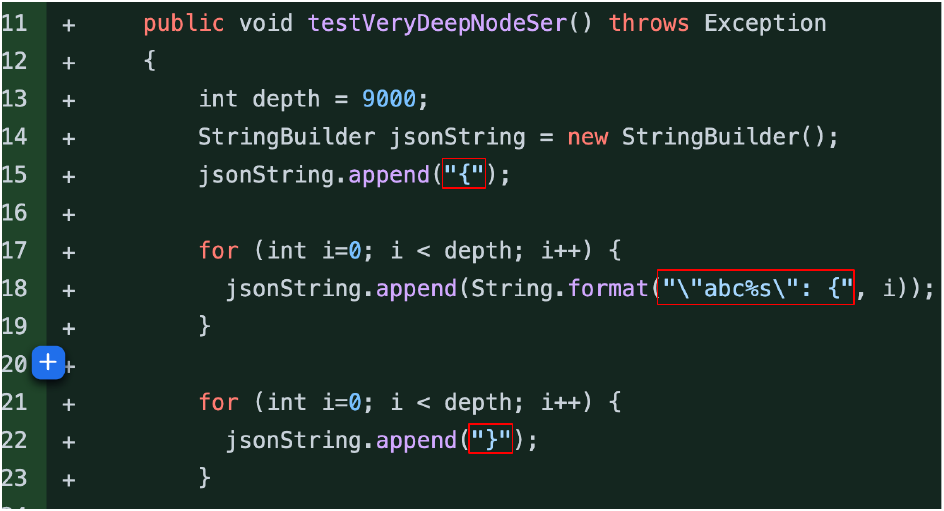}%
    }
    \caption{Input value mentioned in the bug report for issue 3447 in the FasterXML jackson-databind library appears in a test case written by a developer after fixing the bug}
    \label{fig:identic-input}
    \vspace{3mm}
\end{figure*}

\begin{figure*}[!ht]
    \centering
    \flushleft
    \subfigure[Bug report]{
    \includegraphics[scale=0.65]{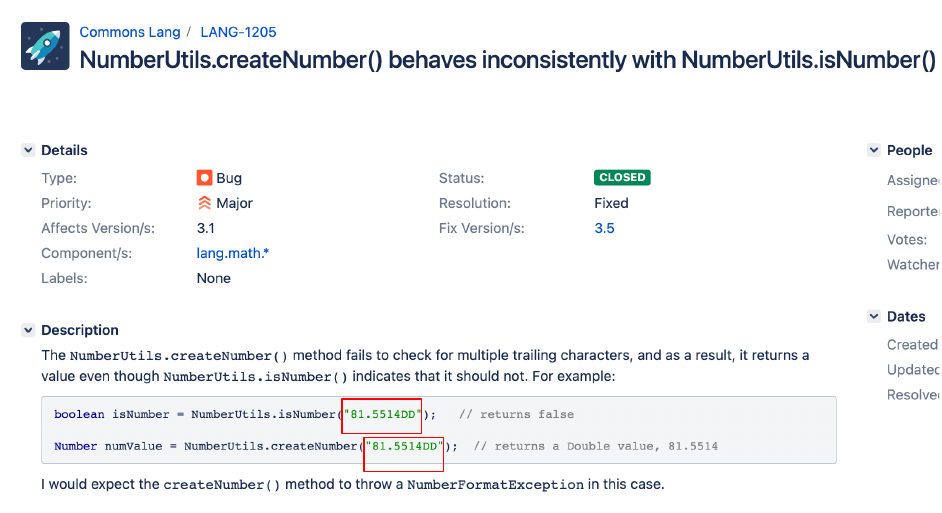}%
    }
    \subfigure[Developer-written test case]{
    \includegraphics[scale=0.65]{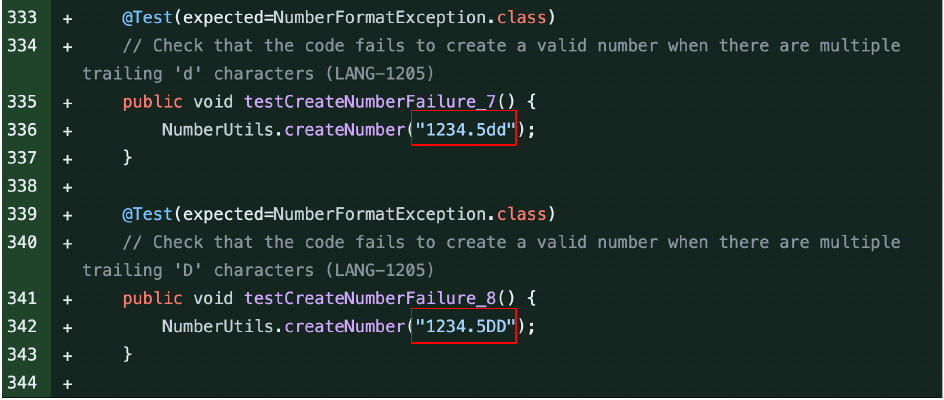}%
    }
    \caption{Pattern of the input value mentioned in the bug report for issue 1205 in the Apache Commons Lang library appears in a test case written by a developer after fixing the bug}
    \label{fig:same-patern}
    \vspace{3mm}
\end{figure*}

A deeper understanding of what constitutes relevant inputs in the context of bug reports requires thorough examination of both bug reports and human-written test cases. Through manual analysis of numerous bug reports, we observed significant variations in the information they contain. Some reports include detailed input values, APIs, expected and actual results, and even complete code segments for reproducing the issue. By comparing these reports with the corresponding human-written test cases that successfully pass after a patch, we discovered a notable overlap in the input values provided. Figure~\ref{fig:identic-input} illustrates instances where inputs from bug reports are directly reflected in the test cases. Additionally, in some cases, the input values in the test cases share a similar pattern with those in the bug reports, though they may not be identical, as shown in Figure~\ref{fig:same-patern}.



Based on these observations, we define inputs from bug reports as relevant when they contribute to the success of a human-written test case after the software has been patched. However, determining the relevance of bug report inputs that differ from those used in test cases presents a unique challenge. It's important to note that not all inputs found in bug reports are necessarily relevant, and conversely, test case inputs that differ from those in the bug report are not automatically irrelevant. To accurately assess the rate of relevant inputs in bug reports and the effectiveness of BRMiner's extraction process, our evaluation focuses specifically on inputs that are identical between the bug reports and the corresponding test cases.

\section{\name}
\label{sec:brminer}

This section introduces a usage scenario and details our approach for mining test input values from bug reports, which has been implemented in the tool \name.


\subsection{Usage scenario}
\label{subsec:usagescenario}
In the context of software testing, bug reports are invaluable resources for capturing real-world inputs that can significantly enhance the testing process. These reports often contain specific inputs that trigger software bugs, but manually extracting and utilizing these inputs, especially across large and complex projects, presents significant challenges. A developer who is responsible for maintaining the project may receive hundreds or even thousands of bug reports, making it time-consuming and impractical to manually sift through each one to identify and extract relevant inputs. Additionally, bug reports often contain a mix of structured and unstructured data, including descriptions, stack traces, error messages, and various literals, which vary in presentation across different reports and add to the complexity of consistently identifying pertinent information without automation. Manual extraction is also prone to human error, such as overlooking critical details or misinterpreting contexts, leading to incomplete or inaccurate test cases that may allow bugs to persist undetected. Furthermore, developers are typically engaged in multiple tasks, including coding, code reviews, and other maintenance activities, making it difficult to allocate sufficient time and resources to manually extract inputs from bug reports without detracting from other essential development activities. As projects grow and the number of bug reports increases, the scalability of manual extraction diminishes, rendering it untenable and necessitating more efficient and scalable solutions.

\begin{figure}[!ht]
    \includegraphics[width=.99\linewidth]{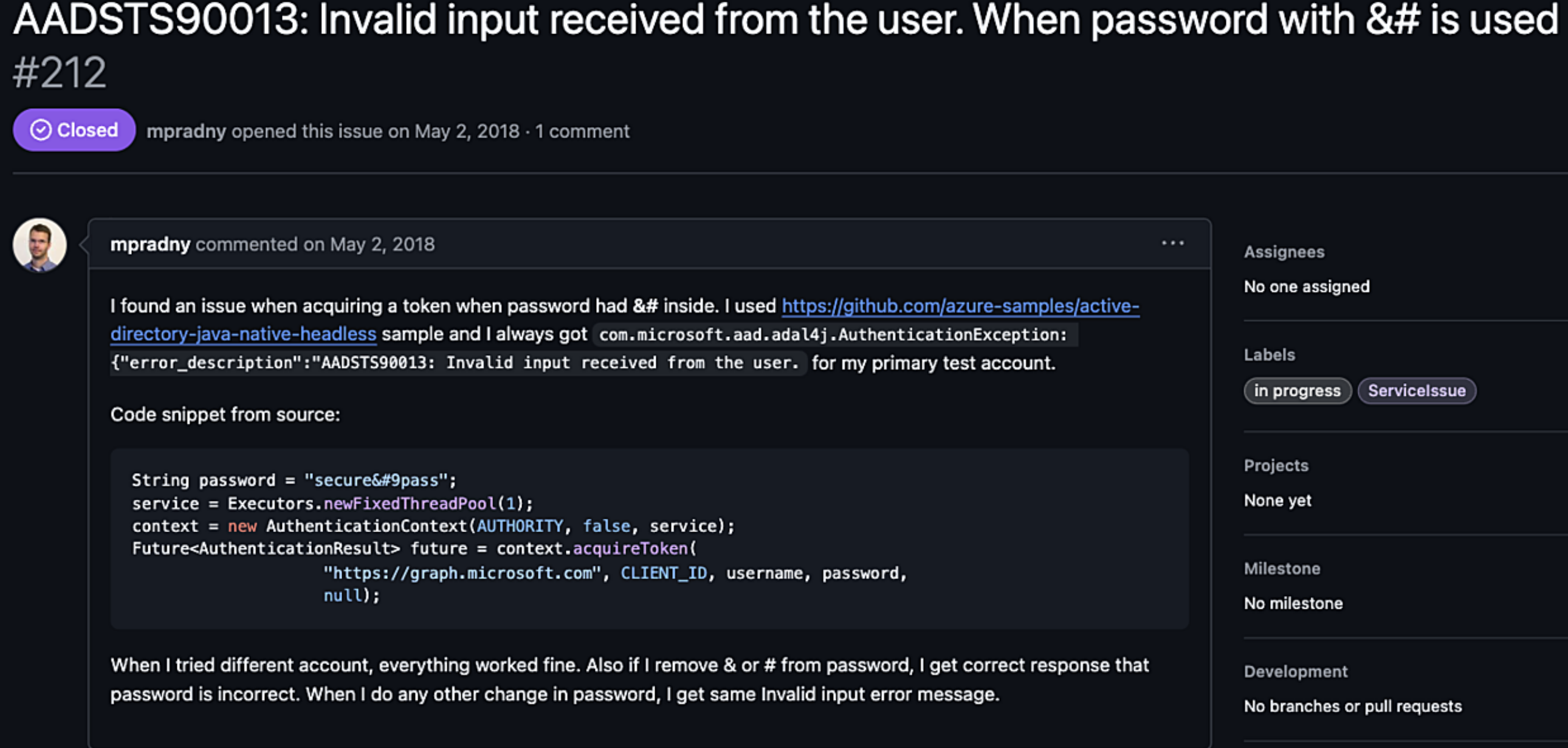}
    \caption{Input value 'secure\&\#9pass' mentioned in the bug report for issue 212 in the Azure AD Authentication Library}
    \label{fig:concrete-example}
    \vspace{2mm}
\end{figure}

Consider a developer named Bob, who is responsible for maintaining the Azure AD Authentication Library. During his routine review of bug reports, Bob encounters a report (Issue \#212\footnote{\url{https://github.com/AzureAD/azure-activedirectory-library-for-java/issues/212}}) submitted by a user as illustrated in Figure~\ref{fig:concrete-example}. This report describes a bug where the system throws an error when a password containing special characters like \textbf{\&} and \textbf{\#} is used. The bug report explicitly mentions that the password \textbf{"secure\&\#9pass"} triggers the error, while other inputs (without these characters) work correctly. Additionally, the report includes contextual details such as the error message \textbf{"error\_description"} and the URL \textbf{"https://graph.microsoft.com"}. 

As the maintainer, Bob needs to extract the relevant inputs, particularly the problematic password, to reproduce and address the bug. While Bob can manually extract inputs from a limited number of bug reports, this approach becomes extremely time-consuming and error-prone when dealing with hundreds or thousands of reports. The extensive text, stack traces, error messages, and other literals within each report compound the difficulty, making manual extraction impractical in large-scale projects where developers like Bob face a significant volume of complex reports. BRMiner addresses this challenge by automating the extraction process. For instance, in the real-world scenario based on Issue \#212 from the Azure AD Authentication Library, Bob encounters a bug report describing an issue triggered by a password containing special characters (specifically “secure\&\#9pass”). The report also includes contextual details such as the error message and the associated URL. Manually extracting these inputs from hundreds of similar reports would be highly inefficient and error-prone. For this particular bug report, BRMiner identifies and extracts the following literals:
\begin{itemize}[leftmargin=*]
    \item {\bf "secure\&\#9pass":} the password causing the error
     \item {\bf "error\_description":} the error message
     \item {\bf "https://graph.microsoft.com":} the URL involved
\end{itemize}

BRMiner leverages java parser and regular expressions to capture these literals and employs a Large Language Model (LLM) to filter out irrelevant details. It prioritizes the most relevant input—\textbf{"secure\&\#9pass"}—for use in the test case generation process. This automation allows Bob to focus on resolving the issue instead of manually sifting through the report. By automating the input extraction process, BRMiner allows Bob to focus on other development tasks while ensuring that critical inputs like \textbf{"secure\&\#9pass"} are accurately captured and used for testing. This automation saves Bob time, reduces the risk of human error, and ensures that test cases are generated efficiently, even when dealing with complex or extensive bug reports.



\subsection{Approach overview}
\label{subsec:overview}
\begin{figure}
    \includegraphics[width=.99\linewidth]{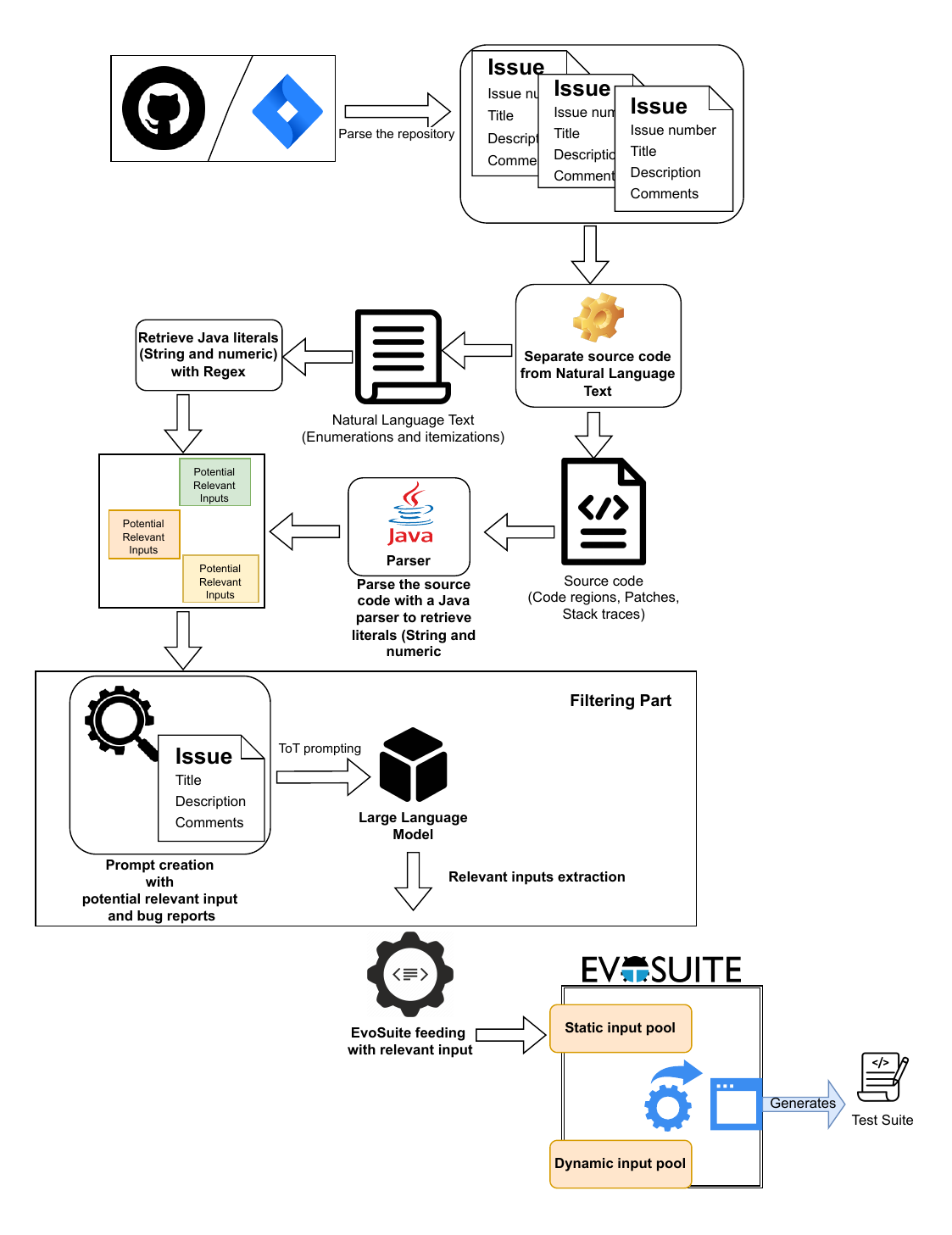}
    \caption{Overview of \name, an automatic approach to extract potential test inputs from bug reports}
    \label{fig:approach}
    \vspace{2mm}
\end{figure}

Figure~\ref{fig:approach} provides a high-level overview of the \name  approach, detailing the key steps involved in extracting and refining relevant test inputs from bug reports for automated test generation. These steps are systematically outlined in Algorithm~\ref{alg:BRMiner}, which details the sequential process of parsing, separating, and extracting literals, followed by the integration of Large Language Models (LLMs) for input refinement. The approach is designed to address key challenges in test input generation, including handling complex literals, ensuring the relevance of extracted inputs  and effectively incorporating these inputs into automated test generation tools like EvoSuite and Randoop.

\begin{algorithm}
\caption{BRMiner Approach}\label{alg:BRMiner}
\begin{algorithmic}[1]

\State \textbf{Input:} Bug reports from GitHub/Jira (Title, Description, Comments)
\State \textbf{Output:} Refined list of relevant inputs for test case generation

\State

\Procedure{BRMiner}{}
    \State \textbf{Step 1: Parse Bug Reports}
    \For{each Bug Report $B_i$ in Repository}
        \State Parse $B_i$ to extract \textit{Title, Description, Comments}
    \EndFor
    
    \State
    
    \State \textbf{Step 2: Separate Code and Text}
    \For{each Bug Report $B_i$}
        \State Use \textit{Infozilla} to separate \textit{source code} from \textit{natural language text}
    \EndFor
    
    \State
    
    \State \textbf{Step 3: Extract Potential Inputs}
    \For{each Bug Report $B_i$}
        \State Use \textit{Javalang} parser to extract literals (strings, numerics) from source code
        \State Use \textit{Regex} to extract literals (strings, numerics) from natural language text
        \State Store extracted inputs as \textit{Potential Relevant Inputs} $P_i$
    \EndFor
    
    \State
    
    \State \textbf{Step 4: Integrate LLM for Filtering}
    \For{each Bug Report $B_i$}
        \State Generate a \textit{Tree of Thoughts (ToT)} prompt based on $B_i$ and $P_i$
        \State Query \textit{GPT-3.5-turbo} with ToT prompt to analyze and filter $P_i$
        \State Retrieve a refined list of relevant inputs $R_i$ from LLM
    \EndFor
    
    \State
    
    \State \textbf{Step 5: Generate Test Cases}
    \State Integrate $R_i$ into \textit{EvoSuite’s} static input pool
    \State Use EvoSuite to generate test cases based on $R_i$
    
    \State \textbf{Return} Generated test cases
\EndProcedure

\end{algorithmic}
\end{algorithm}
\vspace{5mm}

\subsubsection{Technical Challenges and Considerations}
%
When extracting relevant test inputs from bug reports, several technical challenges must be addressed to ensure the accuracy and effectiveness of the generated test cases. 

\vspace{0.2cm}\noindent\textbf{Handling Complex Literals:} One challenge lies in handling complex literals. Bug reports often contain inputs embedded within code snippets, concatenated strings, or method calls, making it difficult to extract values that can be directly used for test generation. BRMiner overcomes this by employing Javalang to parse and extract string and numeric literals from source code, while also applying regular expressions (regex) to natural language text. This dual approach allows BRMiner to capture entire concatenated strings or method arguments, ensuring that the extracted inputs are complete and usable in the test generation process. Another challenge involves the contextual relevance of inputs. 

\vspace{0.2cm}\noindent\textbf{Contextual Relevance of Inputs:} Bug reports may not always explicitly mention the inputs that caused the bug, leaving vague or incomplete descriptions. For example, a report may simply state that "the system crashes when invalid input is provided," without specifying the exact input that triggered the crash. To address this, BRMiner integrates Large Language Models (LLMs) to filter and refine the extracted literals, ensuring that the most relevant inputs are identified based on the context of the bug report. This approach enhances the relevance of the inputs used in test generation, even in cases where the bug report lacks specific details. 

\vspace{0.2cm}\noindent\textbf{Version Discrepancies:} Finally, version discrepancies pose another challenge, as inputs extracted from older bug reports may become outdated due to changes in the software codebase. While BRMiner is not explicitly designed to handle version discrepancies, it mitigates this issue by leveraging automated test generators like EvoSuite and Randoop. These tools explore multiple execution paths using diverse inputs, helping to compensate for version differences and ensuring that the test suite generation remains comprehensive.

As previously illustrated in Section~\ref{subsec:usagescenario}, the Azure AD Authentication Library bug report (Issue \#212) exemplifies how BRMiner extracts relevant inputs like \textbf{"secure\&\#9pass"} from a report that contains multiple literals and contextual data. The report describes an invalid input error when a password contains special characters such as \textbf{\&} or \textbf{\#}. In this case, the report mentions several potential literals, including \textbf{"error\_description"}, \textbf{"secure\&\#9pass"}, and \textbf{"https://graph.microsoft.com"}, which could be relevant to reproducing the issue. For a human, manually analyzing such a detailed report, especially when it includes numerous literals, extensive text, stack traces, or error outputs, would be highly time-consuming. BRMiner automates this process by extracting all potential literals using regular expressions. After identifying the key literals, BRMiner leverages a Large Language Model (LLM) to filter the extracted inputs, determining that \textbf{"secure\&\#9pass"} is the most relevant for test case generation. This filtering process significantly reduces manual effort by focusing only on the necessary inputs for reproducing the bug, while discarding irrelevant details like the error message and the URL. The filtered input, \textbf{"secure\&\#9pass"}, is then used in automated test generation with EvoSuite, ensuring more accurate and focused test cases.


\subsubsection{Initial Steps: Parsing and Extraction}
The \name process begins by leveraging the application programming interfaces (APIs) of GitHub\footnote{https://github.com/sigmavirus24/github3.py} and Jira\footnote{https://github.com/pycontribs/jira} to systematically explore and parse bug reports. This step is crucial for identifying potential relevant inputs embedded within the reports, which may include titles, descriptions, and comments. To accurately segregate different types of information within the bug reports, BRMiner employs Infozilla~\citep{bettenburg2008extracting} to separate source code snippets from natural language text. This segregation allows for a clearer distinction between the software's codebase and the descriptive content provided within the bug reports.

\subsubsection{Literal Extraction from Source Code and Natural Language Text}
Once the text and code have been separated, BRMiner proceeds to extract relevant literals using two complementary methods:
\begin{itemize}[leftmargin=*]
    \item {\bf Source Code Extraction:} Using Javalang\footnote{https://github.com/c2nes/javalang}, a robust parser designed for analyzing Java source code, BRMiner identifies and extracts string and numeric literals embedded within the code snippets associated with the bug reports. These literals might include hardcoded values, constants, or any other relevant data embedded directly in the code.

    \item {\bf Natural Language Text Extraction:} Regular expressions (regex) are employed to extract similar literals from the natural language text found in the bug reports. This includes strings enclosed in single or double quotes and numeric values, which could represent key parameters, user inputs, or other critical data mentioned in the bug descriptions.
\end{itemize}
For instance, regex patterns are used to identify and extract strings that may represent variable names, error messages, or specific input values, as well as numeric literals that describe conditions or thresholds leading to a bug.

\subsubsection{Application of Extracted Inputs in Test Case Generation}
The refined inputs produced through the \name approach are then incorporated into EvoSuite’s static input pool, where they serve as seed inputs for automated test generation. EvoSuite leverages these inputs to explore different execution paths and generate comprehensive test suites, thereby enhancing the robustness and reliability of the software.

To establish a baseline for comparison, experiments were also conducted using an LLM alone, without any pre-extracted inputs. In this scenario, the LLM independently analyzes the bug report content to determine relevant test inputs, which are then used for test generation. This baseline helps assess the effectiveness of \name in improving test input relevance. Furthermore, to address concerns about the generalizability of \name across different testing methodologies, we extended our evaluation by incorporating Randoop, a feedback-directed random testing tool. Randoop was used with the same input scenarios, allowing us to benchmark \name against another well-established testing methodology.

\section{Experimental Setup}
\label{sec:setup}

The goal of our empirical study is to evaluate the effectiveness of \name in extracting relevant inputs from bug reports and to assess whether these inputs enhance the quality of tests generated by automated test generation tools such as EvoSuite and Randoop. BRMiner integrates a Large Language Model (LLM), specifically GPT-3.5-turbo, to refine the extracted inputs through a filtering process, thereby improving the relevance and effectiveness of the generated test cases. Additionally, we employed Randoop as a complementary test generation tool to benchmark our results against those obtained with EvoSuite. The prompts designed for the Tree of Thoughts (ToT) technique used in the LLM filtering process will also be discussed in detail.

In the following subsections, we present the research questions guiding our study, describe the dataset used for evaluation, and detail the experimental setup, including the LLM parameters, prompt design, and the testing environment. The implementation of our approach and all related experimental data is publicly available in a currently anonymous repository at: \url{https://anonymous.4open.science/r/BRMiner-C853}.

\subsection{Research Questions}


\vspace{0.2cm}\noindent\textbf{RQ1: To what extent are relevant inputs from bug reports incorporated into manually written test cases, and how often are inputs that trigger bugs explicitly mentioned in the bug reports?}
This research question investigates two key aspects: \ding{182}~The extent to which developers incorporate relevant inputs from bug reports into their manually written test cases when addressing bugs, and \ding{183}~The proportion of bug reports within the Defects4J dataset that explicitly mention the inputs responsible for triggering bugs. To address these objectives, the study will analyze a large-scale dataset, such as Defects4J, by performing an intersection analysis between tokens extracted from bug reports and literals identified in associated test cases to evaluate alignment. Additionally, an evaluation of bug reports from Defects4J projects hosted on Jira and GitHub will quantify the frequency of explicit versus implicit input mentions, highlighting patterns in how input details are documented within bug reports.

\vspace{0.2cm}\noindent\textbf{RQ2: What is the relevance of actual inputs among the potential inputs extracted by BRMiner?}
This question focuses on evaluating the efficacy of BRMiner in extracting relevant inputs from bug reports within the context of the Defects4J projects. The objective is to measure: \ding{182}~The proportion of extracted inputs that are genuinely useful for generating effective test cases; and \ding{183}~The tool’s effectiveness in minimizing irrelevant inputs. This evaluation is conducted using three configurations: BRMiner, the combination of Regex + Javalang in the BRMiner approach, and the baseline LLM alone.

\vspace{0.2cm}\noindent\textbf{RQ3: Do tests generated using all the relevant inputs extracted by BRMiner exhibit a higher bug detection rate compared to tests generated without utilizing these inputs?}
This research question examines whether the inclusion of inputs extracted by BRMiner enhances the bug detection capability of generated test cases. Specifically, it assesses: \ding{182}~Whether these extracted inputs lead to a higher rate of bug detection when compared to test cases generated without BRMiner's inputs; and \ding{183}~How the use of these inputs impacts the overall effectiveness of tools like EvoSuite and Randoop in identifying software defects.

\vspace{0.2cm}\noindent\textbf{RQ4: Does the utilization of all extracted relevant inputs by BRMiner result in higher code coverage compared to tests generated without using these inputs?}
This question aims to evaluate the impact of BRMiner’s extracted inputs on code coverage metrics in automatically generated test cases. The focus is on: \ding{182}~Determining whether incorporating these inputs results in more comprehensive exploration of the codebase; and \ding{183}~Assessing if these inputs help uncover additional execution paths that might be missed by standard test generation methods. The experiment will compare the code coverage metrics—such as branch, instruction, method, and line coverage—between test cases generated with and without BRMiner’s inputs.

\subsection{Dataset}
\label{subsec:dataset}
To address our research questions, we utilized the version 2.0.0 of Defects4J~\citep{just2014defects4j}, an open-source database designed for software testing and debugging research. Defects4J provides a collection of real software defects described in user-written bug reports, each project consisting of a buggy program version and a fixed program version. Since our study focuses on the use of bug reports, JFreeChart was excluded due to the lack of sufficient and consistent bug report information.

\subsection{Large Language Model}
\label{subsec:large-language-model}
In our study, we utilized OpenAI's GPT-3.5-turbo model to enhance the relevance of extracted inputs from bug reports in the BRMiner approach. GPT-3.5-turbo is a state-of-the-art language model with a token limit of 4,096 tokens, which includes both the input and output tokens in each interaction. To optimize the model's performance, we set the token limit to the maximum allowable, ensuring that as much contextual information as possible could be included in each prompt.

For the filtering process, we maintained the model's default parameters, setting the temperature to 0.7 to balance creativity and coherence in the generated outputs. The temperature setting controls the randomness of the model's predictions, with a value of 0.7 providing a good mix of deterministic and diverse responses.

To manage the token limits effectively, especially given the potential length of bug reports, we employed OpenAI’s tiktoken\footnote{\url{https://github.com/openai/tiktoken}} library. This library allowed us to count and manage tokens within prompts, ensuring that we stayed within the model's constraints while maximizing the information provided to the model. When necessary, we adjusted the content of the prompts by prioritizing key elements such as the title, description, comments, and extracted inputs, while omitting less critical information to avoid exceeding the token limit. This approach enabled us to leverage the full capacity of GPT-3.5-turbo for refining input relevance in our experiments.

\subsection{Prompt Design}
\label{subsec:prompt-design}
Effective prompt design is crucial in leveraging Large Language Models (LLMs) for tasks such as input refinement in test case generation. Drawing inspiration from established research in unit test generation~\citep{siddiq2024using, chen2023chatunitest, tang2024chatgpt}, we carefully crafted prompts to guide the LLMs toward extracting relevant and high-quality inputs. Our approach leverages the Tree of Thoughts (ToT) prompting technique, which helps the LLM explore different reasoning paths and refine its output based on the contextual information provided.
Each prompt consists of two key components: a Natural Language Description (NLD) that outlines the task and a Contextual Data Placeholder (CDP) that includes extracted inputs and relevant bug report details. We will delve into the specifics of each part and explain how they contribute to the overall effectiveness of the BRMiner approach.
\begin{itemize}[leftmargin=0.2cm]

\item \textbf{Natural Language Description:} The NLD is designed to clearly and concisely instruct the LLM on the task at hand. It includes:
\begin{itemize}
    \item (i) A role-playing directive that positions the LLM as a software engineering expert tasked with analyzing bug reports, enhancing the model’s focus on identifying relevant inputs~\citep{chen2023chatunitest, tang2024chatgpt}.
    \item (ii) A task-specific directive to identify and refine potential test inputs from the provided bug report details, guiding the LLM to consider both the extracted literals and the semantic context of the bug.
    \item (iii) A control statement that requires the output to be formatted in a structured JSON format, ensuring the results are easy to parse and integrate into the test generation process.
\end{itemize}

\item \textbf{Contextual Data:} This section of the prompt varies depending on the configuration:
\begin{itemize}
    \item (i) For the BRMiner configuration, the prompt includes the extracted literals (strings and numeric values) from the bug reports, as well as the title, description, and comments. This combination allows the LLM to refine the inputs based on both raw data and the broader context provided by the bug report.
    \item (ii) For the LLM Alone baseline, the prompt consists only of the bug report details without any pre-extracted inputs, allowing the LLM to independently generate potential test inputs.
\end{itemize}

\end{itemize}

The design of these prompts is illustrated in Figure~\ref{fig:prompt-templates}, which provides examples of the ToT prompts used in both the BRMiner approach and the LLM Alone baseline. By structuring the prompts in this manner, we aim to harness the full potential of GPT-3.5-turbo in refining and generating inputs that are both relevant and effective for test case generation.

\begin{figure}[ht]
\centering

\begin{minipage}[b]{\textwidth}
    \centering
    \fbox{\parbox{\textwidth}{
        \captionof{subfigure}{ToT Prompting in LLM Alone Baseline}
        \texttt{
Imagine three different software engineering experts working on maintaining a software project hosted on GitHub or Jira. Their task is to analyze a bug report, specifically its Title, Description, and Comments, to identify potential test inputs that can trigger the described bug. 
Consider the following bug report with title \textbf{[Title]}. Below is the description of the bug report:\textbf{[Description]}.
Here are the comments associated with the bug report: \textbf{[Comments]}
Each expert should propose potential inputs classified as String, Float, Integer, that could be relevant for triggering the bug. They must discuss the relevance of these inputs and discard any that are deemed irrelevant. Finally, they should compile a list of relevant inputs in the following JSON format for ease of use:
\textbf{[JSON\_FORMAT]}
Please ensure the complete JSON file is placed between the three quotes \textbf{``` ```} for easy extraction.
        }
        \label{fig:tot-llm}
    }}
\end{minipage}

\vspace{3mm} 

\begin{minipage}[b]{\textwidth}
    \centering
    \fbox{\parbox{\textwidth}{
        \captionof{subfigure}{ToT Prompting in BRMiner Approach}
        \texttt{
Imagine three different software engineering experts working on maintaining a software project hosted on GitHub or Jira. Their task is to analyze the contents of a bug report, including its Title, Description, and Comments, to determine potential test inputs that could be used to trigger the described bug.
Consider the following bug report with title \textbf{[Title]}. Below is the description of the bug report: \textbf{[Description]}.
Here are the comments associated with the bug report: \textbf{[Comments]}
The following inputs have been automatically extracted from this bug report: \textbf{[INPUTS-LIST]}.
The experts must analyze both the bug report and these extracted inputs, then select relevant inputs classified as String, Float, Integer. They should discuss the relevance of all chosen inputs, discarding any that are deemed irrelevant. Finally, they should compile a complete list of relevant inputs in the following JSON format:
\textbf{[JSON\_FORMAT]}
Please ensure the complete JSON file is placed between the three quotes \textbf{``` ```} for easy extraction.
        }
        \label{fig:tot-brminer}
    }}
\end{minipage}

\caption{Examples of ToT prompt design used for relevant inputs extraction.}
\label{fig:prompt-templates}
\vspace{3mm}
\end{figure}

\subsection{Testing Environment}
\subsubsection{EvoSuite SBST Tool}
For our experiments, we ussed EvoSuite\footnote{\url{https://github.com/EvoSuite/evosuite}} version 1.2.1, the latest version available at the time of our study. EvoSuite is a Search-Based Software Testing (SBST) tool that automatically generates test cases with a focus on maximizing code coverage. To integrate BRMiner, we modified EvoSuite's source code to include inputs extracted by BRMiner into EvoSuite's static input pool (Figure~\ref{fig:approach}). This modification allowed us to use BRMiner’s inputs as seed inputs, enhancing the generation of test cases that are more aligned with the real-world scenarios described in bug reports.

Dynamic Symbolic Execution (DSE) was employed during test generation, with branch, method, and line coverage as the primary criteria. We disabled the minimization option in EvoSuite to maintain comprehensive test generation, ensuring that all possible scenarios are tested. Assertion minimization was also disabled to preserve the thoroughness of the generated tests. All other EvoSuite parameters were left at their default settings to align with standard practices in the research community.

To ensure robustness and reliability, each experiment was repeated five times, with results aggregated across these iterations. This iterative approach helps reduce the impact of random variations, providing a more stable and accurate evaluation of the tool’s performance.

\subsubsection{Randoop}
We also conducted experiments using Randoop\footnote{\url{https://randoop.github.io/randoop/}} version 4.3.1, a feedback-directed random testing tool. Randoop automatically generates unit tests for Java programs by randomly combining methods and constructors, then executing them to detect unexpected behavior or crashes. Randoop produces two types of unit tests: error-revealing tests and regression tests. For this study, we focused on error-revealing tests, which are designed to identify potential errors in the classes under test. These tests are particularly useful for detecting exceptions or errors during the execution of the generated tests.

To handle potential issues such as infinite loops or methods that wait for user input, we configured Randoop to run each test in a separate thread by setting the --usethreads option to true. The --call-timeout was set to its default value of 5000 milliseconds, ensuring that any non-returning method calls are stopped forcefully after the specified time limit. This setup helps prevent Randoop from stalling during test generation, although it does result in a decrease in generation speed.

\subsubsection{Time Budget}
For both EvoSuite and Randoop, we allocated a time budget of three minutes per test case generation. This time budget was selected based on practical considerations, as it aligns with the typical constraints faced by developers. Previous studies~\citep{fraser2015does, arcuri2013parameter, shamshiri2015automatically} have demonstrated that a three-minute time budget is sufficient for effective test case generation in most cases. This consistent time budget across tools ensures comparability between the results obtained from EvoSuite and Randoop.

\subsubsection{Hardware}
All experiments were conducted on a machine equipped with an AMD EPYC 7552 48-Core Processor running at 3.2 GHz, with 640 GB of RAM. While not all of these resources were allocated for the experiments, the hardware provided sufficient computational power to run the experiments efficiently within the allotted time budget. This setup ensures that the results are reliable and reproducible under similar conditions.

\section{Experimental Results}
\label{sec:results}
This section outlines the experimental results and addresses the research questions.

\subsection{[RQ1]: Incorporation of Relevant Bug Report Inputs into Manually Written Test Cases and Frequency of Explicit Input Mentions.}

\noindent\textbf{Experiment Goal:} This experiment aims to assess the consistency with which relevant inputs from bug reports are incorporated into manually written test cases across a large dataset. Initial manual verification identified certain relevant inputs; however, this study seeks to determine their consistent presence in bug reports associated with successfully patched software versions. By following the definition provided in Section~\ref{subsec:rel-inputs} and aggregating results by project, we provide an overall estimate of the incorporation of these inputs into manual test cases.
Additionally, to better understand the nature of these relevant inputs, the experiment will quantify the proportion of bug reports that explicitly mention test inputs, those that describe issues more implicitly, and those that do not mention inputs at all. This detailed categorization highlights the variability in input descriptions and provides insights into the challenges of extracting actionable test inputs across diverse bug report scenarios.


\noindent\textbf{Experiment Design:} This experiment design is structured into two parts, each addressing a distinct aspect of input relevance from bug reports. The first part of the experiment examines the extent to which relevant inputs are incorporated into human-written test cases by following these steps:

\begin{itemize}
    \item \textbf{Grouping by Project:} For each project in Defects4J, we first extracted all the inputs contained in the test cases using the \texttt{Javalang} parser. These extracted inputs were then grouped by project. Similarly, we tokenized each bug report within a project using the \texttt{code-tokenize} tool, developed as part of~\citep{richter2022tssb}, and grouped the resulting tokens by project.
    
    \item \textbf{Intersection Analysis:} Once the inputs from the test cases and the tokens from the bug reports were grouped by project, we performed an intersection analysis for each project. This analysis identified the common inputs shared between the bug reports and the test cases written by human developers. By aggregating the results across all bug reports within each project, we obtained a global estimate of the number of identical inputs found in both sources.
\end{itemize}

Given the variability in the quality and completeness of bug reports, we aggregated the results by project. This aggregation provides a comprehensive estimate that highlights the extent to which inputs found in bug reports are consistently leveraged in human-written test cases, even when individual reports may vary in quality and quantity. Algorithm~\ref{alg:intersection} outlines the process.

\vspace{3mm}
\begin{algorithm}
\vspace{3mm}
\caption{Incorporation of Relevant Inputs from Bug Reports into Test Cases}\label{alg:intersection}
\begin{algorithmic}[1]
    \Procedure{AnalyzeRelevantInputs}{}
        \State \textbf{Input:} BugReports[], TestCases[]
        \State \textbf{Output:} CommonInputs[]
        
        \For{each Project \textit{P} in Defects4J}
            \State \textbf{Step 1: Tokenization}
            \State \textit{P\_Tokens} $\gets$ TokenizeAndGroupByProject(BugReports[P])
            
            \State \textbf{Step 2: Literal Extraction}
            \State \textit{P\_Literals} $\gets$ ExtractAndGroupByProject(TestCases[P])
            
            \State \textbf{Step 3: Intersection Analysis}
            \State \textit{CommonInputs[P]} $\gets$ Intersection(P\_Tokens, P\_Literals)
        \EndFor
        
        \State \textbf{return} CommonInputs
    \EndProcedure
\end{algorithmic}
\end{algorithm}
\vspace{5mm}

The second part of the experiment focuses on quantifying the proportion of bug reports across three categories: (1) those with no inputs mentioned, (2) those with inputs explicitly mentioned, and (3) those with inputs implicitly described. This classification is essential for understanding the variability in input information and ensuring BRMiner’s effectiveness in processing diverse types of bug reports. Conducting this analysis manually for a significant dataset would be prohibitively expensive and time-consuming, given the thousands of bug reports involved. To address this challenge, we employed a Chain-of-Thought (CoT) prompt with a Large Language Model (LLM), enabling systematic and scalable classification across the dataset while maintaining the necessary scale for reliable and generalizable conclusions.

\begin{itemize} 

    \item \textbf{Prompt Design:} The LLM classified bug reports into three distinct categories to ensure a comprehensive analysis of the variability in input information. The first category, No Inputs Mentioned, includes bug reports that do not reference any specific input(s) triggering the bug. These reports often describe issues in vague or general terms without providing actionable details about the inputs involved. The second category, Explicit Input Mention, comprises bug reports that explicitly mention specific input values, such as strings, numbers, or variable values, that directly trigger the bug. These reports provide clear and detailed references to the problematic inputs. Finally, the third category, Implicit Input Description, covers bug reports that lack directly stated inputs but provide descriptions that imply potential input(s). In such cases, the input values are not explicitly mentioned but can be inferred from the context or details provided in the report. This classification ensures a structured approach to understanding and processing bug reports with varying levels of input detail.

    \item \textbf{Data Extraction Format:} The LLM recorded its classification in a structured JSON format. The format included the classification category and a list of specific inputs (if available). If no inputs were mentioned, the JSON entry indicated “None” for inputs and a classification of “No Inputs Mentioned.” This structured approach allows for efficient analysis and ensures consistency across all bug reports.
    
\end{itemize}

Figure~\ref{fig:cot-prompt-templates} illustrates the design of the Chain-of-Thought (CoT) prompt utilized in this experiment. This prompt guides the LLM through a structured process for analyzing each bug report, enabling it to classify reports into three distinct categories: reports with no mentioned inputs, reports with explicitly mentioned inputs, and those with implicit input descriptions. By processing the bug report content—covering titles, descriptions, and comments—the CoT prompt ensures a comprehensive and consistent classification while addressing the variability and nuances inherent in bug report information.

\begin{figure}[ht]
\centering
\begin{minipage}[b]{\textwidth}
    \centering
    \fbox{\parbox{\textwidth}{
        \texttt{
You are a software engineering expert responsible for analyzing bug reports for maintenance purpose. Your task is to analyze bug reports, specifically Title, Description, and Comments, to identify potential test inputs that can trigger the described bug. For each bug report, you must analyze the content, detect the intentions in the bug reports, and classify each bug report into one of the following categories:
\textbf{1.No Inputs Mentioned:} The bug report does not reference any specific input(s) triggering the bug.
\textbf{2. Explicit Input Mention:} Specific input values (e.g., strings, numbers, or variable values) are directly mentioned in the bug report.
\textbf{3. Implicit Input (Implicit Input):} Input values are not directly mentioned or are vague. In this case no specific input value is mentioned, but it implies an issue with invalid input handling.
For each report provide your classification in JSON format. Include the following fields in your response:
\textbf{category:} The classification category ("No Inputs Mentioned," "Explicit Input Mention," or "Implicit Input Description").
\textbf{inputs:} A list of inputs if explicitly mentioned, or "None" if no inputs are referenced.
Here is how you must format the response: \textbf{[JSON\_FORMAT]}.
Consider the following bug report with title \textbf{[Title]}. Below is the description of the bug report: \textbf{[Description]}.
Here are the comments associated with the bug report: \textbf{[Comments]}
Please ensure the complete JSON file is placed between the three quotes \textbf{``` ```} for easy extraction.}
        }
        \label{fig:cot-llm}
    }
\end{minipage}
\caption{Chain-of-Thought Prompt Design for Classifying Input Mentions in Bug Reports}
\label{fig:cot-prompt-templates}
\vspace{2mm}
\end{figure}
To ensure the reliability and accuracy of the LLM's classifications across the three categories, we employed a rigorous sampling methodology. This step was necessary to validate the LLM's performance and establish confidence in the results while addressing the variability inherent in bug report content.
The sampling process was designed to evaluate the effectiveness and consistency of the LLM's classifications without requiring a complete manual review of the entire dataset. By analyzing a representative subset, we ensured the reliability of the LLM’s outputs while optimizing the validation effort. In this study, we applied the sampling methodology described in Algorithm~\ref{alg:sampling-validation}, ensuring proportional representation across all categories.

\begin{algorithm}
\vspace{3mm}
\caption{Sampling for Validation of LLM Classifications}\label{alg:sampling-validation}
\begin{algorithmic}[1]
    \Procedure{ValidateLLMClassifications}{}
        \State \textbf{Input:} BugReports[], Categories = \{NoInputs, ExplicitInputs, ImplicitInputs\}
        \State \textbf{Output:} ValidationMetrics

        \State \textbf{Step 1: Proportional Sampling}
        \For{each Category $C$ in Categories}
            \State $CategoryReports[C] \gets FilterReports(BugReports, C)$
            \State $SampleSize[C] \gets 0.10 \times |CategoryReports[C]|$
            \State $SampledReports[C] \gets RandomSample(CategoryReports[C], SampleSize[C])$
        \EndFor

        \State \textbf{Step 2: Human Validation}
        \For{each $Report$ in $SampledReports$}
            \State $LLMClassification \gets LLMClassify(Report)$
            \State $HumanClassification \gets HumanClassify(Report)$
            \State $ValidationResults \gets Compare(LLMClassification, HumanClassification)$
        \EndFor

        \State \textbf{Step 3: Evaluation Metrics}
        \State $Accuracy \gets CalculateAccuracy(ValidationResults)$
        \State $DiscrepancyRate \gets CalculateDiscrepancyRate(ValidationResults)$

        \State \textbf{Step 4: Threshold Evaluation}
        \If{$Accuracy < Threshold$}
            \State AdjustPromptOrProcess()
        \EndIf

        \State \textbf{return} ValidationMetrics
    \EndProcedure
\end{algorithmic}
\vspace{5mm}
\end{algorithm}

The primary purpose of the sampling methodology was to verify the accuracy and consistency of the LLM’s classifications across the three identified categories of bug reports. This validation was essential to demonstrate the robustness of the LLM in handling diverse and nuanced input information, while also identifying areas where its performance could be improved. For validation, bug reports were divided into three distinct categories: (1) No Inputs Mentioned, which includes bug reports that do not reference any specific input(s) triggering the bug; (2) Explicit Input Mention, comprising bug reports that explicitly state the inputs that trigger the bug, such as strings, numbers, or variable values; and (3) Implicit Input Description, covering bug reports that describe inputs in an implied manner without direct mention, often requiring inference to determine potential input(s).

A representative 10\% sample was drawn from each of these categories, ensuring proportional representation of bug reports across all projects in the dataset. Two independent human reviewers, who are software engineering experts, manually evaluated the sampled bug reports. They classified the reports and validated the inputs identified by the LLM. The classifications provided by the human reviewers were then compared with those generated by the LLM to calculate evaluation metrics. The Accuracy metric was defined as the percentage of correct classifications by the LLM and is computed using the formula:

\begin{itemize}
    \item \textbf{Accuracy:} The percentage of correct classifications by the LLM. It is computed as follows:
    \[
    \text{Accuracy} = \frac{\text{Number of Correct Classifications}}{\text{Total Number of Classifications}} \times 100
    \]

    \item \textbf{Discrepancy Rate:} The percentage of cases where the LLM's classification or input extraction was incorrect. It is computed as follows:
    \[
    \text{Discrepancy Rate} = \frac{\text{Number of Incorrect Classifications}}{\text{Total Number of Classifications}} \times 100
    \]
\end{itemize}

In these formulas, correct classifications refer to cases where the LLM's classification aligns with the human reviewer’s classification, while incorrect classifications denote cases where the LLM's outputs deviate. The total number of classifications represents the total number of bug reports sampled and validated during the experiment.

To ensure the reliability of the LLM's classifications, a theoretical accuracy threshold of 90\% was established. This threshold served as a benchmark to determine whether adjustments to the LLM prompt or validation process were necessary. Specifically, if the accuracy of the LLM’s classifications fell below this threshold, refinements would be implemented to improve performance. Similarly, a high discrepancy rate between the LLM's outputs and human classifications would prompt a review of the classification approach. However, in this study, no adjustments were required as the accuracy met the defined threshold, underscoring the robustness of the initial prompt design.

This validation process ensures that the classifications provided by the LLM are reliable and representative of the entire dataset. The results of this validation are critical in demonstrating the robustness of the LLM’s classifications and identifying areas for refinement.

\noindent\textbf{Experiment Results: }The results of this experiment, shown in Table~\ref{table:1}, indicate that across the Defects4J projects, 1,129 out of the 7,855 inputs extracted from the test cases (approximately 14.37\%) were also present in the bug reports. This finding, which aligns with the initial manual exploration, demonstrates a significant overlap and highlights the importance of bug reports as a source of relevant inputs for automatic test case generation. It is important to note that this analysis provides an aggregated view, ensuring that the results reflect a consistent trend across projects rather than isolated instances.

Concerning the second experiment, which examined the frequency of explicit input mentions, a rigorous sampling methodology was employed to validate the reliability and accuracy of the LLM's classifications across the three identified categories. This process involved calculating representative sample sizes for each category based on the total number of bug reports. Specifically, the sample sizes were determined using a 10\% sampling rate, ensuring proportional representation of bug reports across all projects. For the \textit{No Inputs Mentioned} category, which contained 5,509 bug reports, a sample size of 551 was calculated. Similarly, for the \textit{Explicit Input Mention} category with 10,069 bug reports, a sample size of 1,007 was determined, and for the \textit{Implicit Input Description} category with 12,042 bug reports, a sample size of 1,204 was selected.

Each sampled bug report was evaluated independently by two human reviewers, both software engineering experts. These reviewers manually classified the bug reports and validated the inputs identified by the LLM. The classifications provided by the human reviewers were then compared against the LLM’s classifications to compute evaluation metrics, including \textit{Accuracy} and \textit{Discrepancy Rate}. \textit{Accuracy} was defined as the percentage of correct classifications made by the LLM, while \textit{Discrepancy Rate} measured the percentage of cases where the LLM's classification deviated from the human reviewers' classifications. The evaluation results are summarized in Table~\ref{table:sampling-results}. For the \textit{No Inputs Mentioned} category, the LLM achieved an accuracy of 96.18\% with a discrepancy rate of 3.82\%. For the \textit{Explicit Input Mention} category, the accuracy was 95.34\%, with a discrepancy rate of 4.66\%. For the \textit{Implicit Input Description} category, the LLM’s accuracy was 94.35\%, and the discrepancy rate was 5.65\%. Across all categories, the overall accuracy was 95.07\%, with a discrepancy rate of 4.93\%. These results demonstrate that the LLM consistently provides accurate classifications with minimal errors, even across diverse and nuanced bug report content.

\begin{table}[h!]
    \vspace{2mm}
    \centering
    \caption{Evaluation of LLM Classification Accuracy and Discrepancy Rates Across Categories.}
    \scalebox{0.6}{
    \begin{tabular}{lrrrrr}
    \toprule
    \textbf{Category} & \textbf{Sample Size} & \textbf{LLM Correct} & \textbf{LLM Incorrect} & \textbf{Accuracy (\%)} & \textbf{Discrepancy Rate (\%)} \\
    \midrule
    No Inputs Mentioned & 551 & 530 & 21 & 96.18 & 3.82 \\
    Explicit Input Mention & 1,007 & 960 & 47 & 95.34 & 4.66 \\
    Implicit Input Description & 1,204 & 1,136 & 68 & 94.35 & 5.65 \\
    \midrule
    \textbf{Total/Average} & \textbf{2,762} & \textbf{2,626} & \textbf{136} & \textbf{95.07} & \textbf{4.93} \\
    \bottomrule
    \end{tabular}
    }
    \label{table:sampling-results}
\end{table}


Building upon these validation results, the second experiment quantified the distribution of bug reports across three categories: "No Inputs Mentioned," "Explicit Input Mention," and "Implicit Input Description." Globally, across the entire dataset, 19.95\% of bug reports fell into the "No Inputs Mentioned" category, while 36.46\% explicitly mentioned specific input values, and 43.60\% provided implicit descriptions. These results indicate that nearly one-fifth of bug reports do not reference any input information, presenting challenges for automated tools reliant on such details.

Focusing on the subset of bug reports that explicitly or implicitly mention inputs, the analysis, summarized in Table~\ref{table:cot-result}, reveals that 44.37\% of these reports explicitly mention specific input values, while 55.63\% provide implicit descriptions. This distribution highlights the variability in how inputs are detailed, emphasizing the dual challenges of extracting relevant information from both explicit mentions and less directly stated descriptions. These findings reinforce the importance of tools like BRMiner, which can effectively process and extract actionable test inputs from both explicit and implicit input mentions, ensuring comprehensive and accurate input identification for automated testing.

These combined results highlight not only the relevance of bug reports as a source of test inputs but also the varying degrees of input detail provided, reinforcing the need for automated tools capable of effectively processing both explicit and implicit input mentions.
\begin{table}
\vspace{3mm}
    \centering
    \caption{Number of relevant inputs in bug reports and associated test cases}
    \scalebox{0.8}
{
    \begin{tabular}{@{}l@{}rrr@{}}
        \toprule
        \textbf{Projects} & \textbf{\# of bugs} & \textbf{\# of inputs} & \textbf{Bug reports inputs} \\
        & & \textbf{in test cases} & \textbf{$\cap$ test cases literals}\\\midrule
        Cli & 39 & 484 & 93 \\ 
        Codec & 18 & 119 & 29 \\ 
        Collections & 4 & 20 & 17 \\
        Compress & 47 & 260 & 49 \\ 
        Csv & 16 & 96 & 30 \\ 
        JxPath & 22 & 188 & 23 \\ 
        Lang & 64 & 1,102 & 120 \\ 
        Math & 106 & 1,295 & 408 \\ 
        Closure & 174 & 2,486 & 33 \\
        Gson & 18 & 67 & 11 \\ 
        JacksonCore & 26 & 158 & 30 \\ 
        JacksonDatabind & 112 & 446 & 101 \\
        JacksonXml & 6 & 43 & 8 \\ 
        Jsoup & 93 & 660 & 119 \\ 
        Mockito & 38 & 89 & 12 \\ 
        Time & 26 & 342 & 46 \\ \midrule
        \centering \textbf{Total} & \textbf{809} & \textbf{7,855} & \textbf{1,129} \\
        \bottomrule
    \end{tabular}
    \label{table:1}
}
\vspace{3mm}
\end{table}

\begin{table}[]
\vspace{3mm}
    \centering
    \caption{Distribution of Bug Reports with Explicit and Implicit Input Mentions Across Projects.}

    \scalebox{0.77}
{    
\begin{tabular}{lrrrrr}
\toprule
\textbf{Projects}       & \multicolumn{1}{l}{\textbf{\#Bug reports}} & \multicolumn{1}{l}{\textbf{\#Explicit}} & \multicolumn{1}{l}{\textbf{\#Implicit}} & \multicolumn{1}{l}{\textbf{Explicit(\%)}} & \multicolumn{1}{l}{\textbf{Implicit(\%)}} \\\midrule
Cli                    & 306                                        & 127                                    & 179                                   & 41.50                                        & 58.50                                       \\
Codec                  & 295                                        & 136                                    & 159                                   & 46.10                                        & 53.90                                       \\
Collections            & 770                                        & 329                                    & 441                                   & 42.73                                        & 57.27                                       \\
Compress               & 635                                        & 315                                    & 320                                   & 49.61                                        & 50.39                                       \\
Csv                    & 300                                        & 132                                    & 168                                   & 44.00                                        & 56.00                                       \\
JxPath                  & 194                                        & 88                                     & 106                                   & 45.36                                        & 54.64                                       \\
Lang                   & 1645                                       & 792                                    & 853                                   & 48.15                                        & 51.85                                       \\
Math                   & 1636                                       & 730                                    & 906                                   & 44.62                                        & 55.38                                       \\
Closure                & 3965                                       & 1960                                   & 2005                                  & 49.43                                        & 50.57                                       \\
Gson                   & 2182                                       & 1049                                   & 1133                                  & 48.08                                        & 51.92                                       \\
JacksonCore                    & 805                                        & 335                                    & 470                                   & 41.61                                        & 58.39                                       \\
JacksonDatabind               & 3581                                       & 1666                                   & 1915                                  & 46.52                                        & 53.48                                       \\
JacksonXml             & 543                                        & 192                                    & 351                                   & 35.36                                        & 64.64                                       \\
Jsoup                  & 1829                                       & 777                                    & 1052                                  & 42.48                                        & 57.52                                       \\
Mockito                & 2764                                       & 1161                                   & 1603                                  & 42.00                                        & 58.00                                       \\
Time                   & 661                                        & 280                                    & 381                                   & 42.36                                        & 57.64                                       \\ \midrule
\centering \textbf{Total/Average} & 22111                                      & 10069                                  & \textbf{12042}                                 & 44.37                                        & \textbf{55.63}     \\
        \bottomrule                                 
\end{tabular}
\label{table:cot-result}

}
\end{table}

\vspace{0.2cm}\noindent\highlight{Summary of \textbf{RQ1:} 
The identification of 1,129 relevant inputs, representing 15\% of extracted data, underscores the critical role of bug reports in providing actionable test data through exact matches with test cases. Validation of the LLM’s classifications, with an overall accuracy of 95.07\% and a discrepancy rate of 4.93\%, further demonstrates the reliability of the LLM in categorizing bug reports across diverse content. Globally, 19.95\% of bug reports lacked input mentions, while 36.46\% explicitly mentioned specific inputs and 43.60\% provided implicit descriptions. Among reports with inputs, 44.37\% explicitly mentioned values, and 55.63\% provided implicit descriptions. These findings highlight the dual challenges of extracting inputs from explicit and implicit mentions, reinforcing the importance of tools like BRMiner in enabling comprehensive and effective test case generation.
}

\subsection{[RQ2]: Relevance Rate of \name-Extracted Potential Relevant Inputs}

\vspace{0.2cm}\noindent\textbf{Experiment Goal:}
The goal of this experiment is to evaluate BRMiner’s effectiveness in automatically extracting relevant inputs from bug reports. Specifically, we aim to assess both the Relevant Input Rate (RIR) and the Relevant Input Extraction Accuracy Rate (RIEAR) across various configurations. This analysis seeks to provide empirical evidence of BRMiner’s performance in identifying relevant inputs from real-world bug reports within the Defects4J projects.

\vspace{0.2cm}\noindent\textbf{Experiment Design:}
To achieve this, we conducted experiments using four different configurations:

\begin{itemize}
\item \textbf{Javalang Only:} This configuration uses the Javalang parser to extract inputs from both source code and natural language texts.
\item \textbf{Regex + Javalang:} This combines regular expressions (Regex) with Javalang to enhance input extraction from both source code and natural language texts.
\item \textbf{BRMiner:} This configuration represents our BRMiner approach, which integrates LLM filtering. BRMiner refines the extracted inputs using a ToT prompt and GPT-3.5-turbo to enhance the relevance of the final inputs used for testing.
\item \textbf{LLM Alone:} As a baseline, LLM Alone uses GPT-3.5-turbo to analyze bug reports and identify potential test inputs without any pre-extracted inputs.
\end{itemize}

The performance of these configurations was quantified using two key metrics:

\vspace{0.2cm}\noindent\textbf{Relevant Input Rate (RIR):} This metric measures the proportion of relevant inputs extracted by BRMiner (or any other configuration) relative to the total relevant inputs identified from the intersection of bug reports and test cases. It is calculated as:

\[
\text{RIR(\%)} = \left(\frac{| \text{TC} \cap \text{BRM} |}{| \text{BR} \cap \text{TC} |}\right) \times 100
\]

\vspace{0.2cm}\noindent\textbf{Relevant Input Extraction Accuracy Rate (RIEAR):} This metric indicates the proportion of BRMiner-extracted inputs that are actually relevant, considering the total number of unique inputs extracted by BRMiner. It is calculated as:

\[
\text{RIEAR(\%)} = \left(\frac{| \text{TC} \cap \text{BRM} |}{| \text{BRM} |}\right) \times 100
\]

Where:

\begin{itemize}
\item $| \text{TC} \cap \text{BRM} |$ represents the number of inputs that are present both in the test cases (TC) and those extracted by BRMiner (BRM).
\item $| \text{BR} \cap \text{TC} |$ represents the number of relevant inputs identified from the intersection of bug reports (BR) and test cases (TC).
\item $| \text{BRM} |$ represents the total number of unique inputs extracted by BRMiner.
\end{itemize}

\vspace{0.2cm}\noindent\textbf{Experiment Results:}
The results, summarized in Table~\ref{brminer-extraction-efficacy}(\textbf{All Bug Reports} and \textbf{"Implicit Input" Sample}), provide a comprehensive comparison of \textbf{BRMiner} with alternative methods, including \textbf{Javalang Only}, \textbf{Regex + Javalang}, and \textbf{LLM Alone}. The analysis is based on key metrics such as Relevant Input Rate (RIR) and Relevant Input Extraction Accuracy Rate (RIEAR), offering insights into the balance between input diversity and precision.

The \textbf{Javalang Only} configuration demonstrates moderate performance, achieving an RIR of 50.2\% and an RIEAR of 35.2\% across all bug reports. In the "Implicit Input" subset, its performance declines slightly, with an RIR of 47.35\% and an RIEAR of 29.83\%. This indicates that Javalang Only is moderately effective in identifying relevant inputs, but its reliance on syntactic parsing alone limits its ability to handle implicit inputs, resulting in reduced performance in less detailed bug reports.

The \textbf{Regex + Javalang} configuration exhibits the highest RIR, achieving 68.7\% across all bug reports and 68.16\% in the "Implicit Input" subset. However, its RIEAR is significantly lower, at 8.1\% and 13.15\%, respectively. This disparity highlights the method’s tendency to extract a large volume of inputs, many of which are irrelevant or do not match exactly, leading to increased noise. The lack of contextual filtering undermines its practical utility, especially for scenarios requiring precise input identification.

The \textbf{BRMiner} approach strikes a balanced performance, achieving an RIR of 60.03\% and an RIEAR of 31.71\% across all bug reports. Notably, its performance remains stable in the "Implicit Input" subset, with an RIR of 61.66\% and an identical RIEAR of 31.71\%. This consistency underscores BRMiner’s robustness and adaptability to varying levels of input detail in bug reports. By effectively balancing input diversity and precision, BRMiner proves to be a reliable tool for the automated extraction of relevant inputs from bug reports, supporting automated test generation across diverse scenarios.

Lastly, the \textbf{LLM Alone} configuration showcases the highest RIEAR across both samples, achieving 38.22\% in both cases, reflecting its strength in identifying precise and relevant inputs. However, it exhibits the lowest RIR, at 44.95\% for all bug reports and 40.91\% for the "Implicit Input" subset. This performance drop highlights the LLM’s dependency on explicit mentions or pre-extracted inputs to maintain diversity. Its vulnerability in less detailed contexts limits its applicability in scenarios where explicit inputs are not provided.

These results emphasize the trade-offs among the evaluated methods. While Javalang Only offers moderate diversity and precision, it struggles with implicit inputs. Regex + Javalang excels in diversity but generates significant noise due to low precision. LLM Alone achieves the highest precision but lacks diversity, particularly in implicit contexts. In contrast, BRMiner consistently balances input diversity and precision, maintaining stable performance across varying bug report samples. This robustness highlights its practical applicability in the automated extraction of relevant inputs from bug reports, enabling effective and efficient test generation.

\begin{table*}
    \centering
    \caption{\footnotesize Effectiveness results for \name across all bug reports and the "Implicit Input" subset. BR represents Bug Reports, TC stands for Test Cases, BRM refers to unique inputs extracted by \name approach with the four scenarios, BR $\cap$ TC indicates the intersection of inputs extracted from Bug Reports and Test Cases, BR unique Tokens corresponds to the number of unique tokens in each Bug Report, and \name $\cap$ TC represents the intersection of inputs extracted by \name and Test Cases. "Databind" in the projects column refers to the jacksonDatabind project.}
    \label{brminer-extraction-efficacy}
    \scalebox{0.41}
    {
        \begin{tabular}{@{}l@{}rr|rrrr|rrrr|rrrr|rrrr@{}}
            \toprule
            \multicolumn{19}{c}{\textbf{All the bug reports}} \\
            \midrule
            & & & \multicolumn{4}{c|}{Javalang Only} & \multicolumn{4}{c|}{Regex + Javalang} & \multicolumn{4}{c|}{BRMiner} & \multicolumn{4}{c}{LLM Alone} \\
            \midrule
            & BR $\cap$ TC & BR unique & BRM $\cap$ & BRM  & RIR & RIEAR & BRM $\cap$ & BRM  & RIR & RIEAR & BRM $\cap$ & BRM  & RIR & RIEAR & BRM $\cap$ & BRM  & RIR & RIEAR \\
            Projects & inputs       & tokens    & TC    & inputs & (\%)  & (\%) & TC       & inputs & (\%)  & (\%) & TC       & inputs & (\%)  & (\%) & TC       & inputs & (\%)  & (\%) \\
            \midrule
            Cli & 93 & \num{13146} & 62 & 149 & 66.7 & \textbf{41.6} & \textbf{70} & 628 & \textbf{75.3} & 11.1 & 65 & 165 & 69.89 & 39.39 & 39 & 135 & 41.94 & 28.89 \\
            Codec & 29 & \num{9702} & 13 & 73 & 44.8 & \textbf{17.8} & \textbf{15} & 488 & 51.7 & 3.1 & 11 & 267 & 37.93 & 4.12 & 10 & \textbf{57} & 34.48 & 17.54 \\
            Collections & 17 & \num{2320} & \textbf{15} & 32 &  \textbf{88.2} & \textbf{46.9} & \textbf{15} & 94 & \textbf{88.2} & 15.9 & 10 & 55 & 58.82 & 18.18 & 9 & 24 & 52.94 & 37.50 \\
            Compress & 49 & \num{20002} & 24 & 123 & 49.0 & 19.5 & \textbf{45} & 975 & \textbf{91.8} & 4.6 & 39 & 158 & 79.59 & \textbf{24.68} & 17 & 73 & 34.69 & 23.29 \\
            Csv & 30 & \num{5143} & 13 & 53 & 43.3 & 24.5 & 13 & 186 & 43.0 & 7.0 & 18 & 64 & 60.0 & 28.13 & \textbf{19} & 41 & \textbf{63.33} & \textbf{46.34} \\
            JxPath & 23 & \num{5525} & 12 & 40 & 52.0 & 30.0 & \textbf{15} & 261 & \textbf{65.2} & 5.7 & 10 & 55 & 43.48 & 18.18 & 13 & 27 & 56.52 & \textbf{48.15} \\
            Lang & 120 & \num{17429} & 56 & 135 & 46.7 & 41.5 & 65 & 736 & 54.2 & 8.8 & \textbf{69} & 195 & \textbf{57.50} & 35.38 & 60 & 93 & 50.0 & \textbf{64.52} \\
            Math & 408 & \num{41444} & 161 & 312 &  39.5 & 51.6 & \textbf{257} & 2201 & \textbf{63.0} & 11.7 & 239 & 537 & 58.58 & 44.51 & 100 & 175 & 24.51 & \textbf{57.14} \\
            Closure & 33 & \num{12927} & 0 & 65 & 00.0 & 0.0 & 0 & 881 & 00.0 & 0.0 & 0 & 213 & 0.0 & 0.0 & 0 & 36 & 0.0 & 0.0 \\
            Gson & 11 & \num{2825} & 5 & 23 & 45.4 & 21.7 & 8 & 23 & 72.7 & 34.8 & 5 & 18 & 45.45 & 27.78 & \textbf{9} & 17 & \textbf{81.82} & \textbf{52.94} \\
            JacksonCore & 30 & \num{4239} & 19 & 40 & 63.3 & \textbf{47.5} & \textbf{25} & 256 & \textbf{83.3} & 9.8 & 23 & 105 & 76.67 & 21.90 & 11 & 28 & 36.67 & 39.29 \\
            Databind & 101 & \num{21507} & 56 & 182 & 55.4 & 30.8 & 92 & 1794 & 91.1 & 5.0 & \textbf{95} & 135 & \textbf{94.06} & \textbf{70.37} & 35 & 103 & 34.65 & 33.98 \\
            JacksonXml & 8 & \num{1372} & \textbf{7} & 19 & \textbf{87.5} & 36.8 & \textbf{7} & 41 & \textbf{87.5} & 17.1 & 5 & 21 & 62.50 & 23.81 & \textbf{7} & 19 & \textbf{87.5} & \textbf{36.84} \\
            Jsoup & 119 & \num{11935} & 55 & 154 & 46.2 & 35.7 & 87 & 154 & 73.1 & 56.5 & \textbf{92} & 96 & \textbf{77.31} & \textbf{95.83} & 39 & 99 & 32.77 & 39.39 \\
            Mockito & 12 & \num{5698} & 3 & 26 & 25.0 & 11.5 & \textbf{7} & 445 & 58.3 & 1.6 & \textbf{7} & 93 & \textbf{58.33} & 7.53 & 5 & 26 & 41.67 & \textbf{19.23} \\
            Time & 46 & \num{3010} & 23 & 64 & 50.0 & \textbf{35.9} & \textbf{46} & 270 & \textbf{100} & 17.0 & 37 & 109 & 80.43 & 33.94 & 21 & 78 & 45.65 & 26.92 \\
            \midrule
            Total/Average & \num{1129} & \num{178224} & 524 & 1490 & 50.2 & 35.2 & \textbf{767} & 9433 & \textbf{68.7} & 8.1 & 725 & 2286 & 60.03 & 31.71 & 394 & 1031 & 44.95 & \textbf{38.22} \\
            \bottomrule
        \end{tabular}
    }
    \vspace{1mm}

    \scalebox{0.41}
    {
        \begin{tabular}{@{}l@{}rr|rrrr|rrrr|rrrr|rrrr@{}}
            \toprule
            \multicolumn{19}{c}{\textbf{The "Implicit Input" sample}} \\
            \midrule
            & & & \multicolumn{4}{c|}{Javalang Only} & \multicolumn{4}{c|}{Regex + Javalang} & \multicolumn{4}{c|}{BRMiner} & \multicolumn{4}{c}{LLM Alone} \\
            \midrule
            & BR $\cap$ TC & BR unique & BRM $\cap$ & BRM  & RIR & RIEAR & BRM $\cap$ & BRM  & RIR & RIEAR & BRM $\cap$ & BRM  & RIR & RIEAR & BRM $\cap$ & BRM  & RIR & RIEAR \\
            Projects & inputs       & tokens    & TC    & inputs & (\%)  & (\%) & TC       & inputs & (\%)  & (\%) & TC       & inputs & (\%)  & (\%) & TC       & inputs & (\%)  & (\%) \\
            \midrule
            Cli & 82 & \num{11695} & 52 & 126 & 63.41 & \textbf{41.27} & \textbf{59} & 534 & \textbf{73.17} & 11.21 & 57 & 146 & 69.9 & 39.39 & 30 & 103 & 36.43 & 28.89 \\
            Codec & 25 & \num{8631} & 11 & 62 & 44.0 & \textbf{17.74} & \textbf{11} & 415 & 52.0 & 3.13 & 10 & 238 & 39.26 & 4.12 & 8 & \textbf{45} & 31.38 & 17.54 \\
            Collections & 15 & \num{2063} & \textbf{12} & 27 &  \textbf{80.0} & \textbf{44.44} & \textbf{12} & 80 & \textbf{86.67} & 16.25 & 9 & 49 & 59.03 & 18.18 & 7 & 19 & 46.26 & 37.50 \\
            Compress & 44 & \num{17794} & 20 & 104 & 45.45 & 19.23 & \textbf{38} & 830 & \textbf{86.36} & 4.57 & 35 & 142 & 79.70 & \textbf{24.86} & 14 & 58 & 30.85 & 23.29 \\
            Csv & 26 & \num{4575} & 11 & 45 & 42.31 & 24.44 & 11 & 158 & 42.31 & 6.96 & 16 & 57 & 61.48 & 28.13 & \textbf{15} & 32 & \textbf{56.71} & \textbf{46.34} \\
            JxPath & 20 & \num{4915} & 10 & 34 & 50.00 & 29.41 & \textbf{10} & 222 & \textbf{65.0} & 5.86 & 9 & 49 & 44.54 & 18.18 & 10 & 21 & 50.79 & \textbf{48.15} \\
            Lang & 107 & \num{15505} & 47 & 114 & 43.93 & 41.23 & 55 & 626 & 51.40 & 8.77 & \textbf{61} & 175 & \textbf{57.80} & 35.38 & 48 & 74 & 44.44 & \textbf{64.52} \\
            Math & 362 & \num{36869} & 137 & 265 &  37.85 & 51.70 & \textbf{218} & 1874 & \textbf{60.50} & 11.68 & 213 & 478 & 58.71 & 44.51 & 78 & 136 & 21.50 & \textbf{57.14} \\
            Closure & 29 & \num{11500} & 0 & 55 & 00.0 & 0.0 & 0 & 750 & 00.0 & 0.0 & 0 & 191 & 0.0 & 0.0 & 0 & 29 & 0.0 & 0.0 \\
            Gson & 9 & \num{2513} & 4 & 19 & 44.44 & 21.05 & 6 & 19 & 77.78 & 35.0 & 4 & 16 & 49.79 & 27.78 & \textbf{7} & 13 & \textbf{79.23} & \textbf{52.94} \\
            JacksonCore & 26 & \num{3771} & 16 & 34 & 61.54 & \textbf{47.06} & \textbf{21} & 218 & \textbf{80.77} & 9.63 & 20 & 93 & 78.13 & 21.90 & 8 & 21 & 32.43 & 39.29 \\
            Databind & 90 & \num{19133} & 45 & 155 & 50.00 & 29.03 & 78 & 1528 & 86.67 & 5.10 & \textbf{85} & 121 & \textbf{94.27} & \textbf{70.37} & 28 & 81 & 30.57 & 33.98 \\
            JacksonXml & 6 & \num{1220} & \textbf{5} & 16 & \textbf{83.33} & 31.25 & \textbf{4} & 34 & \textbf{100.0} & 17.14 & 4 & 18 & 72.56 & 23.81 & \textbf{5} & 14 & \textbf{86.5} & \textbf{36.84} \\
            Jsoup & 105 & \num{10617} & 46 & 131 & 43.81 & 35.11 & 74 & 131 & 70.48 & 56.49 & \textbf{81} & 85 & \textbf{77.54} & \textbf{95.83} & 30 & 76 & 28.60 & 39.39 \\
            Mockito & 10 & \num{5069} & 2 & 22 & 20.00 & 9.09 & \textbf{5} & 379 & 60.0 & 1.58 & \textbf{6} & 82 & \textbf{61.88} & 7.53 & 4 & 20 & 38.40 & \textbf{19.23} \\
            Time & 40 & \num{2677} & 19 & 54 & 47.5 & \textbf{35.19} & \textbf{39} & 229 & \textbf{97.50} & 16.96 & 33 & 96 & 81.84 & 33.94 & 16 & 60 & 40.40 & 26.92 \\
            \midrule
            Total/Average & \num{996} & \num{158547} & 437 & 1263 & 47.35 & 29.83 & 641 & 8027 & \textbf{68.16} & 13.15 & \textbf{645} & 2036 & 61.66 & 31.71 & 307 & 803 & 40.91 & \textbf{38.22} \\
            \bottomrule
        \end{tabular}
    }
    \vspace{2mm}
\end{table*}

\vspace{0.2cm}\noindent\highlight{Summary of \textbf{RQ2:} The results, demonstrate BRMiner’s robust performance in extracting relevant inputs across both all bug reports and the "Implicit Input" subset. With stable metrics—RIR of 60.03\% and 61.66\%, and RIEAR of 31.71\% in both cases—BRMiner effectively balances input diversity and precision, showcasing its adaptability to varying levels of detail in bug reports. Compared to alternative methods, Javalang Only shows moderate performance but struggles with implicit inputs, Regex + Javalang achieves high diversity at the cost of precision, and LLM Alone offers exceptional precision but lacks input diversity. BRMiner’s consistent and balanced approach highlights its practical applicability in extracting relevant inputs for automated test generation, even in challenging scenarios.
}

\subsection{[RQ3]: Effect of \name-extracted inputs on bug detection rate}

\noindent\textbf{Experiment Goal: }The goal of this experiment is to evaluate the effectiveness of BRMiner in improving the bug detection capabilities of automated test generation tools, specifically EvoSuite and Randoop. We aim to assess how incorporating various input extraction strategies, including those derived from BRMiner, impacts the number of bugs detected compared to baseline scenarios where no inputs are used.

\noindent\textbf{Experiment Design:}\label{RQ3-experiment-design}
This experiment evaluates the effectiveness of BRMiner in enhancing the bug detection capabilities of automated test generation tools, specifically EvoSuite and Randoop. The objective is to determine whether the inputs extracted by BRMiner from bug reports can improve the performance of these tools in identifying software bugs compared to scenarios where no external inputs are used (NoLit).

\begin{figure*}
  \centering
  \includegraphics[width=0.9\textwidth]{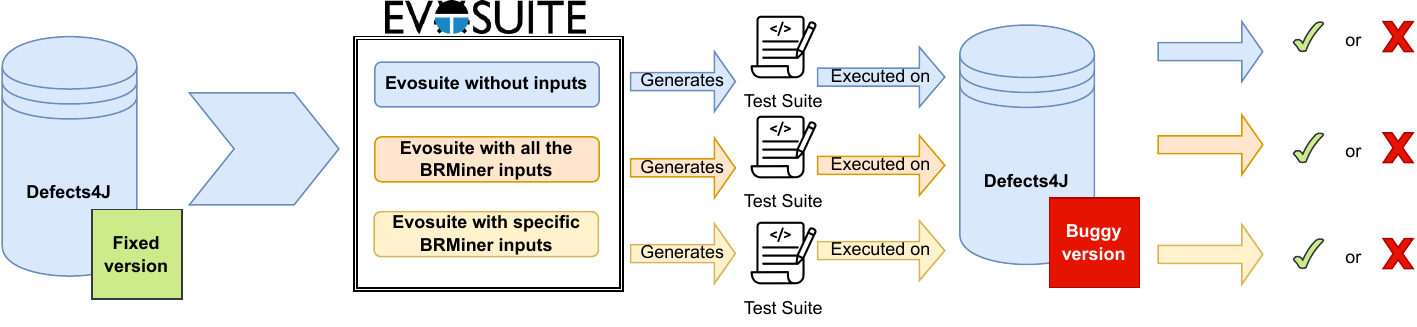}
   \caption{Overview of the methodology used to design the bugs detection experiments}
    \label{fig:bug-detection-methodology}
\end{figure*}

We employed a regression testing scenario, a common practice in software testing, to assess the effectiveness of automatically generated test cases. In this approach, test cases are generated on a fixed version of the software where the bug has been resolved, and then these tests are executed on the buggy version to determine if they can detect the defect. This methodology allows us to use the behavior of the fixed version as a reliable oracle, addressing the challenge of determining whether a test has passed or failed (the oracle problem). While we acknowledge that this approach assumes the availability of a fixed version, it provides a controlled environment for a systematic comparison of different input strategies, making it a practical method for evaluating the contributions of BRMiner. In real-world scenarios, where a fixed version may not always be available, developers often manually verify the results, which is time-consuming. This highlights the importance of further research into automated test oracle generation, which we suggest as a future direction. The regression testing scenario used in this study is inspired by previous work~\citep{shamshiri2015automatically} that successfully applied this methodology to evaluate automatic test generation tools. The overall process, depicted in Figure~\ref{fig:bug-detection-methodology}, involves generating test cases on the fixed version of the software and then executing these tests on the buggy version to measure their bug detection effectiveness.

To assess the impact of BRMiner, we conducted experiments using the following input strategies:
\begin{itemize}
    \item \textbf{NoLit:} A baseline scenario where no specific inputs are provided to EvoSuite or Randoop.
    \item \textbf{ProjLit:} Uses project-specific literals extracted from bug reports.
    \item \textbf{AllLit:} Incorporates all literals from all projects.
    \item \textbf{ProjLitLLM:} Project-specific literals filtered through BRMiner with LLM.
    \item \textbf{AllLitLLM:} All project literals filtered through BRMiner with LLM.
    \item \textbf{ProjLitLLMOnly:} Inputs generated solely by LLM for each project.
    \item \textbf{AllLitLLMOnly:} Inputs generated solely by LLM across all projects.
\end{itemize}

In Table~\ref{table:bug-detection-results}, the terms "ProjLit" and "AllLit" refer to configurations where only the \textbf{Regex + Javalang} approach was used.
These configurations allow us to compare the effectiveness of using literals from different sources, both with and without the filtering and generation capabilities of LLM, to determine the most effective strategy for enhancing bug detection.

The experiments were conducted on the Defects4J dataset, featuring real-world software bugs across multiple projects. Test cases were generated using EvoSuite and Randoop under the various input strategies, with each experiment allocated a three-minute time budget per iteration over five iterations to ensure robust results. Certain projects, such as JacksonXml, Mockito, and Closure, were omitted from some experiments due to limitations in EvoSuite's ability to generate tests or issues with non-compilable test cases. Additionally, Randoop failed to execute tests for Compress, JacksonDatabind and JacksonXml, for Cli, Csv, and Math, it executed but did not generate error-revealing tests, influencing the overall bug detection outcomes.

We analyzed the number of bugs detected by each input strategy, the number of test cases generated, and the overlap of bugs detected across different strategies. This comprehensive analysis provides insights not only into the number of bugs detected but also into the specific contributions of BRMiner and its ability to enhance the effectiveness of automated test generation.

\vspace{1mm}

\noindent\textbf{Experiment Results: }
The results, summarized in Table~\ref{table:bug-detection-results} (All Bug Reports and "Implicit Input" Sample), demonstrate that BRMiner, significantly enhances the bug detection capabilities of automated test generation tools such as EvoSuite and Randoop. Across all bug reports, the \textbf{AllLitLLM} scenario detected the highest number of bugs (\textbf{313 with EvoSuite and 111 with Randoop}), surpassing the baseline \textbf{NoLit} scenario (\textbf{295 with EvoSuite and 93 with Randoop}). The \textbf{ProjLitLLM} configuration also showed strong performance, detecting \textbf{311 bugs with EvoSuite and 107 with Randoop}, underscoring the value of project-specific inputs. These results confirm that integrating BRMiner's comprehensive input extraction, especially when enhanced by LLM filtering, substantially improves bug detection rates compared to scenarios without input extraction (baselines).

In the "Implicit Input" subset, BRMiner's configurations maintained robust performance. The \textbf{AllLitLLM} scenario again achieved the highest detection rates (\textbf{176 with EvoSuite and 59 with Randoop}), demonstrating BRMiner's ability to extract and utilize relevant inputs even when explicit mentions are absent. The \textbf{ProjLitLLM} configuration followed closely, detecting \textbf{174 bugs with EvoSuite and 57 with Randoop}. This consistency highlights BRMiner’s adaptability to varying levels of input detail, reaffirming its effectiveness in real-world testing scenarios.

Among the evaluated configurations, BRMiner consistently achieved the highest bug detection rates across both samples, with minimal performance degradation in the absence of explicit inputs. This reliability emphasizes its robustness in extracting and filtering relevant inputs, even in challenging contexts. The integration of LLM filtering ensures that the most relevant data is prioritized for test generation, significantly boosting the utility of extracted inputs. In contrast, the \textbf{NoLit} baseline detected fewer bugs, underscoring the critical role of extracted inputs in improving automated test generation. The \textbf{ProjLit} and \textbf{AllLit} configurations, while improving detection rates compared to NoLit, were surpassed by BRMiner, illustrating the added value of contextual filtering. Finally, the \textbf{LLMOnly} configurations detected fewer bugs than their BRMiner counterparts, highlighting the importance of combining pre-extraction methods with LLM filtering to maximize the utility of extracted inputs.

\begin{table*}
    \centering
    \caption{\footnotesize Comparison of bug detection effectiveness across different input strategies for different experiments. Definitions of terms used: NoLit refers to experiments where no literals were used (baseline). ProjLit refers to the use of project-specific literals, and AllLit includes all literals from all projects. LLM indicates that filtering was performed using LLM, while LLMOnly means LLM was used alone to extract inputs. Evo denotes experiments conducted with EvoSuite, and Ran refers to those conducted with Randoop. "Databind" in the projects column refers to the jacksonDatabind project.}

    \label{table:bug-detection-results}
    \scalebox{0.58}
    {
        \begin{tabular}{l|c|cccccccccccccc}
            \toprule
            \multicolumn{16}{c}{\textbf{All the bug reports}} \\
            \midrule
            & \textbf{\#Bugs} & \multicolumn{14}{c}{Bug Detection (\#)} \\
            \cmidrule{3-16}
            Projects & \textbf{} & \multicolumn{2}{c|}{NoLit} & \multicolumn{2}{c|}{ProjLit} & \multicolumn{2}{c|}{AllLit} & \multicolumn{2}{c|}{ProjLitLLM} & \multicolumn{2}{c|}{AllLitLLM} & \multicolumn{2}{c|}{ProjLitLLMOnly} & \multicolumn{2}{c}{AllLitLLMOnly} \\
            &  & Evo & Ran & Evo & Ran & Evo & Ran & Evo & Ran & Evo & Ran & Evo & Ran & Evo & Ran \\
            \midrule
            Cli & 39 & 28 & 0 & \textbf{29} & 0 & 28 & 0 & 28 & 0 & \textbf{29} & 0 & 28 & 0 & \textbf{29} & 0 \\
            Codec & 18 & 10 & 9 & 11 & 10 & \textbf{12} & 11 & \textbf{12} & 11 & 11 & \textbf{12} & 11 & 10 & 11 & 10 \\
            Collections & 4 & 1 & 1 & 1 & 1 & 1 & 1 & 1 & 1 & 1 & 1 & 1 & 1 & 1 & 1 \\
            Compress & 47 & 28 & -- & 27 & -- & 27 & -- & 28 & -- & 28 & -- & 28 & -- & 28 & -- \\
            Csv & 16 & 13 & 0 & 14 & 0& \textbf{15} & 0 & \textbf{15} & 0 & 14 & 0 & \textbf{15} & 0 & \textbf{15} & 0 \\
            Gson & 18 & 5 & 3 & \textbf{6} & 4 & 5 & 3 & 5 & 5 & \textbf{6} & 4 & 5 & 3 & 5 & 4 \\
            JacksonCore & 26 & 17 & 8 & 18 & 8 & \textbf{19} & 9 & \textbf{19} & 9 & \textbf{19} & 9 & 17 & 8 & 18 & 8 \\
            Databind & 112 & 4 & -- & 4 & -- & \textbf{5} & -- & \textbf{5} & -- & \textbf{5} & -- & \textbf{5} & -- & \textbf{5} & -- \\
            JacksonXml & 6 & 0 & -- & 0 & -- & 0 & -- & 0 & -- & 0 & -- & 0 & -- & 0 & -- \\
            Jsoup & 93 & 52 & 35 & 55 & 37 & \textbf{59} & 41 & \textbf{59} & 41 & \textbf{59} & 44 & 54 & 35 & 54 & 35 \\
            JxPath & 22 & 17 & 8 & 17 & 8 & \textbf{19} & 9 & \textbf{19} & 8 & \textbf{19} & 9 & 17 & 8 & 17 & 9 \\
            Lang & 64 & 40 & 22 & \textbf{41} & 23 & 40 & 22 & 40 & 23 & \textbf{41} & 23 & 40 & \textbf{41} & \textbf{41} & 23 \\
            Math & 106 & 70 & 0 & \textbf{71} & 0 & 67 & 0 & 70 & 0 & \textbf{71} & 0 & 70 & 0 & \textbf{71} & 0 \\
            Time & 26 & 10 & 7 & 10 & 8 & 10 & 9 & 10 & 9 & 10 & 9 & 10 & 9 & 10 & 9 \\
            \midrule
            \textbf{Total} & 597 & 295 & 93 & 304 & 99 & 307 & 105 & 311 & 107 & \textbf{313} & 111 & 301 & 96 & 305 & 99 \\
            \bottomrule
        \end{tabular}
    }
    \vspace{2mm}

     \label{table:bug-detection-results}
    \scalebox{0.58}
    {
        \begin{tabular}{l|c|cccccccccccccc}
    \toprule
    \multicolumn{16}{c}{\textbf{The "Implicit Input" sample}} \\
    \midrule
    & \textbf{\#Bugs} & \multicolumn{14}{c}{Bug Detection (\#)} \\
    \cmidrule{3-16}
    Projects & \textbf{} & \multicolumn{2}{c|}{NoLit} & \multicolumn{2}{c|}{ProjLit} & \multicolumn{2}{c|}{AllLit} & \multicolumn{2}{c|}{ProjLitLLM} & \multicolumn{2}{c|}{AllLitLLM} & \multicolumn{2}{c|}{ProjLitLLMOnly} & \multicolumn{2}{c}{AllLitLLMOnly} \\
    &  & Evo & Ran & Evo & Ran & Evo & Ran & Evo & Ran & Evo & Ran & Evo & Ran & Evo & Ran \\
    \midrule
    Cli & 25 & 18 & 0 & 18 & 0 & 18 & 0 & 19 & 0 & \textbf{20} & 0 & 18 & 0 & 18 & 0 \\
    Codec & 8 & 4 & 4 & 5 & 4 & 5 & 5 & \textbf{6} & 5 & 5 & \textbf{6} & 5 & 4 & 5 & 4 \\
    Collections & 3 & 1 & 1 & 1 & 1 & 1 & 1 & 1 & 1 & 1 & 1 & 1 & 1 & 1 & 1 \\
    Compress & 19 & 11 & -- & 11 & -- & 11 & -- & \textbf{12} & -- & \textbf{12} & -- & 11 & -- & 11 & -- \\
    Csv & 6 & 5 & 0 & 5 & 0 & 5 & 0 & \textbf{6} & 0 & 5 & 0 & 5 & 0 & 5 & 0 \\
    Gson & 10 & 3 & 2 & 3 & 2 & 3 & 2 & 3 & 3 & \textbf{4} & 2 & 3 & 2 & 3 & 2 \\
    JacksonCore & 14 & 9 & 4 & 9 & 4 & 10 & 5 & \textbf{11} & 5 & \textbf{11} & 5 & 9 & 4 & 9 & 4 \\
    Databind & 54 & 2 & -- & 2 & -- & 2 & -- & \textbf{3} & -- & \textbf{3} & -- & 2 & -- & 2 & -- \\
    JacksonXml & 2 & 0 & -- & 0 & -- & 0 & -- & 0 & -- & 0 & -- & 0 & -- & 0 & -- \\
    Jsoup & 46 & 25 & 17 & 27 & 18 & 29 & 20 & \textbf{31} & 22 & \textbf{31} & 23 & 26 & 17 & 26 & 17 \\
    JxPath & 9 & 7 & 3 & 7 & 3 & 8 & 4 & \textbf{9} & 4 & \textbf{9} & 4 & 7 & 3 & 7 & 4 \\
    Lang & 35 & 22 & 12 & 22 & 12 & 22 & 12 & 23 & 13 & \textbf{24} & 13 & 22 & 12 & 22 & 12 \\
    Math & 65 & 43 & 0 & 43 & 0 & 41 & 0 & 46 & 0 & \textbf{47} & 0 & 43 & 0 & 43 & 0 \\
    Time & 12 & \textbf{5} & 3 & \textbf{5} & 4 & \textbf{5} & 4 & \textbf{5} & 4 & \textbf{5} & 4 & \textbf{5} & 4 & \textbf{5} & 4 \\
    \midrule
    \textbf{Total} & 306 & 154 & 46 & 159 & 49 & 159 & 52 & 174 & 57 & \textbf{176} & 59 & 157 & 48 & 159 & 49 \\
    \bottomrule
\end{tabular}
    }
    \vspace{2mm}
\end{table*}

The results presented in Table~\ref{generated-tests-results} (All Bug Reports and "Implicit Input" Sample) demonstrate a clear relationship between the number of tests generated and the bugs detected across different scenarios. Across both samples, BRMiner configurations that incorporate LLM filtering, particularly the \textbf{AllLitLLM} scenario, consistently achieve the highest bug detection rates while generating fewer tests compared to other configurations.

For all bug reports, the \textbf{AllLitLLM} scenario detected the highest number of bugs (\textbf{313 with EvoSuite and 111 with Randoop}) while generating fewer tests (\textbf{144\,248 with EvoSuite and 6\,294 with Randoop}), reflecting an efficient balance between test generation and bug detection. Similarly, the \textbf{ProjLitLLM} scenario also performed well, detecting \textbf{311 bugs with EvoSuite and 107 with Randoop} while generating slightly more tests (\textbf{145\,291 with EvoSuite and 6\,731 with Randoop}). In contrast, the \textbf{NoLit} baseline generated the most tests (\textbf{148\,246 with EvoSuite and 6\,568 with Randoop}) but detected fewer bugs (\textbf{295 with EvoSuite and 93 with Randoop}), indicating inefficiencies in converting generated tests into detected bugs.

In the "Implicit Input" subset, BRMiner maintained robust performance. The \textbf{AllLitLLM} scenario again detected the highest number of bugs (\textbf{176 with EvoSuite and 59 with Randoop}) while generating fewer tests (\textbf{79\,375 with EvoSuite and 3\,173 with Randoop}), underscoring its capability to extract and leverage relevant inputs effectively even in less detailed bug reports. Similarly, the \textbf{ProjLitLLM} scenario followed closely, detecting \textbf{174 bugs with EvoSuite and 57 with Randoop} while generating slightly more tests (\textbf{69\,411 with EvoSuite and 3\,397 with Randoop}). The baseline \textbf{NoLit} scenario continued to demonstrate lower efficiency, generating the most tests (\textbf{89\,624 with EvoSuite and 4\,201 with Randoop}) while detecting fewer bugs (\textbf{154 with EvoSuite and 46 with Randoop}).

These findings highlight the effectiveness of BRMiner’s LLM-filtered configurations, particularly \textbf{AllLitLLM}, in achieving high bug detection rates with fewer tests. This efficiency stems from the use of well-filtered inputs, which prioritize relevant data and reduce noise during test generation, thus enhancing the overall effectiveness of automated testing tools like EvoSuite and Randoop.

\begin{table*}
    \centering
    \caption{\footnotesize Number of generated tests and number of detected bugs with a time budget of three minutes and five iterations for different experiments. Definitions of terms used: NoLit refers to experiments where no literals were used (baseline). ProjLit refers to the use of project-specific literals, and AllLit includes all literals from all projects. LLM indicates that filtering was performed using LLM, while LLMOnly means LLM was used alone to extract inputs. Evo denotes experiments conducted with EvoSuite, and Ran refers to those conducted with Randoop. "Databind" in the projects column refers to the JacksonDatabind project.}
    \label{generated-tests-results}
    \scalebox{0.51}
    {
        \begin{tabular}{l|cccccccccccccc}
            \toprule
            & \multicolumn{14}{c}{Generated Tests (\#) With All the bug reports} \\
            \cmidrule{2-15}
            Projects & \multicolumn{2}{c|}{NoLit} & \multicolumn{2}{c|}{ProjLit} & \multicolumn{2}{c|}{AllLit} & \multicolumn{2}{c|}{ProjLitLLM} & \multicolumn{2}{c|}{AllLitLLM} & \multicolumn{2}{c|}{ProjLitLLMOnly} & \multicolumn{2}{c}{AllLitLLMOnly} \\
            & Evo & Ran & Evo & Ran & Evo & Ran & Evo & Ran & Evo & Ran & Evo & Ran & Evo & Ran \\
            \midrule
            Cli & 2059 & 0 & 1818 & 0 & 1853 & 0 & 1811 & 0 & 1782 & 0 & 1826 & 0 & 1834 & 0 \\
            Codec & 5946 & 523 & 6112 & 550 & 6113 & 563 & 6068 & 530 & 6050 & 535 & 6106 & 563 & 6108 & 560 \\
            Collections & 19528 & 391 & 18899 & 460 & 19884 & 454 & 19842 & 395 & 19813 & 391 & 19857 & 455 & 19865 & 454 \\
            Compress & 22010 & -- & 21672 & -- & 21887 & -- & 21755 & -- & 21771 & -- & 21793 & -- & 21812 & -- \\
            Csv & 1149 & 0 & 1127 & 0 & 1068 & 0 & 1073 & 0 & 1020 & 0 & 1109 & 0 & 1095 & 0 \\
            Gson & 4568 & 1741 & 4438 & 2285 & 4331 & 2387 & 4360 & 1850 & 4295 & 1800 & 4372 & 1950 & 4364 & 1735 \\
            JacksonCore & 18472 & 376 & 17836 & 384 & 17390 & 489 & 17558 & 323 & 17424 & 350 & 17521 & 370 & 17503 & 365 \\
            Databind & 3001 & -- & 3671 & -- & 2984 & -- & 3303 & -- & 21952 & -- & 3273 & -- & 22011 & -- \\
            JacksonXml & 207 & -- & 184 & -- & 247 & -- & 191 & -- & 169 & -- & 210 & -- & 219 & -- \\
            Jsoup & 9151 & 676 & 10524 & 748 & 10470 & 587 & 10432 & 730 & 10401 & 535 & 10447 & 755 & 10454 & 634 \\
            JxPath & 11783 & 12 & 11373 & 12 & 11041 & 12 & 11182 & 12 & 11062 & 12 & 11237 & 12 & 11179 & 12 \\
            Lang & 23639 & 455 & 22439 & 449 & 21613 & 497 & 21956 & 423 & 21735 & 433 & 21968 & 449 & 21881 & 443 \\
            Math & 4759 & 0 & 3791 & 0 & 3729 & 0 & 3735 & 0 & 3682 & 0 & 3739 & 0 & 3740 & 0 \\
            Time & 21974 & 2797 & 22073 & 3241 & 21995 & 3443 & 22009 & 2875 & 21952 & 2641 & 22011 & 2833 & 22011 & 2739 \\
            \midrule
            \textbf{Generated Tests (\#)} & 148246 & 6568 & 146957 & 7617 & 144605 & 7966 & 145291 & 6731 & \textbf{144248} & 6294 & 145468 & 6920 & 145278 & 6476 \\
            \textbf{Bugs Detected (\#)} & 295 & 138 & 304 & 145 & 307 & 147 & 311 & 153 & \textbf{313} & 157 & 301 & 142 & 305 & 144 \\
            \bottomrule
        \end{tabular}
    }
    \vspace{2mm}

    \label{generated-tests-results}
    \scalebox{0.51}
    {
        \begin{tabular}{l|cccccccccccccc}
    \toprule
    & \multicolumn{14}{c}{Generated Tests (\#) With The ”Implicit Input” sample} \\
    \cmidrule{2-15}
    Projects & \multicolumn{2}{c|}{NoLit} & \multicolumn{2}{c|}{ProjLit} & \multicolumn{2}{c|}{AllLit} & \multicolumn{2}{c|}{ProjLitLLM} & \multicolumn{2}{c|}{AllLitLLM} & \multicolumn{2}{c|}{ProjLitLLMOnly} & \multicolumn{2}{c}{AllLitLLMOnly} \\
    & Evo & Ran & Evo & Ran & Evo & Ran & Evo & Ran & Evo & Ran & Evo & Ran & Evo & Ran \\
    \midrule
    Cli & 1441 & 0 & 1273 & 0 & 1297 & 0 & 1001 & 0 & 985 & 0 & 1112 & 0 & 1095 & 0 \\
    Codec & 3568 & 314 & 3667 & 330 & 3668 & 338 & 2875 & 251 & 2867 & 254 & 3195 & 279 & 3185 & 282 \\
    Collections & 9764 & 196 & 9450 & 230 & 9942 & 227 & 7835 & 156 & 7824 & 154 & 8706 & 173 & 8693 & 172 \\
    Compress & 17608 & -- & 17338 & -- & 17510 & -- & 13745 & -- & 13755 & -- & 15272 & -- & 15283 & -- \\
    Csv & 460 & 0 & 451 & 0 & 427 & 0 & 339 & 0 & 322 & 0 & 377 & 0 & 358 & 0 \\
    Gson & 2284 & 871 & 2219 & 1143 & 2166 & 1194 & 1722 & 731 & 1696 & 711 & 1913 & 812 & 1884 & 790 \\
    JacksonCore & 11083 & 226 & 10702 & 230 & 10434 & 293 & 8320 & 153 & 8256 & 166 & 9244 & 170 & 9174 & 184 \\
    Databind & 2101 & -- & 2570 & -- & 2089 & -- & 1826 & -- & 12136 & -- & 2029 & -- & 13484 & -- \\
    JacksonXml & 104 & -- & 92 & -- & 124 & -- & 75 & -- & 67 & -- & 84 & -- & 74 & -- \\
    Jsoup & 5491 & 406 & 6314 & 449 & 6282 & 352 & 4943 & 346 & 4929 & 254 & 5492 & 384 & 5476 & 282 \\
    JxPath & 4713 & 5 & 4549 & 5 & 4416 & 5 & 3532 & 4 & 3494 & 4 & 3925 & 4 & 3883 & 4 \\
    Lang & 11820 & 228 & 11220 & 225 & 10807 & 249 & 8670 & 167 & 8583 & 171 & 9633 & 186 & 9536 & 190 \\
    Math & 3807 & 0 & 3033 & 0 & 2983 & 0 & 2360 & 0 & 2326 & 0 & 2622 & 0 & 2585 & 0 \\
    Time & 15382 & 1958 & 15451 & 2269 & 15396 & 2410 & 12167 & 1589 & 12136 & 1460 & 13519 & 1766 & 13484 & 1622 \\
    \midrule
    \textbf{Generated Tests (\#)} & 89624 & 4201 & 88327 & 4880 & 87540 & 5067 & 69411 & 3397 & \textbf{79375} & 3173 & 77123 & 3774 & 88194 & 3525 \\
    \textbf{Bugs Detected (\#)} & 154 & 46 & 159 & 49 & 159 & 52 & 174 & 57 & \textbf{176} & 59 & 157 & 48 & 159 & 49 \\
    \bottomrule
\end{tabular}
    }
    \vspace{2mm}
\end{table*}

Table~\ref{evo-randoop-intersections-results} demonstrates the robustness of BRMiner in detecting both unique and intersection-based bugs across different input strategies. BRMiner consistently detected all bugs identified by other configurations, including unique bugs captured by the \textbf{Regex + Javalang} and \textbf{NoLit} baselines. In EvoSuite, the \textbf{AllLitLLM} scenario, leveraging inputs extracted from all projects, was particularly effective, detecting all 13 unique bugs identified by the \textbf{NoLit} baseline in the "All Bug Reports" sample and capturing additional unique bugs missed by other methods. This cross-project input strategy significantly broadened the scope of relevant inputs, allowing BRMiner to address bugs that would otherwise remain undetected. Similarly, in the "Implicit Input" subset, the \textbf{ProjLitLLM} and \textbf{AllLitLLM} configurations maintained their effectiveness, demonstrating resilience and adaptability even when explicit inputs were absent from the bug reports.

For Randoop, BRMiner exhibited similar robustness, consistently detecting all bugs found by other configurations, including the unique bugs identified by the \textbf{NoLit} baseline. The inclusion of cross-project inputs in the \textbf{AllLitLLM} scenario further enhanced its utility, as it allowed the tool to capture a broader range of relevant inputs that improved bug detection coverage. These results underscore the added value of combining project-specific inputs (\textbf{ProjLitLLM}) with a comprehensive cross-project approach (\textbf{AllLitLLM}) to maximize bug detection potential.

These findings highlight BRMiner's ability to enhance bug detection coverage by leveraging both project-specific and cross-project input extraction strategies, augmented with LLM filtering. This comprehensive approach ensures the identification of a wide range of bugs, including those overlooked by traditional methods such as \textbf{Regex + Javalang}. The adaptability of BRMiner across both samples demonstrates its practical utility for real-world testing scenarios, enabling consistent performance regardless of the level of input explicitness in bug reports.
\begin{table*}
    \centering
    \caption{\footnotesize Analysis of unique and intersection-based bug detection across different input strategies. \textbf{NoLit}: Bugs detected without using any input literals. 
\textbf{ProjLit}: Bugs detected using only literals extracted from ProjLit (project literals). 
\textbf{AllLit}: Bugs detected using only literals extracted from AllLit (all literals). 
\textbf{P $\cap$ A}: Bugs detected using literals that intersect between ProjLit and AllLit. 
\textbf{P $\cap$ N}: Bugs detected using literals that intersect between ProjLit and NoLit. 
\textbf{N $\cap$ A}: Bugs detected using literals that intersect between NoLit and AllLit. 
\textbf{N $\cap$ P $\cap$ A}: Bugs detected using literals that intersect among NoLit, ProjLit, and AllLit.}
   
    \label{evo-randoop-intersections-results}
    
    
    \scalebox{0.5}{
        \begin{tabular}{@{}l|ccccccc|ccccccc|ccccccc@{}}
            \toprule
            \multicolumn{22}{c}{\textbf{Intersection and unique bug detection results for EvoSuite With All the bug reports}} \\
            \midrule
            & \multicolumn{7}{c|}{Regex + Javalang} & \multicolumn{7}{c|}{BRMiner} & \multicolumn{7}{c}{LLM Alone} \\
            \cmidrule{2-22}
            Projects & \makecell{No\\Lit} & \makecell{Proj\\Lit} & \makecell{All\\Lit} & \makecell{P\\$\cap$\\A} & \makecell{P\\$\cap$\\N} & \makecell{N\\$\cap$\\A} & \makecell{N\\$\cap$\\P\\$\cap$\\A} & \makecell{No\\Lit} & \makecell{Proj\\Lit} & \makecell{All\\Lit} & \makecell{P\\$\cap$\\A} & \makecell{P\\$\cap$\\N} & \makecell{N\\$\cap$\\A} & \makecell{N\\$\cap$\\P\\$\cap$\\A} & \makecell{No\\Lit} & \makecell{Proj\\Lit} & \makecell{All\\Lit} & \makecell{P\\$\cap$\\A} & \makecell{P\\$\cap$\\N} & \makecell{N\\$\cap$\\A} & \makecell{N\\$\cap$\\P\\$\cap$\\A} \\
            \midrule
            Cli & 1 & 0 & 1 & 2 & 1 & 1 & 25 & 0 & 0 & 1 & 0 & 0 & 0 & 28 & 0 & 0 & 1 & 1 & 0 & 0 & 28 \\
            Codec & 0 & 2 & 1 & 0 & 0 & 0 & 11 & 0 & 1 & 0 & 1 & 0 & 0 & 10 & 0 & 0 & 0 & 1 & 0 & 0 & 11 \\
            Collections & 0 & 0 & 0 & 0 & 0 & 0 & 1 & 0 & 0 & 0 & 0 & 0 & 0 & 1 & 0 & 0 & 0 & 0 & 0 & 0 & 1 \\
            Compress & 1 & 1 & 0 & 1 & 1 & 1 & 23 & 0 & 0 & 0 & 0 & 0 & 0 & 28 & 1 & 0 & 0 & 0 & 0 & 0 & 27 \\
            Csv & 0 & 1 & 0 & 1 & 0 & 0 & 13 & 0 & 1 & 0 & 1 & 0 & 0 & 13 & 0 & 0 & 0 & 2 & 0 & 0 & 13 \\
            Gson & 0 & 0 & 1 & 0 & 0 & 0 & 5 & 0 & 0 & 1 & 0 & 0 & 0 & 5 & 0 & 0 & 0 & 0 & 0 & 0 & 5 \\
            JacksonCore & 1 & 2 & 0 & 1 & 0 & 0 & 13 & 0 & 0 & 0 & 2 & 0 & 0 & 17 & 1 & 0 & 1 & 0 & 0 & 0 & 17 \\
            Databind & 0 & 1 & 0 & 0 & 0 & 0 & 4 & 0 & 0 & 0 & 1 & 0 & 0 & 4 & 0 & 0 & 0 & 1 & 0 & 0 & 4 \\
            JacksonXml & 0 & 0 & 0 & 0 & 0 & 0 & 0 & 0 & 0 & 0 & 0 & 0 & 0 & 0 & 0 & 0 & 0 & 0 & 0 & 0 & 0 \\
            Jsoup & 3 & 2 & 1 & 8 & 4 & 1 & 42 & 0 & 0 & 0 & 7 & 0 & 0 & 52 & 1 & 0 & 0 & 2 & 0 & 0 & 52 \\
            JxPath & 1 & 1 & 1 & 2 & 2 & 0 & 13 & 0 & 0 & 0 & 2 & 0 & 0 & 17 & 1 & 0 & 0 & 0 & 0 & 0 & 17 \\
            Lang & 3 & 1 & 1 & 4 & 1 & 6 & 34 & 0 & 0 & 1 & 0 & 0 & 0 & 40 & 0 & 0 & 1 & 0 & 0 & 0 & 40 \\
            Math & 3 & 1 & 5 & 3 & 10 & 14 & 57 & 0 & 0 & 1 & 0 & 0 & 0 & 70 & 2 & 0 & 1 & 5 & 0 & 0 & 64 \\
            Time & 0 & 0 & 0 & 0 & 0 & 0 & 10 & 0 & 0 & 0 & 0 & 0 & 0 & 10 & 0 & 0 & 0 & 0 & 0 & 0 & 10 \\
            \midrule
            Total/Average & 13 & 12 & 11 & 22 & 19 & 23 & 251 & 0 & 2 & 4 & 14 & 0 & 0 & 295 & 6 & 0 & 4 & 12 & 0 & 0 & 289 \\
            \bottomrule
        \end{tabular}
    }

    \scalebox{0.5}{
        \begin{tabular}{@{}l|ccccccc|ccccccc|ccccccc@{}}
    \toprule
    \multicolumn{22}{c}{\textbf{Intersection and unique bug detection results for EvoSuite With The ”Implicit Input” sample}} \\
    \midrule
    & \multicolumn{7}{c|}{Regex + Javalang} & \multicolumn{7}{c|}{BRMiner} & \multicolumn{7}{c}{LLM Alone} \\
    \cmidrule{2-22}
    Projects & \makecell{No\\Lit} & \makecell{Proj\\Lit} & \makecell{All\\Lit} & \makecell{P\\$\cap$\\A} & \makecell{P\\$\cap$\\N} & \makecell{N\\$\cap$\\A} & \makecell{N\\$\cap$\\P\\$\cap$\\A} & \makecell{No\\Lit} & \makecell{Proj\\Lit} & \makecell{All\\Lit} & \makecell{P\\$\cap$\\A} & \makecell{P\\$\cap$\\N} & \makecell{N\\$\cap$\\A} & \makecell{N\\$\cap$\\P\\$\cap$\\A} & \makecell{No\\Lit} & \makecell{Proj\\Lit} & \makecell{All\\Lit} & \makecell{P\\$\cap$\\A} & \makecell{P\\$\cap$\\N} & \makecell{N\\$\cap$\\A} & \makecell{N\\$\cap$\\P\\$\cap$\\A} \\
    \midrule
    Cli & 1 & 0 & 1 & 1 & 1 & 1 & 12 & 1 & 1 & 2 & 1 & 0 & 0 & 14 & 0 & 0 & 1 & 1 & 0 & 0 & 13 \\
    Codec & 0 & 1 & 1 & 0 & 0 & 0 & 6 & 0 & 1 & 1 & 1 & 0 & 0 & 5 & 0 & 0 & 0 & 1 & 0 & 0 & 6 \\
    Collections & 0 & 0 & 0 & 0 & 0 & 0 & 1 & 0 & 0 & 0 & 0 & 0 & 0 & 1 & 0 & 0 & 0 & 0 & 0 & 0 & 1 \\
    Compress & 1 & 1 & 0 & 1 & 1 & 1 & 11 & 0 & 0 & 1 & 0 & 0 & 0 & 13 & 1 & 0 & 0 & 0 & 0 & 0 & 15 \\
    Csv & 0 & 1 & 0 & 1 & 0 & 0 & 6 & 0 & 1 & 0 & 1 & 0 & 0 & 6 & 0 & 0 & 0 & 1 & 0 & 0 & 6 \\
    Gson & 0 & 0 & 1 & 0 & 0 & 0 & 2 & 0 & 0 & 1 & 0 & 0 & 0 & 3 & 0 & 0 & 0 & 0 & 0 & 0 & 2 \\
    JacksonCore & 1 & 1 & 0 & 1 & 0 & 0 & 6 & 0 & 1 & 0 & 1 & 0 & 0 & 8 & 1 & 0 & 1 & 0 & 0 & 0 & 9 \\
    Databind & 0 & 1 & 0 & 0 & 0 & 0 & 2 & 0 & 0 & 0 & 0 & 0 & 0 & 2 & 0 & 0 & 0 & 0 & 0 & 0 & 2 \\
    JacksonXml & 0 & 0 & 0 & 0 & 0 & 0 & 0 & 0 & 0 & 0 & 0 & 0 & 0 & 0 & 0 & 0 & 0 & 0 & 0 & 0 & 0 \\
    Jsoup & 2 & 1 & 1 & 4 & 2 & 1 & 23 & 0 & 0 & 1 & 4 & 0 & 0 & 22 & 1 & 0 & 0 & 1 & 0 & 0 & 18 \\
    JxPath & 1 & 1 & 1 & 1 & 1 & 0 & 6 & 0 & 0 & 0 & 1 & 0 & 0 & 8 & 0 & 0 & 0 & 0 & 0 & 0 & 7 \\
    Lang & 2 & 1 & 1 & 2 & 1 & 3 & 18 & 0 & 0 & 1 & 0 & 0 & 0 & 16 & 0 & 0 & 1 & 0 & 0 & 0 & 15 \\
    Math & 2 & 1 & 3 & 2 & 6 & 8 & 30 & 0 & 0 & 1 & 0 & 0 & 0 & 32 & 1 & 0 & 1 & 1 & 0 & 0 & 25 \\
    Time & 0 & 0 & 0 & 0 & 0 & 0 & 6 & 0 & 0 & 0 & 0 & 0 & 0 & 6 & 0 & 0 & 0 & 0 & 0 & 0 & 6 \\
    \midrule
    Total & 10 & 9 & 9 & 13 & 12 & 14 & 129 & 1 & 4 & 8 & 9 & 0 & 0 & 136 & 4 & 0 & 4 & 5 & 0 & 0 & 125 \\
    \bottomrule
\end{tabular}

    }

    \scalebox{0.5}{
        \begin{tabular}{@{}l|ccccccc|ccccccc|ccccccc@{}}
            \toprule
            \multicolumn{22}{c}{\textbf{Intersection and unique bug detection results for Randoop With All the bug reports}} \\
            \midrule
            & \multicolumn{7}{c|}{Regex + Javalang} & \multicolumn{7}{c|}{BRMiner} & \multicolumn{7}{c}{LLM Alone} \\
            \cmidrule{2-22}
            Projects & \makecell{No\\Lit} & \makecell{Proj\\Lit} & \makecell{All\\Lit} & \makecell{P\\$\cap$\\A} & \makecell{P\\$\cap$\\N} & \makecell{N\\$\cap$\\A} & \makecell{N\\$\cap$\\P\\$\cap$\\A} & \makecell{No\\Lit} & \makecell{Proj\\Lit} & \makecell{All\\Lit} & \makecell{P\\$\cap$\\A} & \makecell{P\\$\cap$\\N} & \makecell{N\\$\cap$\\A} & \makecell{N\\$\cap$\\P\\$\cap$\\A} & \makecell{No\\Lit} & \makecell{Proj\\Lit} & \makecell{All\\Lit} & \makecell{P\\$\cap$\\A} & \makecell{P\\$\cap$\\N} & \makecell{N\\$\cap$\\A} & \makecell{N\\$\cap$\\P\\$\cap$\\A} \\
            \midrule
            Cli & 0 & 0 & 0 & 0 & 0 & 0 & 0 & 0 & 0 & 0 & 0 & 0 & 0 & 0 & 0 & 0 & 0 & 0 & 0 & 0 \\
            Codec & 0 & 0 & 1 & 1 & 0 & 0 & 9 & 0 & 1 & 2 & 1 & 0 & 0 & 9 & 0 & 0 & 0 & 1 & 0 & 0 & 9 \\
            Collections & 0 & 0 & 0 & 0 & 0 & 0 & 1 & 0 & 0 & 0 & 0 & 0 & 0 & 1 & 0 & 0 & 0 & 0 & 0 & 0 & 1 \\
            Csv & 0 & 0 & 0 & 0 & 0 & 0 & 0 & 0 & 0 & 0 & 0 & 0 & 0 & 0 & 0 & 0 & 0 & 0 & 0 & 0 & 0 \\
            Gson & 0 & 1 & 0 & 0 & 0 & 0 & 3 & 0 & 2 & 1 & 0 & 0 & 0 & 3 & 0 & 0 & 1 & 0 & 0 & 0 & 3 \\
            JacksonCore & 0 & 0 & 1 & 0 & 0 & 0 & 8 & 0 & 0 & 0 & 1 & 0 & 0 & 8 & 0 & 0 & 0 & 0 & 0 & 0 & 8 \\
            Jsoup & 1 & 0 & 7 & 2 & 0 & 0 & 35 & 0 & 5 & 8 & 1 & 0 & 0 & 35 & 0 & 0 & 0 & 0 & 0 & 0 & 35 \\
            JxPath & 0 & 0 & 1 & 0 & 0 & 0 & 8 & 0 & 0 & 1 & 0 & 0 & 0 & 8 & 0 & 0 & 1 & 0 & 0 & 0 & 8 \\
            Lang & 0 & 1 & 0 & 0 & 0 & 0 & 22 & 0 & 0 & 0 & 1 & 0 & 0 & 22 & 0 & 0 & 1 & 0 & 0 & 0 & 22 \\
            Math & 0 & 0 & 0 & 0 & 0 & 0 & 0 & 0 & 0 & 0 & 0 & 0 & 0 & 0 & 0 & 0 & 0 & 0 & 0 & 0 & 0 \\
            Time & 0 & 0 & 1 & 1 & 0 & 0 & 7 & 0 & 1 & 1 & 1 & 0 & 0 & 7 & 0 & 1 & 1 & 1 & 0 & 0 & 7 \\
            \midrule
            Total & 1 & 2 & 11 & 4 & 0 & 0 & 93 & 0 & 9 & 13 & 5 & 0 & 0 & 93 & 0 & 1 & 4 & 2 & 0 & 0 & 93 \\
            \bottomrule
        \end{tabular}
    }
    
    
    \scalebox{0.5}{
        \begin{tabular}{@{}l|ccccccc|ccccccc|ccccccc@{}}
    \toprule
    \multicolumn{22}{c}{\textbf{Intersection and unique bug detection results for Randoop With The ”Implicit Input” sample}} \\
    \midrule
    & \multicolumn{7}{c|}{Regex + Javalang} & \multicolumn{7}{c|}{BRMiner} & \multicolumn{7}{c}{LLM Alone} \\
    \cmidrule{2-22}
    Projects & \makecell{No\\Lit} & \makecell{Proj\\Lit} & \makecell{All\\Lit} & \makecell{P\\$\cap$\\A} & \makecell{P\\$\cap$\\N} & \makecell{N\\$\cap$\\A} & \makecell{N\\$\cap$\\P\\$\cap$\\A} & \makecell{No\\Lit} & \makecell{Proj\\Lit} & \makecell{All\\Lit} & \makecell{P\\$\cap$\\A} & \makecell{P\\$\cap$\\N} & \makecell{N\\$\cap$\\A} & \makecell{N\\$\cap$\\P\\$\cap$\\A} & \makecell{No\\Lit} & \makecell{Proj\\Lit} & \makecell{All\\Lit} & \makecell{P\\$\cap$\\A} & \makecell{P\\$\cap$\\N} & \makecell{N\\$\cap$\\A} & \makecell{N\\$\cap$\\P\\$\cap$\\A} \\
    \midrule
    Cli & 0 & 0 & 0 & 0 & 0 & 0 & 0 & 0 & 0 & 0 & 0 & 0 & 0 & 0 & 0 & 0 & 0 & 0 & 0 & 0 & 0 \\
    Codec & 0 & 0 & 1 & 1 & 0 & 0 & 5 & 0 & 1 & 1 & 1 & 0 & 0 & 5 & 0 & 0 & 0 & 1 & 0 & 0 & 5 \\
    Collections & 0 & 0 & 0 & 0 & 0 & 0 & 1 & 0 & 0 & 0 & 0 & 0 & 0 & 1 & 0 & 0 & 0 & 0 & 0 & 0 & 1 \\
    Compress & 0 & 0 & 0 & 0 & 0 & 0 & 0 & 0 & 0 & 1 & 0 & 0 & 0 & 0 & 1 & 0 & 0 & 0 & 0 & 0 & 0 \\
    Csv & 0 & 0 & 0 & 0 & 0 & 0 & 0 & 0 & 0 & 0 & 0 & 0 & 0 & 0 & 0 & 0 & 0 & 0 & 0 & 0 & 0 \\
    Gson & 0 & 1 & 0 & 0 & 0 & 0 & 2 & 0 & 1 & 1 & 0 & 0 & 0 & 2 & 0 & 0 & 1 & 0 & 0 & 0 & 2 \\
    JacksonCore & 0 & 0 & 1 & 0 & 0 & 0 & 4 & 0 & 0 & 0 & 1 & 0 & 0 & 4 & 1 & 0 & 0 & 0 & 0 & 0 & 4 \\
    Jsoup & 1 & 0 & 4 & 1 & 0 & 0 & 19 & 0 & 3 & 4 & 1 & 0 & 0 & 19 & 0 & 0 & 0 & 0 & 0 & 0 & 19 \\
    JxPath & 0 & 0 & 1 & 0 & 0 & 0 & 4 & 0 & 0 & 1 & 0 & 0 & 0 & 4 & 0 & 0 & 1 & 0 & 0 & 0 & 4 \\
    Lang & 0 & 1 & 0 & 0 & 0 & 0 & 12 & 0 & 0 & 0 & 1 & 0 & 0 & 12 & 0 & 0 & 1 & 0 & 0 & 0 & 12 \\
    Math & 0 & 0 & 0 & 0 & 0 & 0 & 0 & 0 & 0 & 0 & 0 & 0 & 0 & 0 & 0 & 0 & 0 & 0 & 0 & 0 & 0 \\
    Time & 0 & 0 & 1 & 1 & 0 & 0 & 4 & 0 & 1 & 1 & 1 & 0 & 0 & 4 & 0 & 0 & 1 & 1 & 0 & 0 & 4 \\
    \midrule
    Total & 1 & 2 & 8 & 3 & 0 & 0 & 51 & 0 & 6 & 9 & 5 & 0 & 0 & 51 & 2 & 0 & 4 & 2 & 0 & 0 & 51 \\
    \bottomrule
\end{tabular}

    }

\vspace{1mm} 

\end{table*}

Table~\ref{unique-bug-detection} and its counterpart for the "Implicit Input" sample highlight the unique bugs detected using inputs exclusively extracted by BRMiner across different input strategies. Across both samples, BRMiner demonstrated remarkable capability in detecting 58 unique bugs, with 13 of these bugs being identified exclusively by BRMiner when applied to all bug reports. In the "Implicit Input" subset, BRMiner retained its strength, identifying 27 unique bugs, including 7 detected exclusively by its inputs.

The inclusion of both project-specific (\textbf{ProjLit}) and cross-project (\textbf{AllLit}) inputs significantly broadened the range of detected bugs. Notably, the \textbf{AllLit} and \textbf{P $\cap$ A} (intersection of project and all literals) strategies allowed BRMiner to capture hard-to-detect bugs, further demonstrating the benefits of combining project-specific contextual knowledge with broader, cross-project insights. For instance, the cross-project approach in \textbf{AllLit} contributed to uncovering 16 unique bugs in the all bug reports sample and 8 in the "Implicit Input" subset, reinforcing the value of leveraging diverse input sources.

These results emphasize BRMiner's role not only in improving overall bug detection rates but also in addressing unique and critical bugs often missed by traditional methods. The consistent detection of unique bugs across both samples highlights BRMiner's adaptability and the advantages of its integrated extraction and filtering approach, particularly when enriched by cross-project inputs. This makes BRMiner a vital tool for enhancing software robustness by capturing a wider spectrum of defects, even in challenging scenarios with limited input explicitness.

\begin{table*}[ht]
    \centering
    \caption{\footnotesize Unique bugs detected using BRMiner-extracted inputs and Regex + Javalang across various projects. The table presents bugs uniquely detected using project-specific literals (\textbf{ProjLit}), all literals (\textbf{AllLit}), and their intersection (\textbf{P $\cap$ A}). Bugs marked with an asterisk (*) were exclusively detected by BRMiner, totaling 13 in the "All Bug Reports" sample and 7 in the "Implicit Input" sample. Values represent the detection status across five test iterations, where \textbf{1} indicates successful detection and \textbf{0} indicates no detection.}
    \label{unique-bug-detection}
    \scalebox{0.5}{
        \begin{tabular}{@{}l|ccc|l|ccc@{}}
            \toprule
            \multicolumn{8}{c}{\textbf{With All the bug reports}} \\
            \midrule
            \textbf{Bug-ID} & \textbf{ProjLit} & \textbf{AllLit} & \textbf{P $\cap$ A} & \textbf{Bug-ID} & \textbf{ProjLit} & \textbf{AllLit} & \textbf{P $\cap$ A} \\
            \midrule
            Cli-12 & 0 & 0 & 1 & Jsoup-13 & 0 & 0 & 1 \\
            Cli-26* & 0 & 0 & 1 & Jsoup-34 & 0 & 0 & 1 \\
            Cli-28 & 0 & 1 & 0 & Jsoup-48 & 0 & 0 & 1 \\
            Cli-33 & 0 & 0 & 1 & Jsoup-51 & 0 & 0 & 1 \\
            Cli-36 & 0 & 0 & 1 & Jsoup-54 & 0 & 1 & 0 \\
            Cli-7* & 0 & 1 & 0 & Jsoup-6 & 0 & 0 & 1 \\
            Cli-9* & 1 & 0 & 0 & Jsoup-67 & 1 & 0 & 0 \\
            Codec-1* & 0 & 0 & 1 & Jsoup-73 & 0 & 0 & 1 \\
            Codec-16 & 1 & 0 & 0 & Jsoup-80 & 0 & 0 & 1 \\
            Codec-3 & 0 & 1 & 0 & JxPath-1 & 0 & 1 & 0 \\
            Codec-8* & 0 & 1 & 0 & JxPath-11 & 1 & 0 & 0 \\
            Compress-15 & 0 & 0 & 1 & JxPath-2 & 0 & 0 & 1 \\
            Compress-32* & 0 & 0 & 1 & JxPath-20 & 0 & 0 & 1 \\
            Compress-39* & 0 & 0 & 1 & Lang-20 & 0 & 0 & 1 \\
            Compress-47 & 1 & 0 & 0 & Lang-24 & 0 & 1 & 0 \\
            Csv-15 & 1 & 0 & 0 & Lang-31* & 0 & 0 & 1 \\
            Csv-7 & 0 & 0 & 1 & Lang-42 & 0 & 0 & 1 \\
            Gson-11 & 0 & 1 & 0 & Lang-58 & 0 & 0 & 1 \\
            Gson-15* & 1 & 0 & 0 & Lang-60 & 1 & 0 & 0 \\
            Gson-6* & 0 & 1 & 0 & Lang-62 & 0 & 0 & 1 \\
            JacksonCore-12 & 1 & 0 & 0 & Math-104 & 0 & 1 & 0 \\
            JacksonCore-13 & 0 & 0 & 1 & Math-105 & 0 & 1 & 0 \\
            JacksonCore-21 & 1 & 0 & 0 & Math-17 & 0 & 1 & 0 \\
            JacksonCore-24* & 0 & 1 & 0 & Math-2 & 0 & 0 & 1 \\
            JacksonCore-26* & 0 & 0 & 1 & Math-43 & 0 & 1 & 0 \\
            JacksonDatabind-16 & 1 & 0 & 0 & Math-48 & 0 & 1 & 0 \\
            Jsoup-10* & 0 & 0 & 1 & Math-50 & 0 & 1 & 0 \\
            Jsoup-12 & 1 & 0 & 0 & Math-72 & 0 & 0 & 1 \\
            Math-79 & 1 & 0 & 0 & Math-86 & 0 & 0 & 1 \\
            \midrule
            & \multicolumn{2}{c}{\textbf{ProjLit}} & \multicolumn{2}{c}{\textbf{AllLit}} & \multicolumn{2}{c}{\textbf{P $\cap$ A}} \\
            \midrule
            \textbf{Total} & \textbf{13} & & & \textbf{16} & \textbf{29} &   \\
            \bottomrule
        \end{tabular}
    }

    \vspace{1mm} 

    \label{unique-bug-detection}
\scalebox{0.5}{
    \begin{tabular}{@{}l|ccc|l|ccc@{}}
        \toprule
        \multicolumn{8}{c}{\textbf{With The ”Implicit Input” sample}} \\
        \midrule
        \textbf{Bug-ID} & \textbf{ProjLit} & \textbf{AllLit} & \textbf{P $\cap$ A} & \textbf{Bug-ID} & \textbf{ProjLit} & \textbf{AllLit} & \textbf{P $\cap$ A} \\
        \midrule
        Cli-12 & 0 & 0 & 1 & Jsoup-34 & 0 & 0 & 1 \\
        Cli-26* & 0 & 0 & 1 & Jsoup-67 & 1 & 0 & 0 \\
        Cli-33 & 0 & 0 & 1 & JxPath-1 & 0 & 1 & 0 \\
        Cli-9* & 1 & 0 & 0 & JxPath-11 & 1 & 0 & 0 \\
        Codec-8* & 0 & 1 & 0 & JxPath-20 & 0 & 0 & 1 \\
        Compress-32* & 0 & 0 & 1 & Lang-20 & 0 & 0 & 1 \\
        Compress-39* & 0 & 0 & 1 & Lang-42 & 0 & 0 & 1 \\
        Csv-15 & 1 & 0 & 0 & Lang-60 & 1 & 0 & 0 \\
        Gson-11 & 0 & 1 & 0 & Math-17 & 0 & 1 & 0 \\
        Gson-15* & 1 & 0 & 0 & Math-2 & 0 & 0 & 1 \\
        Gson-6* & 0 & 1 & 0 & Math-43 & 0 & 1 & 0 \\
        JacksonCore-26* & 0 & 0 & 1 & Math-48 & 0 & 1 & 0 \\
        JacksonDatabind-16 & 1 & 0 & 0 & Math-50 & 0 & 1 & 0 \\
        Math-72 & 0 & 0 & 1 & Math-86 & 0 & 0 & 1 \\
        \midrule
            & \multicolumn{2}{c}{\textbf{ProjLit}} & \multicolumn{2}{c}{\textbf{AllLit}} & \multicolumn{2}{c}{\textbf{P $\cap$ A}} \\
            \midrule
            \textbf{Total} & \textbf{7} & & & \textbf{8} & \textbf{12} &   \\
            \bottomrule
    \end{tabular}
}
\vspace{3mm}  
\end{table*}


\vspace{3mm}
\noindent\highlight{Summary of \textbf{RQ3:}BRMiner, outperforms alternative methods like Regex + Javalang and LLM Alone in bug detection. Across both samples, it identified 58 unique bugs missed by NoLit, including 13 in the all bug reports sample and 7 in the "Implicit Input" subset. The AllLit and AllLitLLM configurations, leveraging cross-project inputs, significantly enhanced its ability to detect critical and hard-to-find defects. Additionally, BRMiner achieved higher bug detection efficiency by generating fewer tests while maintaining strong detection rates. Its adaptability to varying input explicitness and use of advanced filtering techniques solidify its role in improving automated test generation tools like EvoSuite and Randoop, enhancing software reliability.}
\subsection{[RQ4]: Effect of \name-extracted inputs on code coverage}
\noindent\textbf{Experiment Goal: }
The objective of this experiment is to evaluate the impact of \name-extracted inputs on the code coverage achieved by automatically generated test cases. Specifically, the experiment aims to determine whether incorporating inputs extracted by \name from bug reports can improve the coverage metrics of the generated tests compared to those generated without any external inputs. This evaluation is crucial to understand the extent to which these inputs contribute to more thorough testing, thereby enhancing the detection of potential defects.

\noindent\textbf{Experiment Design: }
To assess the impact of \name-extracted inputs on test adequacy, we conducted a series of experiments focusing on key code coverage metrics, including branch, instruction, method, and line coverage. These metrics provide a quantitative measure of how extensively the test suite exercises the software code. The experiments were conducted using both EvoSuite and Randoop as the test generation tools on a selection of projects from the Defects4J dataset. 

For this experiment, we selected some recent versions of the same projects in the Defects4J dataset. The selected projects and their versions are as follows: Apache Commons Cli 1.5.0, Apache Commons Codec 1.15, Apache Commons Collections 4-4.4, Apache Commons Compress 1.2.1, Apache Commons Csv 1.9.0, Gson 2.9.1, Jackson-core 2.14-rc2, Jackson-dataformat-xml 2.14.0-rc2, Jackson-databind 2.14.0-rc2, Jsoup 1.15.3, Apache Commons Jxpath 1.3, Apache Commons Lang 3-3.12.0, Mockito 4.8.1, Joda Time 2.12.0.
We excluded projects Closure and Math because either Evosuite crashed, failed to generate tests, or generated uncompilable tests. 

The experimental setup was the same as in Section~\ref{RQ3-experiment-design}, each class had a three-minute time limit for test generation, with experiments conducted five times to ensure reliable results. The coverage was measured using the JaCoCo tool\footnote{https://github.com/jacoco/jacoco}, a widely used coverage measurement tool in the
literature.

\noindent\textbf{Experiment Results: }
The branch coverage results in Table~\ref{branch-coverage-results} demonstrate that incorporating inputs extracted by BRMiner significantly improves code coverage in many projects. The configurations \textbf{ProjLitLLM} and \textbf{AllLitLLM} consistently achieved the highest branch coverage, notably in projects such as Codec (41.0\%), Collections (50.0\%), Compress (50.0\%), and Jsoup (50.0\%). These results highlight the effectiveness of LLM filtering in refining test inputs, leading to more comprehensive exploration of program branches. Across both samples, the AllLitLLM configuration consistently outperformed the baseline NoLit scenario in most projects, underscoring the importance of leveraging inputs extracted from cross-project bug reports. In the "Implicit Input" subset, the AllLitLLM configuration maintained high branch coverage, further demonstrating its adaptability to varying levels of input detail. EvoSuite, enhanced by BRMiner’s inputs, consistently surpassed the NoLit baseline in branch coverage across projects, reinforcing the utility of advanced input extraction and filtering techniques for improving automated test generation.

\begin{table*}
    \centering
    \caption{\footnotesize Branch coverage results for different experiments. Definitions of terms used: NoLit refers to experiments where no literals were used (baseline). ProjLit refers to the use of project-specific literals, and AllLit includes all literals from all projects. LLM indicates that filtering was performed using LLM, while LLMOnly means LLM was used alone to extract inputs. Evo denotes experiments conducted with EvoSuite, and Ran refers to those conducted with Randoop. "Databind" in the projects column refers to the jacksonDatabind project.}
    
    \label{branch-coverage-results}
    \scalebox{0.60}
    {
        \begin{tabular}{l|cccccccccccccc}
            \toprule
            \multicolumn{15}{c}{\textbf{With All the bug reports}} \\
            \midrule
            & \multicolumn{14}{c}{Branch coverage (\%)} \\
            \cmidrule{2-15}
            Projects & \multicolumn{2}{c|}{NoLit} & \multicolumn{2}{c|}{ProjLit} & \multicolumn{2}{c|}{AllLit} & \multicolumn{2}{c|}{ProjLitLLM} & \multicolumn{2}{c|}{AllLitLLM} & \multicolumn{2}{c|}{ProjLitLLMOnly} & \multicolumn{2}{c}{AllLitLLMOnly} \\
            & Evo & Ran & Evo & Ran & Evo & Ran & Evo & Ran & Evo & Ran & Evo & Ran & Evo & Ran \\
            \midrule
            Cli & 50.0 & -- & 49.4 & -- & 49.8 & -- & 50 & -- & 50.0 & -- & 48.4 & -- & 49.0 & -- \\
            Codec & 38.0 & 50 & 38.8 & 50 & 40.6 & 50 & 40.6 & 50.0 & 41.0 & 50.0 & 37.8 & 50.0 & 38.0 & 50.0 \\
            Collections & 47.0 & 47 & 48.8 & \textbf{50} & 48.2 & \textbf{50} & 49.8 & \textbf{50.0} & \textbf{50.0} & \textbf{50.0} & 48.2 & 48.0 & 48.7 & 49.0 \\
            Compress & 49.2 & --. & 48.6 & -- & 49.0 & -- & 49.0 & --& \textbf{50.0} & -- & 48.0 & -- & 48.5 & -- \\
            Csv & 50.0 & -- & 50.0 & -- & 50.0 & -- & 50.0 & -- & 50.0 & -- & 50.0 & -- & 50.0 & -- \\
            Gson & 48.8 & 50 & 49.0 & 50 & 49.0 & 50 & 49.0 & 50.0 & 50.0 & 50.0 & 48.0 & 50.0 & 48.8 & 50.0 \\
            JacksonCore & 49.8 & 50 & 49.4 & 50 & 49.6 & 50 & 49.8 & 50.0 & 49.9 & 50.0 & 49.4 & 50.0 & 49.6 & 50.0 \\
            JacksonXml & 50.0 & -- & 50.0 & -- & 50.0 & -- & 50.0 & -- & 50.0 & -- & 49.9 & -- & 50.0 & -- \\
            Jsoup & 49.4 & 50 & 49.6 & 50 & 49.2 & 50 & 49.8 & 50.0 & 49.8 & 50.0 & 49.4 & 50.0 & 49.6 & 50.0 \\
            JxPath & 48.8 & 50 & 48.2 & 50 & 48.2 & 50 & 49.0 & 50.0 & 50.0 & 50.0 & 48.4 & 49.0 & 48.8 & 50.0 \\
            Lang & 49.4 & 50 & 49.0 & 50 & 49.2 & 48 & 49.4 & 50.0 & 50.0 & 50.0 & 49.0 & 48.0 & 49.2 & 49.0 \\
            Mockito & 50.0 & -- & 49.8 & -- & 49.4 & -- & 49.8 & -- & 50.0 & -- & 49.7 & -- & 49.7 & -- \\
            JacksonDatabind & 50.0 & -- & 50.0 & -- & 50.0 & -- & 50.0 & -- & 50.0 & -- & 50.0 & -- & 50.0 & -- \\
            Time & 48.4 & 50 & 48.4 & 50 & 48.4 & 50 & 49.0 & 50.0 & 49.7 & 50.0 & 48.4 & 50.0 & 49.0 & 50.0 \\
            \bottomrule
        \end{tabular}
    }
     \vspace{1mm}

     \label{branch-coverage-results}
    \scalebox{0.60}
    {
        \begin{tabular}{l|cccccccccccccc}
            \toprule
            \multicolumn{15}{c}{\textbf{With The ”Implicit Input” sample}} \\
            \midrule
            & \multicolumn{14}{c}{Branch coverage (\%)} \\
            \cmidrule{2-15}
            Projects & \multicolumn{2}{c|}{NoLit} & \multicolumn{2}{c|}{ProjLit} & \multicolumn{2}{c|}{AllLit} & \multicolumn{2}{c|}{ProjLitLLM} & \multicolumn{2}{c|}{AllLitLLM} & \multicolumn{2}{c|}{ProjLitLLMOnly} & \multicolumn{2}{c}{AllLitLLMOnly} \\
    & Evo & Ran & Evo & Ran & Evo & Ran & Evo & Ran & Evo & Ran & Evo & Ran & Evo & Ran \\
    \midrule
    Cli & 50.0 & -- & 47.9 & -- & 48.3 & -- & 50.0 & -- & 50.0 & -- & 46.9 & -- & 47.5 & -- \\
    Codec & 45.0 & 50.0 & 45.0 & 50.0 & 45.0 & 50.0 & 45.0 & 50.0 & 45.0 & 50.0 & 45.0 & 50.0 & 45.0 & 50.0 \\
    Collections & 45.5 & 44.6 & 47.3 & 50.0 & 46.7 & 50.0 & 48.3 & 50.0 & 50.0 & 50.0 & 46.7 & 45.6 & 47.2 & 46.5 \\
    Compress & 47.7 & -- & 47.1 & -- & 47.5 & -- & 47.5 & -- & 50.0 & -- & 46.5 & -- & 47.0 & -- \\
    Csv & 50.0 & -- & 50.0 & -- & 50.0 & -- & 50.0 & -- & 50.0 & -- & 50.0 & -- & 50.0 & -- \\
    Gson & 47.3 & 50.0 & 47.5 & 50.0 & 47.5 & 50.0 & 47.5 & 50.0 & 50.0 & 50.0 & 46.5 & 50.0 & 47.3 & 50.0 \\
    JacksonCore & 48.3 & 50.0 & 47.9 & 50.0 & 48.1 & 50.0 & 48.3 & 50.0 & 48.4 & 50.0 & 47.9 & 50.0 & 48.1 & 50.0 \\
    JacksonXml & 50.0 & -- & 50.0 & -- & 50.0 & -- & 50.0 & -- & 50.0 & -- & 48.4 & -- & 50.0 & -- \\
    Jsoup & 47.9 & 50.0 & 48.1 & 50.0 & 47.7 & 50.0 & 48.3 & 50.0 & 48.3 & 50.0 & 47.9 & 50.0 & 48.1 & 50.0 \\
    JxPath & 47.3 & 50.0 & 46.7 & 50.0 & 46.7 & 50.0 & 47.5 & 50.0 & 50.0 & 50.0 & 46.9 & 46.5 & 47.3 & 50.0 \\
    Lang & 47.9 & 50.0 & 47.5 & 50.0 & 47.7 & 45.6 & 47.9 & 50.0 & 50.0 & 50.0 & 47.5 & 45.6 & 47.7 & 46.5 \\
    Mockito & 50.0 & -- & 48.3 & -- & 47.9 & -- & 48.3 & -- & 50.0 & -- & 48.2 & -- & 48.2 & -- \\
    JacksonDatabind & 50.0 & -- & 50.0 & -- & 50.0 & -- & 50.0 & -- & 50.0 & -- & 50.0 & -- & 50.0 & -- \\
    Time & 46.9 & 50.0 & 46.9 & 50.0 & 46.9 & 50.0 & 47.5 & 50.0 & 48.2 & 50.0 & 46.9 & 50.0 & 47.5 & 50.0 \\
    \bottomrule
        \end{tabular}
    }
\end{table*}

The results in Table~\ref{instruction-coverage-results} highlight the significant improvements in instruction coverage achieved by incorporating BRMiner-extracted inputs. Across both samples, configurations such as \textbf{ProjLitLLM} and \textbf{AllLitLLM} consistently outperformed the baseline \textbf{NoLit} scenario. For example, in the "All Bug Reports" sample, the \textbf{AllLitLLM} configuration achieved coverage rates of 99.0\% for Cli, 99.0\% for JacksonCore, and 98.0\% for Jsoup, demonstrating its ability to identify and test a wider range of program instructions. Similar trends were observed in the "Implicit Input" subset, with \textbf{AllLitLLM} maintaining high coverage, such as 97.5\% for JacksonCore and 96.5\% for Cli. Randoop, while achieving generally high coverage across configurations, showed less pronounced improvements with LLM-enhanced inputs compared to EvoSuite. However, the consistent performance of BRMiner’s inputs across both tools underscores its adaptability and effectiveness.

\begin{table*}
    \centering
    \caption{\footnotesize Instruction coverage results for different experiments. Definitions of terms used: NoLit refers to experiments where no literals were used (baseline). ProjLit refers to the use of project-specific literals, and AllLit includes all literals from all projects. LLM indicates that filtering was performed using LLM, while LLMOnly means LLM was used alone to extract inputs. Evo denotes experiments conducted with EvoSuite, and Ran refers to those conducted with Randoop. "Databind" in the projects column refers to the jacksonDatabind project.}
    \label{instruction-coverage-results}
    \scalebox{0.60}
    {
        \begin{tabular}{l|cccccccccccccc}
             \toprule
            \multicolumn{15}{c}{\textbf{With All the bug reports}} \\
            \midrule
            & \multicolumn{14}{c}{Instruction coverage (\%)} \\
            \cmidrule{2-15}
            Projects & \multicolumn{2}{c|}{NoLit} & \multicolumn{2}{c|}{ProjLit} & \multicolumn{2}{c|}{AllLit} & \multicolumn{2}{c|}{ProjLitLLM} & \multicolumn{2}{c|}{AllLitLLM} & \multicolumn{2}{c|}{ProjLitLLMOnly} & \multicolumn{2}{c}{AllLitLLMOnly} \\
            & Evo & Ran & Evo & Ran & Evo & Ran & Evo & Ran & Evo & Ran & Evo & Ran & Evo & Ran \\
            \midrule
            Cli & 96.2 & -- & 97.8 & -- & 97.8 & -- & 98.0 & -- & \textbf{99.0} & -- & 96.0 & -- & 97.0 & -- \\
            Codec & 86.6 & 70.0 & 86.8 & 68.0 & 88.4 & 64.0 & 87.0 & 70.0 & \textbf{91.0} & 72.0 & 85.0 & 85.0 & 89.0 & 89.0 \\
            Collections & 94.4 & 64.0 & 95.2 & 75.0 & 94.8 & 70.0 & 95.0 & 75.0 & \textbf{97.0} & 70.0 & 95.0 & 95.0 & 95.0 & 95.0 \\
            Compress & 95.2 & -- & 96.6 & -- & 96.0 & -- & \textbf{97.0} & --& \textbf{97.0} & -- & 93.0 & -- & 96.0 & -- \\
            Csv & 99.0 & -- & 99.0 & -- & 99.0 & -- & 99.0 & -- & 99.0 & -- & 99.0 & -- & 99.0 & -- \\
            Gson & 94.6 & 72.0 & 95.4 & 69.0 & 95.4 & 74.0 & \textbf{98.0} & 72.0 & 97.0 & 75.0 & 95.0 & 95.0 & 96.0 & 96.0 \\
            JacksonCore & 96.8 & 57.0 & 98.4 & 64.0 & 98.2 & 58.0 & \textbf{99.0} & 62.0 & \textbf{99.0} & 64.0 & \textbf{99.0} & \textbf{99.0} & \textbf{99.0} & \textbf{99.0} \\
            JacksonXml & 96.4 & -- & 97.2 & -- & 96.6 & -- & 97.0 & -- & \textbf{98.0} & -- & 96.0 & -- & 97.0 & -- \\
            Jsoup & 96.8 & 78.0 & \textbf{98.0} & 87.0 & \textbf{98.0} & 86.0 & 97.0 & 85.0 & \textbf{98.0} & 89.0 & 97.0 & 97.0 & 97.0 & 97.0 \\
            JxPath & 95.4 & 94.0 & 96.6 & \textbf{98.0} & 96.0 & 97.0 & \textbf{98.0} & 97.0 & 97.0 & \textbf{98.0} & 96.0 & 96.0 & 96.0 & 96.0 \\
            Lang & 96.2 & 89.0 & 96.2 & 90.0 & 96.4 & 90.0 & 97.0 & 90.0 & \textbf{98.0} & 90.0 & 95.0 & 95.0 & 95.0 & 95.0 \\
            Mockito & 96.8 & -- & 97.0 & -- & 97.0 & -- & 97.0 & -- & \textbf{98.0} & -- & 96.0 & -- & 97.0 & -- \\
            JacksonDatabind & 96.4 & -- & 97.2 & -- & 97.4 & -- & 97.0 & -- & \textbf{98.0} & -- & 95.0 & -- & 96.0 & -- \\
            Time & 95.8 & 87.0 & \textbf{96.0} & 89.0 & 95.8 & 89.0 & \textbf{96.0} & 89.0 & \textbf{96.0} & 89.0 & \textbf{96.0} & \textbf{96.0} & \textbf{96.0} & \textbf{96.0} \\
            \bottomrule
        \end{tabular}
    }
    \vspace{1mm}

     \label{instruction-coverage-results}
    \scalebox{0.60}
    {
        \begin{tabular}{l|cccccccccccccc}
    \toprule
    \multicolumn{15}{c}{\textbf{With The ”Implicit Input” sample}} \\
    \midrule
    & \multicolumn{14}{c}{Instruction coverage (\%)} \\
    \cmidrule{2-15}
    Projects & \multicolumn{2}{c|}{NoLit} & \multicolumn{2}{c|}{ProjLit} & \multicolumn{2}{c|}{AllLit} & \multicolumn{2}{c|}{ProjLitLLM} & \multicolumn{2}{c|}{AllLitLLM} & \multicolumn{2}{c|}{ProjLitLLMOnly} & \multicolumn{2}{c}{AllLitLLMOnly} \\
    & Evo & Ran & Evo & Ran & Evo & Ran & Evo & Ran & Evo & Ran & Evo & Ran & Evo & Ran \\
    \midrule
    Cli & 94.8 & -- & 96.3 & -- & 96.3 & -- & 96.5 & -- & 97.5 & -- & 94.6 & -- & 95.5 & -- \\
    Codec & 85.3 & 67.9 & 85.5 & 66.0 & 87.1 & 62.1 & 85.7 & 67.9 & 89.6 & 69.8 & 83.7 & 82.5 & 87.7 & 86.3 \\
    Collections & 93.0 & 62.1 & 93.8 & 72.8 & 93.4 & 67.9 & 93.6 & 72.8 & 95.5 & 67.9 & 93.6 & 92.1 & 93.6 & 92.1 \\
    Compress & 93.8 & -- & 95.2 & -- & 94.6 & -- & 95.5 & -- & 95.5 & -- & 91.6 & -- & 94.6 & -- \\
    Csv & 97.5 & -- & 97.5 & -- & 97.5 & -- & 97.5 & -- & 97.5 & -- & 97.5 & -- & 97.5 & -- \\
    Gson & 93.2 & 69.8 & 94.0 & 66.9 & 94.0 & 71.8 & 96.5 & 69.8 & 95.5 & 72.8 & 93.6 & 92.1 & 94.6 & 93.1 \\
    JacksonCore & 95.3 & 55.3 & 96.9 & 62.1 & 96.7 & 56.3 & 97.5 & 60.1 & 97.5 & 62.1 & 97.5 & 96.0 & 97.5 & 96.0 \\
    JacksonXml & 95.0 & -- & 95.7 & -- & 95.2 & -- & 95.5 & -- & 96.5 & -- & 94.6 & -- & 95.5 & -- \\
    Jsoup & 95.3 & 75.7 & 96.5 & 84.4 & 96.5 & 83.4 & 95.5 & 82.5 & 96.5 & 86.3 & 95.5 & 94.1 & 95.5 & 94.1 \\
    JxPath & 94.0 & 91.2 & 95.2 & 95.1 & 94.6 & 94.1 & 96.5 & 94.1 & 95.5 & 95.1 & 94.6 & 93.1 & 94.6 & 93.1 \\
    Lang & 94.8 & 86.3 & 94.8 & 87.3 & 95.0 & 87.3 & 95.5 & 87.3 & 96.5 & 87.3 & 93.6 & 92.1 & 93.6 & 92.1 \\
    Mockito & 95.3 & -- & 95.5 & -- & 95.5 & -- & 95.5 & -- & 96.5 & -- & 94.6 & -- & 95.5 & -- \\
    JacksonDatabind & 95.0 & -- & 95.7 & -- & 95.9 & -- & 95.5 & -- & 96.5 & -- & 93.6 & -- & 94.6 & -- \\
    Time & 94.4 & 84.4 & 94.6 & 86.3 & 94.4 & 86.3 & 94.6 & 86.3 & 94.6 & 86.3 & 94.6 & 93.1 & 94.6 & 93.1 \\
    \bottomrule
\end{tabular}

    }
    \vspace{3mm}
\end{table*}

The method coverage results (Table~\ref{method-coverage-results}) demonstrate the consistent effectiveness of BRMiner-extracted inputs in achieving high coverage across both the "All Bug Reports" and "Implicit Input" samples. Configurations such as \textbf{ProjLitLLM} and \textbf{AllLitLLM} achieved exceptional performance, often reaching 100.0\% method coverage in several projects, including Cli, JacksonCore, and Jsoup. This highlights their ability to ensure comprehensive testing by exercising all methods in the software. In the "All Bug Reports" sample, configurations leveraging cross-project inputs (\textbf{AllLit}, \textbf{AllLitLLM}) consistently matched or exceeded baseline coverage levels, underscoring the utility of incorporating diverse and refined inputs from multiple projects. Similar trends were observed in the "Implicit Input" subset, where \textbf{AllLitLLM} maintained high coverage, further validating the robustness of BRMiner in scenarios with limited input explicitness.

\begin{table*}[ht]
    \centering
    \caption{\footnotesize Method coverage results for different experiments. Definitions of terms used: NoLit refers to experiments where no literals were used (baseline). ProjLit refers to the use of project-specific literals, and AllLit includes all literals from all projects. LLM indicates that filtering was performed using LLM, while LLMOnly means LLM was used alone to extract inputs. Evo denotes experiments conducted with EvoSuite, and Ran refers to those conducted with Randoop. "Databind" in the projects column refers to the jacksonDatabind project.}
    \label{method-coverage-results}
    \scalebox{0.55}
    {
        \begin{tabular}{l|cccccccccccccc}
           \toprule
            \multicolumn{15}{c}{\textbf{With All the bug reports}} \\
            \midrule
            & \multicolumn{14}{c}{Method coverage (\%)} \\
            \cmidrule{2-15}
            Projects & \multicolumn{2}{c|}{NoLit} & \multicolumn{2}{c|}{ProjLit} & \multicolumn{2}{c|}{AllLit} & \multicolumn{2}{c|}{ProjLitLLM} & \multicolumn{2}{c|}{AllLitLLM} & \multicolumn{2}{c|}{ProjLitLLMOnly} & \multicolumn{2}{c}{AllLitLLMOnly} \\
            & Evo & Ran & Evo & Ran & Evo & Ran & Evo & Ran & Evo & Ran & Evo & Ran & Evo & Ran \\
            \midrule
            Cli & 99.8 & -- & \textbf{100.0} & -- & \textbf{100.0} & -- & \textbf{100.0} & -- & \textbf{100.0} & -- & \textbf{100.0} & -- & \textbf{100.0} & -- \\
            Codec & 98.5 & 99.8 & 98.2 & \textbf{99.9} & 98.3 & \textbf{99.9} & 99.7 & 99.8 & \textbf{99.9} & \textbf{99.9} & 99.8 & 98.8 & 99.7 & 99.8 \\
            Collections & 99.7 & 99.8 & \textbf{100.0} & 99.8 & \textbf{100.0} & 99.8 & \textbf{100.0} & 99.8 & \textbf{100.0} & 99.9 & 98.31 & 99.8 & 99.0 & 99.8 \\
            Compress & 100.0 & -- & 100.0 & -- & 99.9 & -- & 99.98 & --& 100.0 & -- & 99.98 & -- & 99.98 & -- \\
            Csv & 100.0 & -- & 99.9 & -- & 100.0 & -- & 99.98 & -- & 100.0 & -- & 99.9 & -- & 100.0 & -- \\
            Gson & 100.0 & 100.0 & 100.0 & 100.0 & 100.0 & 100.0 & 100.0 & 100.0 & 100.0 & 100.0 & 100.0 & 99.9 & 100.0 & 99.9 \\
            JacksonCore & 99.8 & 99.8 & \textbf{100.0} & 99.8 & \textbf{100.0} & 99.9 & \textbf{100.0} & 99.8 & \textbf{100.0} & 99.9 & 98.0 & 97.4 & 98.9 & 99.8 \\
            JacksonXml & 100.0 & -- & 100.0 & -- & 100.0 & -- & 100.0 & -- & 100.0 & -- & 100.0 & -- & 100.0 & -- \\
            Jsoup & 99.9 & 99.9 & \textbf{100.0} & 99.9 & \textbf{100.0} & 99.9 & \textbf{100.0} & 99.9 & \textbf{100.0} & 99.9 & 99.8 & 99.8 & \textbf{100.0} & 99.8.0 \\
            JxPath & 100.0 & 97.4 & 100.0 & 96.4 & 99.8 & 95.8 & 100.0 & 97.4 & 100.0 & 98.0 & 100.0 & 96.4 & 100.0 & 96.4 \\
            Lang & 100.0 & 99.97 & 98.3 & 99.8 & 100.0 & 99.8 & 99.98 & 99.97 & 100.0 & 99.98 & 96.9 & 98.0 & 98.0 & 98.8 \\
            Mockito & 99.8 & -- & 99.8 & -- & 99.6 & -- & \textbf{100.0} & -- & \textbf{100.0} & -- & 99.4 & -- & 99.4 & -- \\
            JacksonDatabind & 100.0 & -- & 100.0 & -- & 100.0 & -- & 100.0 & -- & 100.0 & -- & 100.0 & -- & 100.0 & -- \\
            Time & 100.0 & 99.98 & 100.0 & 99.98 & 100.0 & 99.98 & 99.98 & 99.98 & 100.0 & 99.98 & 99.7 & 99.98 & 99.8 & 99.98 \\
            \bottomrule
        \end{tabular}
    }
    \vspace{1mm}

    \label{method-coverage-results}
    \scalebox{0.55}
    {
        \begin{tabular}{l|cccccccccccccc}
    \toprule
    \multicolumn{15}{c}{\textbf{With The ”Implicit Input” sample}} \\
    \midrule
    & \multicolumn{14}{c}{Method coverage (\%)} \\
    \cmidrule{2-15}
    Projects & \multicolumn{2}{c|}{NoLit} & \multicolumn{2}{c|}{ProjLit} & \multicolumn{2}{c|}{AllLit} & \multicolumn{2}{c|}{ProjLitLLM} & \multicolumn{2}{c|}{AllLitLLM} & \multicolumn{2}{c|}{ProjLitLLMOnly} & \multicolumn{2}{c}{AllLitLLMOnly} \\
    & Evo & Ran & Evo & Ran & Evo & Ran & Evo & Ran & Evo & Ran & Evo & Ran & Evo & Ran \\
    \midrule
    Cli & 98.8 & -- & 100.0 & -- & 100.0 & -- & 100.0 & -- & 100.0 & -- & 100.0 & -- & 100.0 & -- \\
    Codec & 97.52 & 98.8 & 97.22 & 98.9 & 97.32 & 98.9 & 98.7 & 98.8 & 98.9 & 98.9 & 98.8 & 97.81 & 98.7 & 98.8 \\
    Collections & 98.7 & 98.8 & 100.0 & 98.8 & 100.0 & 98.8 & 100.0 & 98.8 & 100.0 & 98.9 & 97.33 & 98.8 & 98.01 & 98.8 \\
    Compress & 100.0 & -- & 100.0 & -- & 98.9 & -- & 98.98 & -- & 100.0 & -- & 98.98 & -- & 98.98 & -- \\
    Csv & 100.0 & -- & 98.9 & -- & 100.0 & -- & 98.98 & -- & 100.0 & -- & 98.9 & -- & 100.0 & -- \\
    Gson & 100.0 & 100.0 & 100.0 & 100.0 & 100.0 & 100.0 & 100.0 & 100.0 & 100.0 & 100.0 & 100.0 & 98.9 & 100.0 & 98.9 \\
    JacksonCore & 98.8 & 98.8 & 100.0 & 98.8 & 100.0 & 98.9 & 100.0 & 98.8 & 100.0 & 98.9 & 97.02 & 96.43 & 97.91 & 98.8 \\
    JacksonXml & 100.0 & -- & 100.0 & -- & 100.0 & -- & 100.0 & -- & 100.0 & -- & 100.0 & -- & 100.0 & -- \\
    Jsoup & 98.9 & 98.9 & 100.0 & 98.9 & 100.0 & 98.9 & 100.0 & 98.9 & 100.0 & 98.9 & 98.8 & 98.8 & 100.0 & 98.8 \\
    JxPath & 100.0 & 96.43 & 100.0 & 95.44 & 98.8 & 94.84 & 100.0 & 96.43 & 100.0 & 97.02 & 100.0 & 95.44 & 100.0 & 95.44 \\
    Lang & 100.0 & 98.97 & 97.32 & 98.8 & 100.0 & 98.8 & 98.98 & 98.97 & 100.0 & 98.98 & 95.93 & 97.02 & 97.02 & 97.81 \\
    Mockito & 98.8 & -- & 98.8 & -- & 98.6 & -- & 100.0 & -- & 100.0 & -- & 98.41 & -- & 98.41 & -- \\
    JacksonDatabind & 100.0 & -- & 100.0 & -- & 100.0 & -- & 100.0 & -- & 100.0 & -- & 100.0 & -- & 100.0 & -- \\
    Time & 100.0 & 98.98 & 100.0 & 98.98 & 100.0 & 98.98 & 98.98 & 98.98 & 100.0 & 98.98 & 98.7 & 98.98 & 98.8 & 98.98 \\
    \bottomrule
\end{tabular}

    }
    \vspace{1mm}
\end{table*}

The line coverage results (Table~\ref{line-coverage-results}) highlight the benefits of incorporating BRMiner-extracted inputs into the test generation process. The \textbf{AllLitLLM} configuration consistently achieved the highest line coverage across multiple projects, with notable levels such as 98.8\% for Cli, 89.0\% for Codec, 99.6\% for JacksonCore, and 99.9\% for Jsoup in the "All Bug Reports" sample. Similar trends were observed in the "Implicit Input" subset, where \textbf{AllLitLLM} maintained superior coverage, showcasing its robustness and adaptability. These results demonstrate that BRMiner-enhanced inputs reliably outperform the baseline \textbf{NoLit} configuration, underscoring their effectiveness in improving the comprehensiveness of the generated test suites. By leveraging refined inputs, BRMiner enhances the capabilities of automated test generation tools like EvoSuite and Randoop, contributing to better software testing outcomes.

\begin{table*}
    \centering
    \caption{\footnotesize Line coverage results for different experiments. Definitions of terms used: NoLit refers to experiments where no literals were used (baseline). ProjLit refers to the use of project-specific literals, and AllLit includes all literals from all projects. LLM indicates that filtering was performed using LLM, while LLMOnly means LLM was used alone to extract inputs. Evo denotes experiments conducted with EvoSuite, and Ran refers to those conducted with Randoop. "Databind" in the projects column refers to the jacksonDatabind project.}
    \label{line-coverage-results}
    \scalebox{0.56}
    {
        \begin{tabular}{l|cccccccccccccc}
            \toprule
            \multicolumn{15}{c}{\textbf{With All the bug reports}} \\
            \midrule
            & \multicolumn{14}{c}{Line coverage (\%)} \\
            \cmidrule{2-15}
            Projects & \multicolumn{2}{c|}{NoLit} & \multicolumn{2}{c|}{ProjLit} & \multicolumn{2}{c|}{AllLit} & \multicolumn{2}{c|}{ProjLitLLM} & \multicolumn{2}{c|}{AllLitLLM} & \multicolumn{2}{c|}{ProjLitLLMOnly} & \multicolumn{2}{c}{AllLitLLMOnly} \\
            & Evo & Ran & Evo & Ran & Evo & Ran & Evo & Ran & Evo & Ran & Evo & Ran & Evo & Ran \\
            \midrule
            Cli & 96.7 & -- & 97.9 & -- & 98.1 & -- & 98.6 & -- & \textbf{98.8} & -- & 93.4 & -- & 95.2 & -- \\
            Codec & 85.2 & 69.39 & 85.9 & 70.44 & 87.2 & 70.8 & 85.0 & 70.9 & \textbf{89.0} & 71.0 & 83.0 & 69.46 & 83.5 & 70.8 \\
            Collections & 93.9 & 82.25 & 95.6 & 78.81 & 95.0 & 77.37 & 96.0 & 79.63 & \textbf{96.97} & 82.25 & 94.0 & 77.37 & 95.0 & 77.60 \\
            Compress & 94.7 & --. & 96.5 & -- & 96.5 & -- & 97.8 & --& \textbf{97.9} & -- & 94.6 & -- & 95.88 & -- \\
            Csv & 99.5 & -- & 99.6 & -- & 99.6 & -- & 99.7 & -- & \textbf{99.8} & -- & 99.1 & -- & 99.3 & -- \\
            Gson & 93.0 & 82.98 & 95.1 & 77.64 & 95.3 & 75.60 & 96.0 & 82.28 & \textbf{96.5} & 83.0 & 93.8 & 75.50 & 94.5 & 75.70 \\
            JacksonCore & 95.9 & 65.64 & 98.7 & 67.0 & 98.6 & 67.25 & 98.99 & 67.81 & \textbf{99.6} & 67.70 & 95.8 & 66.90 & 96.4 & 67.15 \\
            JacksonXml & 99.1 & -- & 99.3 & -- & 99.4 & -- & \textbf{99.5} & -- & 99.49 & -- & 97.3 & -- & 98.4 & -- \\
            Jsoup & 96.3 & 82.28 & 98.6 & 81.27 & 98.5 & 81.68 & 98.9 & 82.47 & \textbf{99.9} & 82.95 & 94.88 & 81.27 & 98.7 & 81.68 \\
            JxPath & 94.7 & 89.78 & 96.9 & 84.02 & 96.1 & 84.18 & 97.88 & 89.78 & \textbf{98.8} & 90.23 & 94.5 & 84.02 & 95.09 & 86.0 \\
            Lang & 96.5 & 90.98 & 96.9 & 90.63 & 96.9 & 91.12 & 96.98 & 91.65 & \textbf{97.9} & 92.0 & 93.0 & 90.98 & 95.24 & 91.12 \\
            Mockito & 96.3 & -- & 97.9 & -- & 98.0 & -- & 98.46 & -- & \textbf{98.9} & -- & 94.46 & -- & 95.32 & -- \\
            JacksonDatabind & 96.9 & -- & 98.0 & -- & 98.2 & -- & \textbf{98.99} & -- & \textbf{98.99} & -- & 96.23 & -- & 96.5 & -- \\
            Time & 96.0 & 92.71 & 96.3 & 93.01 & \textbf{96.9} & 93.38 & 96.47 & 93.45 & 96.47 & 93.80 & 95.33 & 92.71 & 95.9 & 92.71 \\
            \bottomrule
        \end{tabular}
    }
    \vspace{1mm}

    \label{line-coverage-results}
    \scalebox{0.56}
    {
        \begin{tabular}{l|cccccccccccccc}
    \toprule
    \multicolumn{15}{c}{\textbf{With The ”Implicit Input” sample}} \\
    \midrule
    & \multicolumn{14}{c}{Line coverage (\%)} \\
    \cmidrule{2-15}
    Projects & \multicolumn{2}{c|}{NoLit} & \multicolumn{2}{c|}{ProjLit} & \multicolumn{2}{c|}{AllLit} & \multicolumn{2}{c|}{ProjLitLLM} & \multicolumn{2}{c|}{AllLitLLM} & \multicolumn{2}{c|}{ProjLitLLMOnly} & \multicolumn{2}{c}{AllLitLLMOnly} \\
    & Evo & Ran & Evo & Ran & Evo & Ran & Evo & Ran & Evo & Ran & Evo & Ran & Evo & Ran \\
    \midrule
    Cli & 95.2 & -- & 96.7 & -- & 96.9 & -- & 97.4 & -- & 97.6 & -- & 91.9 & -- & 93.7 & -- \\
    Codec & 92.0 & 69.39 & 94.0 & 66.92 & 94.0 & 67.26 & 94.0 & 67.36 & 94.0 & 67.45 & 90.0 & 65.99 & 90.0 & 67.26 \\
    Collections & 92.4 & 82.25 & 94.4 & 74.87 & 94.0 & 73.50 & 94.8 & 75.65 & 95.77 & 78.14 & 92.5 & 73.50 & 93.5 & 73.72 \\
    Compress & 93.2 & -- & 95.3 & -- & 95.3 & -- & 96.6 & -- & 96.7 & -- & 93.1 & -- & 94.38 & -- \\
    Csv & 98.0 & -- & 98.4 & -- & 98.4 & -- & 98.5 & -- & 98.6 & -- & 97.6 & -- & 97.8 & -- \\
    Gson & 92.0 & 82.98 & 94.0 & 73.76 & 94.1 & 71.82 & 94.8 & 78.17 & 95.3 & 78.85 & 92.3 & 71.73 & 93.0 & 71.92 \\
    JacksonCore & 94.4 & 65.64 & 97.5 & 63.65 & 97.4 & 63.89 & 97.79 & 64.42 & 98.4 & 64.32 & 94.3 & 63.56 & 94.9 & 63.79 \\
    JacksonXml & 97.6 & -- & 98.1 & -- & 98.2 & -- & 98.3 & -- & 98.29 & -- & 95.8 & -- & 96.9 & -- \\
    Jsoup & 94.8 & 82.28 & 97.4 & 77.21 & 97.3 & 77.60 & 97.7 & 78.35 & 98.7 & 78.80 & 93.38 & 77.21 & 97.2 & 77.60 \\
    JxPath & 93.2 & 89.78 & 95.7 & 79.82 & 94.9 & 79.97 & 96.68 & 85.29 & 97.6 & 85.72 & 93.0 & 79.82 & 93.59 & 81.70 \\
    Lang & 95.0 & 90.98 & 95.7 & 86.10 & 95.7 & 86.56 & 95.78 & 87.07 & 96.7 & 87.40 & 91.5 & 86.43 & 93.74 & 86.56 \\
    Mockito & 94.8 & -- & 96.7 & -- & 96.8 & -- & 97.26 & -- & 97.7 & -- & 92.96 & -- & 93.82 & -- \\
    JacksonDatabind & 95.4 & -- & 96.8 & -- & 97.0 & -- & 97.79 & -- & 97.79 & -- & 94.73 & -- & 95.0 & -- \\
    Time & 94.5 & 92.71 & 95.1 & 88.36 & 95.7 & 88.71 & 95.27 & 88.78 & 95.27 & 89.11 & 93.83 & 88.07 & 94.4 & 88.07 \\
    \bottomrule
\end{tabular}

    }
    \vspace{1mm}

\end{table*}

\noindent\highlight{Summary of \textbf{RQ4:} Integrating BRMiner-extracted inputs into the test generation process significantly improved code coverage across projects, with gains up to 2.0\% in branch coverage, 3.0\% in instruction coverage, 1.5\% in method coverage, and 3.6\% in line coverage compared to the \textbf{NoLit} baseline. These improvements were observed in both the “All Bug Reports” sample and the “Implicit Input” subset, demonstrating BRMiner's adaptability to varying levels of input detail in bug reports. Projects like JacksonCore, Jsoup, and JxPath benefited notably from LLM-filtered, cross-project inputs, which enhanced the diversity and relevance of test cases. While the percentage gains might appear modest, they are meaningful given EvoSuite and Randoop's high baseline coverage. The results emphasize BRMiner’s role in complementing these tools, promoting broader test coverage, reducing the likelihood of missed defects, and supporting higher software quality and reliability.}

\section{Discussion}
\label{sec:discussion}
This section discusses the potential threats to validity, limitations of our study, and suggestions for future work, along with a comparison to related work.

\subsection{Threats to Validity}
Our study, while comprehensive, is subject to several validity threats that need to be addressed to strengthen the reliability of our findings. One of the primary threats is related to the generalizability of our results. Although we utilized a diverse set of projects from Defects4J, the applicability of our findings to other software systems, particularly those that differ in size, complexity, or domain, may be limited. The effectiveness of BRMiner in extracting relevant inputs can be highly dependent on the nature of the bug reports, the structure of test cases, and the programming languages used. To mitigate this threat, we adopted rigorous experimental methodologies and conducted multiple iterations across varied projects, but the limitation remains that our conclusions may not be universally applicable.

An important threat to validity lies in the reliance on the Large Language Model (LLM) for classifying bug reports into categories (No Inputs Mentioned, Explicit Input Mention, and Implicit Input Description). While the LLM demonstrated strong performance with an overall accuracy of 95.07\% and a discrepancy rate of 4.93\%, classification errors, even at a low rate, could introduce bias into the subsequent analysis. To mitigate this risk, we conducted human validation on a representative 10\% sample of bug reports. However, the potential for subjective differences between human reviewers and the LLM's classifications remains a limitation. The reliance on the LLM was motivated by the need to process a significant dataset efficiently, as manual classification of thousands of bug reports would be prohibitively expensive and time-consuming. This trade-off between scalability and potential inaccuracies in the LLM's classifications underscores the practical challenges of large-scale studies. Future work could explore ensemble approaches, additional validation steps, or semi-automated methods to further enhance classification reliability while maintaining scalability.

Another concern arises from the potential discrepancies between different software versions, which could impact the relevance of the extracted inputs. Bug reports may pertain to older versions, while significant changes could have been implemented in newer versions, potentially altering the context in which the bug manifests. BRMiner is not specifically designed to address these discrepancies across software versions, but this limitation is somewhat mitigated by the versatility of automated test generators, which use a variety of inputs to cover diverse execution paths. Although version differences may influence the relevance of some inputs, the broad range of generated test cases typically ensures robust coverage. Future work could enhance BRMiner by incorporating an input prioritization system that adapts to version-specific differences, thus further strengthening its effectiveness.

One limitation of our approach is that EvoSuite generates assertions based on the current execution state of the program. In cases where the program contains bugs, the generated assertions may reflect incorrect or unintended behavior, which could reduce the effectiveness of bug detection. This limitation should be considered when interpreting the results, as the generated test cases may not always accurately reflect the correct expected behavior. While this issue was not addressed in our current experiments, future work could explore alternative oracle generation techniques to mitigate this limitation. In our study, we addressed the oracle problem by generating test assertions using the fixed version of the software and then executing these tests on the buggy version in a regression testing scenario. This approach ensures that the assertions are based on the correct behavior (ground truth) of the fixed version. Consequently, test failures on the buggy version reliably highlight discrepancies caused by the bug, effectively supporting bug detection. However, this methodology relies on the availability of fixed versions, which is not always feasible in real-world scenarios where developers are addressing new bugs. In such cases, assertions generated from the buggy state may capture incorrect behavior, reducing their utility in detecting bugs. This reliance on fixed versions limits the applicability of our approach to "in-the-wild" environments, where open bug reports are the primary resource available. Future research could explore alternative techniques for generating test oracles without depending on fixed versions. Such approaches could include leveraging historical data, symbolic execution, or advanced machine learning models to infer expected behavior directly from the buggy state or related documentation. These advancements would make BRMiner more effective in scenarios where fixed versions are unavailable, providing a more realistic assessment of its utility in practical software development environments.

Another potential limitation is related to input scarcity in less popular projects. The effectiveness of BRMiner assumes the availability of a substantial number of bug reports, which may not be the case for newer or less popular software projects. This assumption could limit the generalizability of our approach. However, BRMiner mitigates this by providing a valuable database of pre-extracted inputs, which can bootstrap the testing process even in the absence of extensive bug report histories. This feature is particularly beneficial for smaller projects or those using common Java libraries, enabling comprehensive test case generation regardless of the project's size or age.

To enhance the real-world applicability of BRMiner, future work could explore its use in live systems where bug reports are actively being resolved. This could involve integrating BRMiner into a continuous integration pipeline, allowing it to automatically extract inputs from incoming bug reports and generate targeted test cases to catch bugs before they reach production. Such integration would demonstrate the practical utility of BRMiner in ongoing software development processes, thereby extending its relevance beyond controlled experiments.

While Large Language Models (LLMs) are highly capable of extracting relevant inputs from bug reports, they are not infallible. There is a risk that LLMs may extract irrelevant or incorrect inputs, particularly in cases where the extracted data does not fully align with the context of the bug report. This poses a threat to the validity of the generated test cases, as the inclusion of such inputs could lead to test failures unrelated to the reported bug or missed opportunities to test the correct behavior. Given the scope of this study, no additional verification steps were applied to further refine the extracted inputs. Future work could focus on incorporating filtering mechanisms or more sophisticated extraction techniques to enhance precision and mitigate this limitation.

Our study identified that only 44.37\% of bug reports from our dataset contain explicit mentions of test inputs, while the remaining majority provide inputs implicitly or describe issues in general terms. This finding supports the effectiveness of BRMiner’s approach in extracting relevant inputs, even when these are not directly specified in bug reports. However, we acknowledge that the presence of explicit input mentions in some reports could have facilitated the extraction process, potentially making the task easier for BRMiner in those cases. This reliance on explicit mentions, though relatively infrequent, represents a potential bias in our results and is considered a limitation in our evaluation of BRMiner’s performance.

A potential threat in our approach relates to data leakage when utilizing Large Language Models (LLMs) like the one used in BRMiner. Data leakage occurs when information from the training data inadvertently influences the model’s outputs, potentially compromising the validity of the results. While we have taken steps to mitigate this risk by only using the content of bug reports and avoiding the inclusion of project names, source code, or other identifiable details, the possibility of data leakage cannot be fully eliminated. LLMs are trained on vast datasets that may include portions of publicly available code repositories, raising the risk that the model might recall specific details from its training. This could lead to biased input extractions that do not reflect genuine discovery of relevant data. Although the inputs extracted in our study are abstract and generic, and further processed to minimize exposure of sensitive information, this limitation should still be considered when interpreting the results. Future studies could further mitigate this risk by anonymizing project-specific identifiers or employing obfuscation techniques during input extraction, ensuring that LLMs work with data more representative of unseen scenarios.

\subsection{Limitations and Future Work}

Our study has several limitations that need to be acknowledged. One of the key limitations is the focus on a limited set of coverage metrics, primarily bug detection and code coverage. While these metrics are crucial, they do not provide a complete picture of test suite quality. Future studies could incorporate additional metrics such as fault localization, test suite efficiency, and maintainability to provide a more comprehensive assessment of BRMiner's impact.

Another limitation lies in the tokenization and input extraction process used in our experiments. The tokenizer may not have captured all relevant inputs from bug reports and test cases, potentially leading to an underestimation of BRMiner's effectiveness. Conventional tokenizers may miss certain inputs, especially those with complex patterns. While BRMiner is designed to handle complex scenarios where literals may be involved in concatenations or method calls, ensuring that the entire concatenated string or argument values are extracted and used in test generation, there is room for improvement. Future research could explore advanced tokenization techniques or machine learning-based approaches to enhance input extraction accuracy.

Our study was also constrained by the time budget and the number of iterations conducted. Due to resource limitations, we imposed a time budget of three minutes per iteration and conducted a limited number of iterations. This constraint may have impacted the completeness of our results. Increasing the time budget and the number of iterations in future experiments could yield more comprehensive data and potentially reveal additional insights into BRMiner's effectiveness.

The applicability of our findings across diverse software contexts is another area of concern. Our evaluation was limited to projects in the Defects4J dataset, which, while diverse, may not represent all possible software contexts. Future research should expand the evaluation to include a broader range of software systems, programming languages, and domains. This would help validate the generalizability of BRMiner across different types of software and explore any language-specific challenges that may arise.

While BRMiner demonstrates the potential to leverage inputs from bug reports across different projects, the relevance of these inputs may diminish when applied to projects that do not share common libraries, frameworks, or domains. For example, inputs derived from a project utilizing a web application framework may not translate effectively to a project based on entirely different architectural principles. However, in cases where projects share widely used libraries or dependencies—such as Log4J\footnote{\url{https://logging.apache.org/log4j/2.x/index.html}}—the extracted inputs might have utility, as suggested by the AllLit configurations. Although this aspect was not explicitly investigated in this study, it warrants further exploration. Future work could systematically examine the validation and contextual matching of cross-project inputs. Techniques such as dependency analysis, functionality mapping, or domain similarity scoring could help identify relevant inputs and filter out those less applicable. These advancements would enhance BRMiner’s ability to extend its benefits to diverse projects, particularly those with limited bug report histories, and promote more robust and comprehensive testing across varied software contexts.

Additionally, while our study focused on integrating BRMiner with EvoSuite, there is value in exploring how BRMiner can be integrated with other test generation tools and techniques. Broadening the scope of integration could enhance the applicability of our approach and provide insights into its benefits across different test generation paradigms.

Lastly, addressing the challenge of input relevance across different software versions could involve developing an input prioritization system that adapts to changes in software versions. This would help maintain the relevance of test inputs and ensure that BRMiner remains effective even as software evolves. Such advancements could further strengthen BRMiner's utility in dynamic and evolving software environments.
\subsection{Related Work}
\label{subsec:related-work}

\subsubsection{Automated Test Generation}
In this section, we will discuss general approaches to automated test generation, focusing on the tools and methodologies that are most relevant to our study.

\vspace{0.2cm}\noindent\textbf{General Approaches to Automated Test Generation.}
\citet{shamshiri2015automatically} evaluated the effectiveness of automatic test case generators like EvoSuite, Randoop, and AgitarOne for bug detection using the Defects4J dataset in a regression scenario. This study forms a foundation for understanding the efficacy of these tools in generating test cases but did not integrate specific inputs for testing. The study's use of older versions of EvoSuite and Defects4J limits direct comparisons with our work. \citet{almasi2017industrial} conducted experiments with EvoSuite for bug detection on a proprietary financial application, further exploring the tool's capabilities. However, the lack of access to their source code and bug reports prevents direct comparison with our approach.

\vspace{0.2cm}\noindent\textbf{Enhancing Test Generation with Domain-Specific Knowledge.}
TestMiner by ~\citet{ASE-2017-ToffolaSP} proposed enhancing test generation by mining literals from existing tests to identify domain-specific values, particularly for challenging classes. Although this method differs from ours, which extracts literals from bug reports rather than existing tests, it provides valuable insights into improving test generation by incorporating relevant inputs.

\subsubsection{Bug Report-Based Test Generation}
This section will focus on methodologies that utilize bug reports to enhance the test generation process.

\vspace{0.2cm}\noindent\textbf{Compiler Testing with Bug Reports.}
K-Config by \citet{rabin2021configuring} utilized code snippets from GCC bug reports to improve the configuration of the Csmith test generator for compiler testing. By analyzing bug reports, K-Config generated test programs that effectively uncovered compiler bugs. While this study focused on compiler testing rather than general software testing, it highlights the value of integrating bug report information into test generation. LeRe by \citet{zhong2022enriching} took a similar approach by extracting real programs from bug reports to enhance compiler testing. The study introduced differential testing techniques, which could enrich test programs and improve test quality, especially for compilers. Although focused on compilers, this research emphasizes the importance of leveraging bug reports for test generation, aligning with our approach but in a different domain.

\vspace{0.2cm}\noindent\textbf{Performance Testing from Bug Reports.}
PerfLearner by \citet{han2018perflearner} targeted performance bugs by extracting execution commands and input parameters from bug reports. This method improved the detection and understanding of performance bugs by using the extracted information to guide the generation of performance test cases. Although it does not directly involve generating tests for bug detection in the same way our study does, it demonstrates the broader applicability of bug report-based approaches in various testing contexts.

\subsubsection{LLM-Based Inputs for Test Generation}
\vspace{0.2cm}\noindent\textbf{LLMs for Input Generation.}
\citet{liu2024testing} explored the potential of LLMs in generating unusual text inputs that can be used to detect crashes in mobile apps. Their study demonstrates how LLMs can be employed to create diverse and complex inputs, which aligns with our use of LLMs for generating inputs in BRMiner. The insights from this work could inform future enhancements of BRMiner, particularly in generating inputs that cover a broader range of edge cases.

\vspace{0.2cm}\noindent\textbf{White-Box Testing Empowered by LLMs.}
\citet{yang2023white} examine the use of LLMs in white-box compiler fuzzing. Their study shows how LLMs can assist in creating test inputs that expose deeper and more intricate bugs in compilers. This approach shares similarities with our work, where LLMs are used to filter and enhance the relevance of inputs extracted from bug reports for test generation, though applied in a different domain.

\vspace{0.2cm}\noindent\textbf{LLMs for Variable Discovery in Metamorphic Testing.}
\citet{tsigkanos2023variable} focuse on the use of LLMs to discover variables that can be used in metamorphic testing of scientific software. This research highlights the ability of LLMs to understand and manipulate domain-specific knowledge, which can be parallelized with our work's use of LLMs to identify and utilize relevant inputs from bug reports for more effective test generation.

\section{Conclusion}
\label{sec:conclusion}
In this paper, we presented \name, an innovative approach that enhances automated test generation by extracting and leveraging relevant inputs from bug reports. By integrating these inputs into tools like EvoSuite and Randoop, \name aims to improve bug detection and code coverage. Through comprehensive experiments on the Defects4J dataset, we demonstrated \name's effectiveness across a wide range of projects and configurations. Our findings revealed that \name not only detected up to 18 additional bugs compared to baseline configurations but also achieved significant improvements in branch, instruction, method, and line coverage metrics.

Notably, \name showed remarkable adaptability to varying levels of input explicitness in bug reports. In the "Implicit Input" subset, where bug reports lacked direct mentions of inputs, \name still achieved substantial gains in both bug detection and coverage. These results highlight the robustness of \name's input extraction and filtering mechanisms, particularly when leveraging cross-project input diversity and LLM-based filtering. The ability to perform well even in less descriptive bug report scenarios underscores its potential applicability in real-world contexts where bug reports vary in quality and detail.

While \name has shown promise in enhancing test generation, further research is needed to evaluate its utility in dynamic, real-time development environments. Future work will focus on integrating \name into continuous integration pipelines and applying it to live systems where bug reports are actively being resolved. Additionally, exploring \name's capability to manage discrepancies across software versions and adapt to projects with limited or sparse bug report histories will broaden its applicability.

In conclusion, \name represents a significant advancement in bridging the gap between bug reports and automated test generation. By addressing the limitations of traditional input strategies and adapting to diverse bug reporting scenarios, \name demonstrates its potential to improve software quality and reliability. The findings from both the comprehensive and "Implicit Input" subsets highlight \name's versatility and effectiveness, paving the way for its integration into modern software development workflows. This approach sets a foundation for more robust, scalable, and context-aware software testing practices.


\section*{Declarations}

\subsection*{Funding}
This research was funded in whole, or in part, by the Luxembourg National Research Fund (FNR), grant reference AFR PhD bilateral, project reference 17185670. For the purpose of open access, and in fulfilment of the obligations arising from the grant agreement, the author has applied a Creative Commons Attribution 4.0 International (CC BY 4.0) license to any Author Accepted Manuscript version arising from this submission.

\subsection*{Conflict of Interest/Competing Interests}
The authors declare that they have no conflicts of interest or competing interests.

\subsection*{Ethical Approval}
This article does not contain any studies with human participants or animals performed by any of the authors.

\subsection*{Informed Consent}
Not applicable.

\subsection*{Author Contributions}
All authors contributed equally to this work. They jointly conceived the study, conducted the experiments, analyzed the results, and wrote the manuscript. All authors reviewed and approved the final version of the manuscript.

\subsection*{Data Availability}
To promote transparency and facilitate reproducibility, all artifacts related to this study are publicly available at the following repository: 
\begin{center}
    \url{https://anonymous.4open.science/r/BRMiner-C853}
\end{center}
The repository includes the mined test inputs, along with all code and scripts used for the mining approach.






\bibliographystyle{spbasic.bst}       
\bibliography{references}

\begin{thebibliography}{73}
\providecommand{\natexlab}[1]{#1}
\providecommand{\url}[1]{{#1}}
\providecommand{\urlprefix}{URL }
\expandafter\ifx\csname urlstyle\endcsname\relax
  \providecommand{\doi}[1]{DOI~\discretionary{}{}{}#1}\else
  \providecommand{\doi}{DOI~\discretionary{}{}{}\begingroup
  \urlstyle{rm}\Url}\fi
\providecommand{\eprint}[2][]{\url{#2}}

\bibitem[{Almasi et~al.(2017)Almasi, Hemmati, Fraser, Arcuri, and
  Benefelds}]{almasi2017industrial}
Almasi MM, Hemmati H, Fraser G, Arcuri A, Benefelds J (2017) An industrial
  evaluation of unit test generation: Finding real faults in a financial
  application. In: 2017 IEEE/ACM 39th International Conference on Software
  Engineering: Software Engineering in Practice Track (ICSE-SEIP), IEEE, pp
  263--272

\bibitem[{Amatriain(2024)}]{amatriain2024prompt}
Amatriain X (2024) Prompt design and engineering: Introduction and advanced
  methods. arXiv preprint arXiv:240114423

\bibitem[{Arcuri and Fraser(2013)}]{arcuri2013parameter}
Arcuri A, Fraser G (2013) Parameter tuning or default values? an empirical
  investigation in search-based software engineering. Empirical Software
  Engineering 18:594--623

\bibitem[{Artzi et~al.(2011)Artzi, Dolby, Jensen, M{\o}ller, and
  Tip}]{artzi2011framework}
Artzi S, Dolby J, Jensen SH, M{\o}ller A, Tip F (2011) A framework for
  automated testing of javascript web applications. In: Proceedings of the 33rd
  International Conference on Software Engineering, pp 571--580

\bibitem[{Bai et~al.(2022)Bai, Jones, Ndousse, Askell, Chen, DasSarma, Drain,
  Fort, Ganguli, Henighan et~al.}]{bai2022training}
Bai Y, Jones A, Ndousse K, Askell A, Chen A, DasSarma N, Drain D, Fort S,
  Ganguli D, Henighan T, et~al. (2022) Training a helpful and harmless
  assistant with reinforcement learning from human feedback. arXiv preprint
  arXiv:220405862

\bibitem[{Baldoni et~al.(2018)Baldoni, Coppa, D’elia, Demetrescu, and
  Finocchi}]{baldoni2018survey}
Baldoni R, Coppa E, D’elia DC, Demetrescu C, Finocchi I (2018) A survey of
  symbolic execution techniques. ACM Computing Surveys (CSUR) 51(3):1--39

\bibitem[{Bettenburg et~al.(2008)Bettenburg, Premraj, Zimmermann, and
  Kim}]{bettenburg2008extracting}
Bettenburg N, Premraj R, Zimmermann T, Kim S (2008) Extracting structural
  information from bug reports. In: Proceedings of the 2008 international
  working conference on Mining software repositories, pp 27--30

\bibitem[{Bozkurt and Harman(2011)}]{bozkurt2011automatically}
Bozkurt M, Harman M (2011) Automatically generating realistic test input from
  web services. In: Proceedings of 2011 IEEE 6th International Symposium on
  Service Oriented System (SOSE), IEEE, pp 13--24

\bibitem[{Brown et~al.(2020)Brown, Mann, Ryder, Subbiah, Kaplan, Dhariwal,
  Neelakantan, Shyam, Sastry, Askell et~al.}]{brown2020language}
Brown T, Mann B, Ryder N, Subbiah M, Kaplan JD, Dhariwal P, Neelakantan A,
  Shyam P, Sastry G, Askell A, et~al. (2020) Language models are few-shot
  learners. Advances in neural information processing systems 33:1877--1901

\bibitem[{Cadar and Sen(2013)}]{cadar2013symbolic}
Cadar C, Sen K (2013) Symbolic execution for software testing: three decades
  later. Communications of the ACM 56(2):82--90

\bibitem[{Cadar et~al.(2008)Cadar, Ganesh, Pawlowski, Dill, and
  Engler}]{cadar2008exe}
Cadar C, Ganesh V, Pawlowski PM, Dill DL, Engler DR (2008) Exe: Automatically
  generating inputs of death. ACM Transactions on Information and System
  Security (TISSEC) 12(2):1--38

\bibitem[{Cedric~Richter(2022)}]{richter2022tssb}
Cedric~Richter HW (2022) Tssb-3m: Mining single statement bugs at massive
  scale. In: MSR

\bibitem[{Chen et~al.(2023)Chen, Hu, Zhi, Han, Deng, and
  Yin}]{chen2023chatunitest}
Chen Y, Hu Z, Zhi C, Han J, Deng S, Yin J (2023) Chatunitest: A framework for
  llm-based test generation. arXiv e-prints pp arXiv--2305

\bibitem[{Devlin et~al.(2018)Devlin, Chang, Lee, and
  Toutanova}]{devlin2018bert}
Devlin J, Chang MW, Lee K, Toutanova K (2018) Bert: Pre-training of deep
  bidirectional transformers for language understanding. arXiv preprint
  arXiv:181004805

\bibitem[{Elbaum et~al.(2003)Elbaum, Karre, and
  Rothermel}]{elbaum2003improving}
Elbaum S, Karre S, Rothermel G (2003) Improving web application testing with
  user session data. In: 25th International Conference on Software Engineering,
  2003. Proceedings., IEEE, pp 49--59

\bibitem[{Fan et~al.(2023)Fan, Gokkaya, Harman, Lyubarskiy, Sengupta, Yoo, and
  Zhang}]{fan2023large}
Fan A, Gokkaya B, Harman M, Lyubarskiy M, Sengupta S, Yoo S, Zhang JM (2023)
  Large language models for software engineering: Survey and open problems. In:
  2023 IEEE/ACM International Conference on Software Engineering: Future of
  Software Engineering (ICSE-FoSE), IEEE, pp 31--53

\bibitem[{Fazzini et~al.(2018)Fazzini, Prammer, d'Amorim, and
  Orso}]{fazzini2018automatically}
Fazzini M, Prammer M, d'Amorim M, Orso A (2018) Automatically translating bug
  reports into test cases for mobile apps. In: Proceedings of the 27th ACM
  SIGSOFT International Symposium on Software Testing and Analysis, pp 141--152

\bibitem[{Fraser and Arcuri(2011)}]{fraser2011evosuite}
Fraser G, Arcuri A (2011) Evosuite: automatic test suite generation for
  object-oriented software. In: Proceedings of the 19th ACM SIGSOFT symposium
  and the 13th European conference on Foundations of software engineering, pp
  416--419

\bibitem[{Fraser et~al.(2015)Fraser, Staats, McMinn, Arcuri, and
  Padberg}]{fraser2015does}
Fraser G, Staats M, McMinn P, Arcuri A, Padberg F (2015) Does automated unit
  test generation really help software testers? a controlled empirical study.
  ACM Transactions on Software Engineering and Methodology (TOSEM) 24(4):1--49

\bibitem[{Galeotti et~al.(2013)Galeotti, Fraser, and
  Arcuri}]{galeotti2013improving}
Galeotti JP, Fraser G, Arcuri A (2013) Improving search-based test suite
  generation with dynamic symbolic execution. In: 2013 ieee 24th international
  symposium on software reliability engineering (issre), IEEE, pp 360--369

\bibitem[{Galeotti et~al.(2014)Galeotti, Fraser, and
  Arcuri}]{galeotti2014extending}
Galeotti JP, Fraser G, Arcuri A (2014) Extending a search-based test generator
  with adaptive dynamic symbolic execution. In: Proceedings of the 2014
  international symposium on software testing and analysis, pp 421--424

\bibitem[{Godefroid et~al.(2005)Godefroid, Klarlund, and
  Sen}]{godefroid2005dart}
Godefroid P, Klarlund N, Sen K (2005) Dart: Directed automated random testing.
  In: Proceedings of the 2005 ACM SIGPLAN conference on Programming language
  design and implementation, pp 213--223

\bibitem[{Han et~al.(2018)Han, Yu, and Lo}]{han2018perflearner}
Han X, Yu T, Lo D (2018) Perflearner: Learning from bug reports to understand
  and generate performance test frames. In: Proceedings of the 33rd ACM/IEEE
  international conference on automated software engineering, pp 17--28

\bibitem[{Harman and McMinn(2010)}]{5342440}
Harman M, McMinn P (2010) A theoretical and empirical study of search-based
  testing: Local, global, and hybrid search. IEEE Transactions on Software
  Engineering 36(2):226--247, \doi{10.1109/TSE.2009.71}

\bibitem[{Just et~al.(2014)Just, Jalali, and Ernst}]{just2014defects4j}
Just R, Jalali D, Ernst MD (2014) Defects4j: A database of existing faults to
  enable controlled testing studies for java programs. In: Proceedings of the
  2014 international symposium on software testing and analysis, pp 437--440

\bibitem[{King(1976)}]{king1976symbolic}
King JC (1976) Symbolic execution and program testing. Communications of the
  ACM 19(7):385--394

\bibitem[{Kochhar et~al.(2013{\natexlab{a}})Kochhar, Bissyand{\'e}, Lo, and
  Jiang}]{kochhar2013adoption}
Kochhar PS, Bissyand{\'e} TF, Lo D, Jiang L (2013{\natexlab{a}}) Adoption of
  software testing in open source projects--a preliminary study on 50,000
  projects. In: 2013 17th european conference on software maintenance and
  reengineering, IEEE, pp 353--356

\bibitem[{Kochhar et~al.(2013{\natexlab{b}})Kochhar, Bissyand{\'e}, Lo, and
  Jiang}]{kochhar2013empirical}
Kochhar PS, Bissyand{\'e} TF, Lo D, Jiang L (2013{\natexlab{b}}) An empirical
  study of adoption of software testing in open source projects. In: 2013 13th
  International Conference on Quality Software, IEEE, pp 103--112

\bibitem[{Kochhar et~al.(2015)Kochhar, Thung, Nagappan, Zimmermann, and
  Lo}]{kochhar2015understanding}
Kochhar PS, Thung F, Nagappan N, Zimmermann T, Lo D (2015) Understanding the
  test automation culture of app developers. In: 2015 IEEE 8th International
  Conference on Software Testing, Verification and Validation (ICST), IEEE, pp
  1--10

\bibitem[{Kojima et~al.(2022)Kojima, Gu, Reid, Matsuo, and
  Iwasawa}]{kojima2022large}
Kojima T, Gu SS, Reid M, Matsuo Y, Iwasawa Y (2022) Large language models are
  zero-shot reasoners. Advances in neural information processing systems
  35:22199--22213

\bibitem[{Liu et~al.(2017)Liu, Zhang, Pistoia, Zheng, Marques, and
  Zeng}]{liu2017automatic}
Liu P, Zhang X, Pistoia M, Zheng Y, Marques M, Zeng L (2017) Automatic text
  input generation for mobile testing. In: 2017 IEEE/ACM 39th International
  Conference on Software Engineering (ICSE), IEEE, pp 643--653

\bibitem[{Liu et~al.(2024)Liu, Chen, Wang, Chen, Wu, Tian, Huang, Hu, and
  Wang}]{liu2024testing}
Liu Z, Chen C, Wang J, Chen M, Wu B, Tian Z, Huang Y, Hu J, Wang Q (2024)
  Testing the limits: Unusual text inputs generation for mobile app crash
  detection with large language model. In: Proceedings of the IEEE/ACM 46th
  International Conference on Software Engineering, pp 1--12

\bibitem[{Long(2023)}]{long2023large}
Long J (2023) Large language model guided tree-of-thought. arXiv preprint
  arXiv:230508291

\bibitem[{Macedo et~al.(2024)Macedo, Tian, Cogo, and
  Adams}]{macedo2024exploring}
Macedo M, Tian Y, Cogo FR, Adams B (2024) Exploring the impact of the output
  format on the evaluation of large language models for code translation. arXiv
  preprint arXiv:240317214

\bibitem[{Majumdar and Xu(2007)}]{majumdar2007directed}
Majumdar R, Xu RG (2007) Directed test generation using symbolic grammars. In:
  Proceedings of the 22nd IEEE/ACM International Conference on Automated
  Software Engineering, pp 134--143

\bibitem[{Mariani et~al.(2014)Mariani, Pezz{\`e}, Riganelli, and
  Santoro}]{mariani2014link}
Mariani L, Pezz{\`e} M, Riganelli O, Santoro M (2014) Link: exploiting the web
  of data to generate test inputs. In: Proceedings of the 2014 International
  Symposium on Software Testing and Analysis, pp 373--384

\bibitem[{McMinn et~al.(2012)McMinn, Shahbaz, and Stevenson}]{mcminn2012search}
McMinn P, Shahbaz M, Stevenson M (2012) Search-based test input generation for
  string data types using the results of web queries. In: 2012 IEEE Fifth
  International Conference on Software Testing, Verification and Validation,
  IEEE, pp 141--150

\bibitem[{Milani~Fard et~al.(2014)Milani~Fard, Mirzaaghaei, and
  Mesbah}]{milani2014leveraging}
Milani~Fard A, Mirzaaghaei M, Mesbah A (2014) Leveraging existing tests in
  automated test generation for web applications. In: Proceedings of the 29th
  ACM/IEEE international conference on Automated software engineering, pp
  67--78

\bibitem[{Naveed et~al.(2023)Naveed, Khan, Qiu, Saqib, Anwar, Usman, Barnes,
  and Mian}]{naveed2023comprehensive}
Naveed H, Khan AU, Qiu S, Saqib M, Anwar S, Usman M, Barnes N, Mian A (2023) A
  comprehensive overview of large language models. arXiv preprint
  arXiv:230706435

\bibitem[{Pacheco et~al.(2007)Pacheco, Lahiri, Ernst, and
  Ball}]{pacheco2007feedback}
Pacheco C, Lahiri SK, Ernst MD, Ball T (2007) Feedback-directed random test
  generation. In: 29th International Conference on Software Engineering
  (ICSE'07), IEEE, pp 75--84

\bibitem[{Panichella et~al.(2015)Panichella, Kifetew, and
  Tonella}]{panichella2015reformulating}
Panichella A, Kifetew FM, Tonella P (2015) Reformulating branch coverage as a
  many-objective optimization problem. In: 2015 IEEE 8th international
  conference on software testing, verification and validation (ICST), IEEE, pp
  1--10

\bibitem[{Panichella et~al.(2017)Panichella, Kifetew, and
  Tonella}]{panichella2017automated}
Panichella A, Kifetew FM, Tonella P (2017) Automated test case generation as a
  many-objective optimisation problem with dynamic selection of the targets.
  IEEE Transactions on Software Engineering 44(2):122--158

\bibitem[{Perera et~al.(2020)Perera, Aleti, B{\"o}hme, and
  Turhan}]{perera2020defect}
Perera A, Aleti A, B{\"o}hme M, Turhan B (2020) Defect prediction guided
  search-based software testing. In: Proceedings of the 35th IEEE/ACM
  International Conference on Automated Software Engineering, pp 448--460

\bibitem[{Pradel and Gross(2012)}]{pradel2012fully}
Pradel M, Gross TR (2012) Fully automatic and precise detection of thread
  safety violations. In: Proceedings of the 33rd ACM SIGPLAN conference on
  Programming Language Design and Implementation, pp 521--530

\bibitem[{Rabin and Alipour(2021)}]{rabin2021configuring}
Rabin MRI, Alipour MA (2021) Configuring test generators using bug reports: a
  case study of gcc compiler and csmith. In: Proceedings of the 36th Annual ACM
  Symposium on Applied Computing, pp 1750--1758

\bibitem[{Raffel et~al.(2020)Raffel, Shazeer, Roberts, Lee, Narang, Matena,
  Zhou, Li, and Liu}]{raffel2020exploring}
Raffel C, Shazeer N, Roberts A, Lee K, Narang S, Matena M, Zhou Y, Li W, Liu PJ
  (2020) Exploring the limits of transfer learning with a unified text-to-text
  transformer. Journal of machine learning research 21(140):1--67

\bibitem[{Reynolds and McDonell(2021)}]{reynolds2021prompt}
Reynolds L, McDonell K (2021) Prompt programming for large language models:
  Beyond the few-shot paradigm. In: Extended Abstracts of the 2021 CHI
  Conference on Human Factors in Computing Systems, pp 1--7

\bibitem[{Sahoo et~al.(2024)Sahoo, Singh, Saha, Jain, Mondal, and
  Chadha}]{sahoo2024systematic}
Sahoo P, Singh AK, Saha S, Jain V, Mondal S, Chadha A (2024) A systematic
  survey of prompt engineering in large language models: Techniques and
  applications. arXiv preprint arXiv:240207927

\bibitem[{Sen and Agha(2006)}]{sen2006cute}
Sen K, Agha G (2006) Cute and jcute: Concolic unit testing and explicit path
  model-checking tools: (tool paper). In: Computer Aided Verification: 18th
  International Conference, CAV 2006, Seattle, WA, USA, August 17-20, 2006.
  Proceedings 18, Springer, pp 419--423

\bibitem[{Sen et~al.(2005)Sen, Marinov, and Agha}]{sen2005cute}
Sen K, Marinov D, Agha G (2005) Cute: A concolic unit testing engine for c. ACM
  SIGSOFT Software Engineering Notes 30(5):263--272

\bibitem[{Shahbaz et~al.(2012)Shahbaz, McMinn, and
  Stevenson}]{shahbaz2012automated}
Shahbaz M, McMinn P, Stevenson M (2012) Automated discovery of valid test
  strings from the web using dynamic regular expressions collation and natural
  language processing. In: 2012 12th International Conference on Quality
  Software, IEEE, pp 79--88

\bibitem[{Shamshiri et~al.(2015)Shamshiri, Just, Rojas, Fraser, McMinn, and
  Arcuri}]{shamshiri2015automatically}
Shamshiri S, Just R, Rojas JM, Fraser G, McMinn P, Arcuri A (2015) Do
  automatically generated unit tests find real faults? an empirical study of
  effectiveness and challenges (t). In: 2015 30th IEEE/ACM International
  Conference on Automated Software Engineering (ASE), IEEE, pp 201--211

\bibitem[{Shanahan(2024)}]{shanahan2024talking}
Shanahan M (2024) Talking about large language models. Communications of the
  ACM 67(2):68--79

\bibitem[{Shelke and Nagpure(2014)}]{shelke2014generation}
Shelke S, Nagpure S (2014) Generation of string test input from web using
  regular expression. International Journal of Computer Applications 975:8887

\bibitem[{Si et~al.(2022)Si, Gan, Yang, Wang, Wang, Boyd-Graber, and
  Wang}]{si2022prompting}
Si C, Gan Z, Yang Z, Wang S, Wang J, Boyd-Graber J, Wang L (2022) Prompting
  gpt-3 to be reliable. arXiv preprint arXiv:221009150

\bibitem[{Siddiq et~al.(2024{\natexlab{a}})Siddiq, Dristi, Saha, and
  Santos}]{siddiq2024quality}
Siddiq ML, Dristi S, Saha J, Santos J (2024{\natexlab{a}}) Quality assessment
  of prompts used in code generation. arXiv preprint arXiv:240410155

\bibitem[{Siddiq et~al.(2024{\natexlab{b}})Siddiq, Santos, Tanvir, Ulfat,
  Al~Rifat, and Lopes}]{siddiq2024using}
Siddiq ML, Santos JC, Tanvir RH, Ulfat N, Al~Rifat F, Lopes VC
  (2024{\natexlab{b}}) Using large language models to generate junit tests: An
  empirical study

\bibitem[{Tang et~al.(2024)Tang, Liu, Zhou, and Luo}]{tang2024chatgpt}
Tang Y, Liu Z, Zhou Z, Luo X (2024) Chatgpt vs sbst: A comparative assessment
  of unit test suite generation. IEEE Transactions on Software Engineering

\bibitem[{Toffola et~al.(2017)Toffola, Staicu, and Pradel}]{ASE-2017-ToffolaSP}
Toffola LD, Staicu CA, Pradel M (2017) {Saying 'hi!' is not enough: mining
  inputs for effective test generation}. In: {Proceedings of the 32nd
  International Conference on Automated Software Engineering}, {IEEE Computer
  Society}, pp 44--49, \doi{10.1109/ASE.2017.8115617}

\bibitem[{Tsigkanos et~al.(2023)Tsigkanos, Rani, M{\"u}ller, and
  Kehrer}]{tsigkanos2023variable}
Tsigkanos C, Rani P, M{\"u}ller S, Kehrer T (2023) Variable discovery with
  large language models for metamorphic testing of scientific software. In:
  International Conference on Computational Science, Springer, pp 321--335

\bibitem[{Valle-G{\'o}mez et~al.(2022)Valle-G{\'o}mez,
  Garc{\'\i}a-Dom{\'\i}nguez, Delgado-P{\'e}rez, and
  Medina-Bulo}]{valle2022mutation}
Valle-G{\'o}mez KJ, Garc{\'\i}a-Dom{\'\i}nguez A, Delgado-P{\'e}rez P,
  Medina-Bulo I (2022) Mutation-inspired symbolic execution for software
  testing. IET Software 16(5):478--492

\bibitem[{Vaswani et~al.(2017)Vaswani, Shazeer, Parmar, Uszkoreit, Jones,
  Gomez, Kaiser, and Polosukhin}]{vaswani2017attention}
Vaswani A, Shazeer N, Parmar N, Uszkoreit J, Jones L, Gomez AN, Kaiser {\L},
  Polosukhin I (2017) Attention is all you need. Advances in neural information
  processing systems 30

\bibitem[{Vogelsang and Fischbach(2024)}]{vogelsang2024using}
Vogelsang A, Fischbach J (2024) Using large language models for natural
  language processing tasks in requirements engineering: A systematic
  guideline. arXiv preprint arXiv:240213823

\bibitem[{Wang et~al.(2024)Wang, Huang, Chen, Liu, Wang, and
  Wang}]{wang2024software}
Wang J, Huang Y, Chen C, Liu Z, Wang S, Wang Q (2024) Software testing with
  large language models: Survey, landscape, and vision. IEEE Transactions on
  Software Engineering

\bibitem[{Weeratunge et~al.(2010)Weeratunge, Zhang, and
  Jagannathan}]{weeratunge2010analyzing}
Weeratunge D, Zhang X, Jagannathan S (2010) Analyzing multicore dumps to
  facilitate concurrency bug reproduction. In: Proceedings of the fifteenth
  International Conference on Architectural support for programming languages
  and operating systems, pp 155--166

\bibitem[{Wei et~al.(2022)Wei, Wang, Schuurmans, Bosma, Xia, Chi, Le, Zhou
  et~al.}]{wei2022chain}
Wei J, Wang X, Schuurmans D, Bosma M, Xia F, Chi E, Le QV, Zhou D, et~al.
  (2022) Chain-of-thought prompting elicits reasoning in large language models.
  Advances in neural information processing systems 35:24824--24837

\bibitem[{Xie et~al.(2005)Xie, Marinov, Schulte, and Notkin}]{xie2005symstra}
Xie T, Marinov D, Schulte W, Notkin D (2005) Symstra: A framework for
  generating object-oriented unit tests using symbolic execution. In: Tools and
  Algorithms for the Construction and Analysis of Systems: 11th International
  Conference, TACAS 2005, Held as Part of the Joint European Conferences on
  Theory and Practice of Software, ETAPS 2005, Edinburgh, UK, April 4-8, 2005.
  Proceedings 11, Springer, pp 365--381

\bibitem[{Yang et~al.(2023)Yang, Deng, Lu, Yao, Liu, Jabbarvand, and
  Zhang}]{yang2023white}
Yang C, Deng Y, Lu R, Yao J, Liu J, Jabbarvand R, Zhang L (2023) White-box
  compiler fuzzing empowered by large language models. arXiv preprint
  arXiv:231015991

\bibitem[{Yao et~al.(2024)Yao, Yu, Zhao, Shafran, Griffiths, Cao, and
  Narasimhan}]{yao2024tree}
Yao S, Yu D, Zhao J, Shafran I, Griffiths T, Cao Y, Narasimhan K (2024) Tree of
  thoughts: Deliberate problem solving with large language models. Advances in
  Neural Information Processing Systems 36

\bibitem[{Yenduri et~al.(2024)Yenduri, Ramalingam, Selvi, Supriya, Srivastava,
  Maddikunta, Raj, Jhaveri, Prabadevi, Wang et~al.}]{yenduri2024gpt}
Yenduri G, Ramalingam M, Selvi GC, Supriya Y, Srivastava G, Maddikunta PKR, Raj
  GD, Jhaveri RH, Prabadevi B, Wang W, et~al. (2024) Gpt (generative
  pre-trained transformer)--a comprehensive review on enabling technologies,
  potential applications, emerging challenges, and future directions. IEEE
  Access

\bibitem[{Yu et~al.(2017)Yu, Zaman, and Wang}]{yu2017descry}
Yu T, Zaman TS, Wang C (2017) Descry: reproducing system-level concurrency
  failures. In: Proceedings of the 2017 11th Joint Meeting on Foundations of
  Software Engineering, pp 694--704

\bibitem[{Zhong(2022)}]{zhong2022enriching}
Zhong H (2022) Enriching compiler testing with real program from bug report.
  In: Proceedings of the 37th IEEE/ACM International Conference on Automated
  Software Engineering, pp 1--12

\bibitem[{Ziegler et~al.(2019)Ziegler, Stiennon, Wu, Brown, Radford, Amodei,
  Christiano, and Irving}]{ziegler2019fine}
Ziegler DM, Stiennon N, Wu J, Brown TB, Radford A, Amodei D, Christiano P,
  Irving G (2019) Fine-tuning language models from human preferences. arXiv
  preprint arXiv:190908593

\end{thebibliography}

\end{document}